\documentclass[aps,reprint,nofootinbib,nobibnotes,notitlepage,superscriptaddress,onecolumn,prd,
 amsmath,amssymb
]{revtex4-2}

\usepackage[caption=false]{subfig}
\usepackage{braket}
\usepackage{float}
\usepackage{lipsum}
\usepackage{graphicx}% include figure files
\usepackage{dcolumn}% align table columns on decimal point
\usepackage{bm}% bold math
\usepackage{natbib}
\usepackage{hyperref}
\hypersetup{
	colorlinks = true,
    linkcolor = Red,
    urlcolor  = Red,
    citecolor = Red
}

\usepackage{microtype}

\usepackage{verbatim}
\usepackage[amssymb]{SIunits}
\usepackage{tabularx}
\usepackage[dvipsnames]{xcolor}
\usepackage{wasysym}

\usepackage{tikz}

\def\be{\begin{equation}}
\def\ee{\end{equation}}

\usepackage{color}
\definecolor{darkgreen}{RGB}{0,120,0}

\definecolor{darkgreen}{RGB}{0,120,0}

\newcommand{\resub}[1]{{#1}}%\textcolor{darkgreen}{#1}}}

\newcommand{\av}[1]{\left\langle{#1}\right\rangle}

\newcommand{\vx}{\vec x}

\newcommand{\C}{\mathsf{C}}

\newcommand{\B}{\mathsf{B}}
\newcommand{\T}{\mathsf{T}}

\newcommand{\Tr}[1]{\operatorname{Tr}\left[{#1}\right]}

\newcommand{\hn}{\hat{\vec n}}

\newcommand{\tjo}[3]{\begin{pmatrix} {#1} & {#2} & {#3}\\ 0 & 0 & 0\end{pmatrix}}
\newcommand{\tj}[6]{\begin{pmatrix} {#1} & {#2} & {#3}\\ {#4} & {#5} & {#6}\end{pmatrix}}

\def\beq{\begin{eqnarray}}
\def\eeq{\end{eqnarray}}

\let\vec\mathbf

\usepackage{empheq}

\newcommand*\widefbox[1]{\fbox{\hspace{2em}#1\hspace{2em}}}

\def\mnras{\rm{MNRAS}}

\def\aap{\rm{A\&A}}

\def\apj{\rm{ApJ}}                 % Astrophysical Journal
\def\jcap{\rm{JCAP}}

\begin{document}

\title{Optimal Estimation of the Binned Mask-Free Power Spectrum, Bispectrum, and Trispectrum on the Full Sky: \resub{Scalar Edition}}
%\date{\today}

\author{Oliver~H.\,E.~Philcox}
\email{ohep2@cantab.ac.uk}
\affiliation{Center for Theoretical Physics, Department of Physics,
Columbia University, New York, NY 10027, USA}
\affiliation{Simons Society of Fellows, Simons Foundation, New York, NY 10010, USA}

\begin{abstract} 
    \noindent We derive optimal estimators for the two-, three-, and four-point correlators of statistically isotropic scalar fields defined on the sphere, such as the Cosmic Microwave Background temperature fluctuations, allowing for arbitrary (linear) masking and inpainting schemes. In each case, we give the optimal unwindowed estimator (obtained via a maximum-likelihood prescription, with an associated Fisher deconvolution matrix), and an idealized form, and pay close attention to their efficient computation. For the trispectrum, we include both parity-even and parity-odd contributions, as allowed by symmetry. The estimators can include arbitrary weighting of the data (and remain unbiased), but are shown to be optimal in the limit of inverse-covariance weighting and Gaussian statistics. The normalization of the estimators is computed via Monte Carlo methods, with the rate-limiting steps (involving spherical harmonic transforms) scaling linearly with the number of bins. An accompanying code package, \href{https://github.com/oliverphilcox/PolyBin}{\textsc{PolyBin}}, implements these estimators in \textsc{python}, and we demonstrate the estimators' efficacy via a suite of validation tests.
\end{abstract}

\maketitle

\section{Introduction}
\label{sec: intro}

%\oliver{make code public}
From statistical chemistry to cosmology, the physical sciences abound with examples of random fields. In many instances, the particular realization of the field (\textit{i.e.}\ its microstate) is not of physical relevance but the distribution from which it is drawn (\textit{i.e.}\ its macrostate) contains valuable information. For example, the precise positions of atoms in a solid are rarely of use, but their distribution encodes the physical properties of the material; likewise, whilst we care not about individual galaxy positions, their ensemble statistics can teach us about the primordial Universe. To understand such systems, we therefore require robust ways of characterizing the statistical properties of random fields.

Perhaps the most well-known statistics are the \textit{correlation functions}. These encode the correlations between the (continuous or discrete) field at different points in space, and, if the system is sufficiently large, can be estimated via spatial averaging. Mathematically, an $N$-point correlator of some field $a$ takes the form \citep[e.g.][]{peebles80}
\beq
    \zeta^{(N)}(\vx_1,\ldots,\vx_N) = \av{a(\vx_1)\cdots a(\vx_N)}, 
\eeq
where $\{\vx_i\}$ are some positions of relevance, and $\av{\cdot}$ represents an ensemble average over realizations of $a$. The simplest statistic is the two-point function (or, in Fourier-space, the power spectrum); this has been used to characterize phenomena as diverse as cell biology, quantum chemistry, and astrophysics. Though this is sufficient for some applications, the rich phenomenology of nature often leads us to consider also the higher-order correlators \citep[e.g.,][]{Philcox:2022yri,OH-3pcf,Philcox:2021hbm,berryman1988,dotsenko1991,hwang1993,sanda2005}.

In this work, we will focus on random fields defined on the two-sphere. We will further specialize to \textit{statistically isotropic} phenomena, \textit{i.e.}\ those whose correlators are invariant under global rotations. Such fields can be conveniently described by working in a spherical harmonic basis: efficient measurement of the harmonic-space correlators, or \textit{polyspectra}, will be the subject of the next thirty pages. Geophysics and cosmology provide a number of examples of statistically isotropic fields: for example, the set of galaxies at some fixed distance from Earth naturally lies on a spherical shell, and, by the Copernican principle, there are no special directions on large scales \citep[e.g.,][]{geoP,1999coph.book.....P}. In the cosmological case, polyspectra with $N>2$ are of particular relevance, since they are predicted to vanish in the simplest models of inflation (barring a number of secondary effects arising at late times) \citep{Guth:1980zm,Linde:1981mu}, thus we will focus primarily on the extraterrestrial case in this work. By measuring the cosmic bispectrum, trispectrum, and beyond, we can thus probe primordial physics, which operates at energy scales vastly in excess of those encountered on Earth.

Measuring correlators beyond the power spectrum is, in general, a difficult task, with na\"ive estimation of an $N$-point correlator scaling exponentially with $N$. For this reason, most cosmological analyses have opted not to measure the full statistic, but to constrain a small number of parameters, corresponding to the amplitudes of specific physical templates \citep[e.g.,][]{Komatsu:2003iq,Komatsu:2001wu,Planck:2018jri,Duivenvoorden:2019ses,Planck:2019kim,Heavens:1998jb,Yadav:2007ny,Munshi:2009wy,Munshi:2009ik}. Whilst this is useful for some studies, it does not facilitate general exploration of the statistic, nor probing physical effects whose forms are not \textit{a priori} known \citep[e.g.,][]{Coulton:2019bnz}. To this end, it is useful to also \resub{measure} the full spectra, projected onto some set of $\ell$-space bins \citep[e.g.,][]{Bucher:2009nm,Bucher:2015ura}, or via some modal decomposition (which parametrize the space using smooth mode functions rather than discrete bins) \citep[e.g.,][]{2009PhRvD..80d3510F,Fergusson:2010gn,Regan:2010cn,2011arXiv1105.2791F,Fergusson:2014gea,2012JCAP...12..032F}.

A second complication arises from observational effects. Usually, one cannot measure the field at all points on the sphere, thus the observed field is modulated by some \textit{mask} (also known as a \textit{window function}), $W$, depending on the galactic plane, experimental limitations, bright stars, noise-dominated regions, \textit{et cetera}. Conventional polyspectrum estimators measure the correlators (known as \textit{pseudo}-spectra) of the masked field $W\,a$ rather than the true correlators of $a$, which complicates their interpretation, particularly given that $W$ is rarely isotropic. To robustly compare measurement and model, one must either deconvolve the measurement (also taking into account spatially-varying noise), or convolve the theory \citep[e.g.,][]{Hivon:2001jp,1997PhRvD..55.5895T}. Both of these are complex for statistics beyond the power spectrum \citep[e.g.,][]{2012JCAP...12..032F}, leading to a number of analyses ignoring the window effects, with potentially dangerous consequences.

The above discussion motivates the development of robust and efficient estimators for binned polyspectra on the two-sphere. To this end, we will build on a variety of tools developed throughout the last thirty years. First, efficient estimation of (windowed) polyspectra has been considered in a number of works, including \citep[e.g.,][]{1997PhRvD..55.5895T,1997ApJ...480...22T,Hivon:2001jp,Madhavacheril:2020ido,Gruetjen:2015sta} for the power spectrum, \citep{2011MNRAS.417....2S,Bucher:2009nm,2009PhRvD..80d3510F,2012JCAP...12..032F,Shiraishi:2014roa,Coulton:2019bnz,Bucher:2015ura,Komatsu:2001wu,Yadav:2007ny,2003MNRAS.341..623S,Duivenvoorden:2019ses,Komatsu:2003iq,2000PhRvD..62j3004G} for the bispectrum and \citep{2015arXiv150200635S,2011MNRAS.412.1993M,Regan:2010cn,2011arXiv1105.2791F,Fergusson:2010gn,Marzouk:2022utf,Smidt:2010ra,Kamionkowski:2010me,Mizuno:2010by,Fergusson:2014gea,Smith:2012ty,Troja:2014yqa} for the trispectrum, primarily for the amplitudes of specific separable shapes. Other works have considered modal approaches to measuring the bispectrum (often for the purpose of estimating specific non-separable templates) and higher-order statistics, and some works have considered binned polyspectra directly \citep[e.g.,][]{Coulton:2019bnz,Shiraishi:2014roa}. Here, we opt to use bins rather than modal decompositions for general interpretability (given that we are not concerned with individual models); the latter may provide a more efficient compressed basis in practice however, though we caution that the various modes are not independent, even in ideal scenarios, and, furthermore, it may be non-trivial to project the theory models onto the relevant basis. 

Self-consistent treatment of the mask in higher-order polyspectra is a novel feature of this work: to achieve this, we will use maximum-likelihood prescriptions, whereupon one first writes down the likelihood for the observed field (which depends both on $W$ and the statistical properties of $a$), then maximizes analytically to find an optimal estimator for the statistic of interest. Such estimators are unbiased (\textit{i.e.}\ their mean is not affected by the window) and avoid the need to include the mask in the theory model. This approach has been previously considered for the power spectrum \citep{1997PhRvD..55.5895T,Chen:2021fum,Bond:1998zw,Oh:1998sr,Hamilton:1999uw,Hamilton:2005kz,Hamilton:2005ma}, as well as the two- and three-point statistics of three-dimensional fields \citep{Philcox:2021ukg,2021PhRvD.103j3504P,1997ApJ...480...22T,1998ApJ...499..555T,Philcox:2021eeh,Beutler:2021eqq}, but, to our knowledge, ours is the first such treatment for higher-order spectra on the sphere. In this work, we will pay particular attention to trispectra, which have been rarely measured directly. Unlike the lower-order correlators, these can be decomposed into two pieces, which are even and odd under parity transformations: the latter has not been previously measured in two-dimensional cosmology, and, in our accompanying work \citep{PhilcoxCMB}, we will use it to test the recent claims of parity-violation in large scale structure \citep{Philcox:2022hkh,Hou:2022wfj,Cabass:2022oap,Cahn:2021ltp}. Finally, we release public code alongside this manuscript which implements all the above estimators (both in full generality, and a simplified form); we envisage that this will facilitate robust analysis of general higher-order correlators in cosmology and beyond.

\vskip 8 pt
The remainder of this work is as follows. In \S\ref{sec: ideal-binned} we set out our definitions for the binned polyspectra, before giving a general discussion of ideal estimators in \S\ref{sec: lik}. In \S\ref{sec: Cl},\,\ref{sec: bl}\,\&\,\ref{sec: tl}, we derive estimators for the binned power spectrum, bispectrum, and trispectrum, giving both the idealized form and the optimal unwindowed estimator in each case. Finally, we verify the estimators numerically in \S\ref{sec: testing} before concluding in \S\ref{sec: conclusion}. To guide the reader through this (necessarily dense) paper, we indicate key equations with boxes, and summarize the relevant estimators at the end of each section. Each estimator is implemented in the public code \textsc{PolyBin}: an extensive tutorial can be found on GitHub.\footnote{\href{https://github.com/oliverphilcox/PolyBin}{GitHub.com/oliverphilcox/PolyBin}}

\section{Ideal Binned Polyspectra}\label{sec: ideal-binned}
We begin by defining our conventions for the fields and polyspectra used in this work, and present a number of results used in the remainder of this work. In general, we will work with scalar fields defined on the two-sphere, such as the atmospheric pressure on Earth or the CMB temperature fluctuations. A general zero-mean signal, labelled $a(\hn)$ can be expanded in spherical harmonics thus:
\beq
    a(\hn) \equiv \sum_{\ell=0}^\infty\sum_{m=-\ell}^{\ell} a_{\ell m}Y_{\ell m}(\hn) \qquad \Leftrightarrow \qquad a_{\ell m} = \int_{\mathbb{S}^2} d\hn\,a(\hn)Y_{\ell m}^*(\hn)
\eeq
where we will keep the summation limits and integration domains implicit henceforth. In many cases, we perform noisy observations of this signal, yielding the observed field, $\tilde a$, defined as
\beq\label{eq: windowed-field-def}
    \tilde a(\hn) \equiv W(\hn)a(\hn)+n(\hn)
\eeq
where $W(\hn)$ is some deterministic window (or mask), defining how various regions of the sphere are observed and $n(\hn)$ is a stochastic noise field. In general, we will denote masked fields with a tilde. Note that we assume $W$ and $a$ to be uncorrelated, such that $\av{Wa}=\av{W}\av{a}=0$; violation of this assumption will significantly complicate the estimators \citep[e.g.,][]{Lembo:2021kxc}.

\subsection{Ideal Correlators}\label{subsec: ideal-correlators}
The power spectrum, $C$, of the signal field can be written
\beq\label{eq: C-l-def}
    \boxed{\av{a_{\ell_1m_1}a_{\ell_2m_2}} \equiv C^{\ell_1\ell_2}_{m_1m_2} \to (-1)^{m_1}\delta^{\rm K}_{\ell_1\ell_2}\delta^{\rm K}_{m_1(-m_2)}C_\ell,}
\eeq
where the angle brackets indicate an average over statistical realizations of the signal, and we have statistical isotropy and homogeneity to obtain the second expression. Whilst this assumption is usually valid for the underlying signal, $a$, realistic noise and masks are often anisotropic, thus the diagonal approximation cannot be used.

Similarly, the bispectrum, $B$, takes the form
\beq\label{eq: B-l-def}
    \boxed{\av{a_{\ell_1m_1}a_{\ell_2m_2}a_{\ell_3m_3}} \equiv B^{\ell_1\ell_2\ell_3}_{m_1m_2m_3} \to \mathcal{G}^{\ell_1\ell_2\ell_3}_{m_1m_2m_3}\,b_{\ell_1\ell_2\ell_3},}
\eeq
where the RHS holds under isotropic and homogeneous assumptions, as before, and we have defined the \textit{reduced} bispectrum $b_{\ell_1\ell_2\ell_3}$. This is symmetric under any permutation of indices, and requires $|\ell_1-\ell_2|\leq \ell_3\leq \ell_1+\ell_2$. \eqref{eq: B-l-def} involves the Gaunt function, defined as the average over three spherical harmonics
\beq\label{eq: gaunt-def}
    \mathcal{G}^{\ell_1\ell_2\ell_3}_{m_1m_2m_3} &\equiv &\sqrt{\frac{(2\ell_1+1)(2\ell_2+1)(2\ell_3+1)}{4\pi}}\tj{\ell_1}{\ell_2}{\ell_3}{m_1}{m_2}{m_3}\tjo{\ell_1}{\ell_2}{\ell_3}\\\nonumber
    &\equiv& \int d\hn\,Y_{\ell_1m_1}(\hn)Y_{\ell_2m_2}(\hn)Y_{\ell_3m_3}(\hn),
\eeq
where the $3\times 2$ matrices are Wigner $3j$ symbols. For an isotropic real scalar field, $a$, the bispectrum is parity-even, and thus vanishes unless $\ell_1+\ell_2+\ell_3$ is even (which is enforced by the Gaunt integral).\footnote{For anisotropic signals, such as galactic dust, non-zero parity-odd bispectra can exist. To compute these, one can use a modified definition of the reduced bispectrum, as discussed in \citep{Shiraishi:2014roa,Coulton:2019bnz}.}

Finally, we can define a trispectrum, $T$, of $a$ via 
\beq\label{eq: T-l-def}
    \av{a_{\ell_1m_1}a_{\ell_2m_2}a_{\ell_3m_3}a_{\ell_4m_4}}_c \equiv T^{\ell_1\ell_2\ell_3\ell_4}_{m_1m_2m_3m_4}.
\eeq
where we take only the \textit{connected} part of the correlator. In this case, the rotationally invariant decomposition is less straightforward, since the reduced trispectrum cannot be fully described by four $\ell$-modes: rather we must introduce also a diagonal element, $L$. As discussed in \citep{Regan:2010cn}, we can introduce the non-redundant function $T^{\ell_1\ell_2}_{\ell_3\ell_4}(L)$, via
\beq\label{eq: t-l-def-regan}
    T^{\ell_1\ell_2\ell_3\ell_4}_{m_1m_2m_3m_4}\to \sum_{L=0}^\infty\sum_{M=-L}^L(-1)^M\tj{\ell_1}{\ell_2}{L}{m_1}{m_2}{-M}\tj{\ell_3}{\ell_4}{L}{m_3}{m_4}{M}\,T^{\ell_1\ell_2}_{\ell_3\ell_4}(L),
\eeq
summing over the diagonal, $L$, and its azimuthal component, and noting that $M=m_1+m_2=-m_3-m_4$. This has a number of non-trivial symmetries, in particular:
\beq
    T^{\ell_2\ell_1}_{\ell_3\ell_4}(L) = (-1)^{\ell_1+\ell_2+L}T^{\ell_1\ell_2}_{\ell_3\ell_4}(L), \qquad T^{\ell_3\ell_4}_{\ell_1\ell_2}(L) = T^{\ell_1\ell_2}_{\ell_3\ell_4}(L).
\eeq
For our purpose, it will be useful to introduce a new trispectrum, $t^{\ell_1\ell_2}_{\ell_3\ell_4}$, via the symmetric definition (similar to \citep{Shiraishi:2014roa} for the bispectrum):
\beq\label{eq: t-l-def}
    \boxed{\av{a_{\ell_1m_1}a_{\ell_2m_2}a_{\ell_3m_3}a_{\ell_4m_4}}_c\equiv T^{\ell_1\ell_2\ell_3\ell_4}_{m_1m_2m_3m_4}\to \sum_{L=0}^\infty\sum_{M=-L}^L(-1)^Mw^{L(-M)}_{\ell_1\ell_2m_1m_2}w^{LM}_{\ell_3\ell_4m_3m_4}\,t^{\ell_1\ell_2}_{\ell_3\ell_4}(L)+\text{23 perms},}
\eeq
summing over twenty-four permutations of $\{\ell_1,\ell_2,\ell_3,\ell_4\}$. This involves a new weighting function, akin to the Gaunt function:
\beq\label{eq: gaunt-spin-def}
    w^{LM}_{\ell_1\ell_2m_1m_2} &\equiv &\resub{\sqrt{\frac{(2\ell_1+1)(2\ell_2+1)(2L+1)}{4\pi}}\tj{\ell_1}{\ell_2}{L}{m_1}{m_2}{M}\tj{\ell_1}{\ell_2}{L}{-1}{-1}{2}}\\\nonumber
    &\equiv& \int d\hn\,{}_{+1}Y_{\ell_1m_1}(\hn){}_{+1}Y_{\ell_2m_2}(\hn){}_{-2}Y_{LM}(\hn), 
\eeq
which we have written as an integral over three spin-weighted spherical harmonics, ${}_{s}Y_{\ell m}(\hn)$, in the final line. Note that this is symmetric under interchange of \resub{$(\ell_1,m_1)$ and $(\ell_2,m_2)$}. The reason for the spin-weighting adopted in $w$ will be explained in \S\ref{subsubsec: t4-ideal}.

The reduced trispectrum obeys the symmetries
\beq\label{eq: t-syms}
    t^{\ell_2\ell_1}_{\ell_3\ell_4}(L) = t^{\ell_1\ell_2}_{\ell_3\ell_4}(L), \qquad t^{\ell_3\ell_4}_{\ell_1\ell_2}(L)=t^{\ell_1\ell_2}_{\ell_3\ell_4}(L);
\eeq
to fully specify the trispectrum, we thus require only values with $\ell_1\leq \ell_2$, $\ell_3\leq \ell_4$, $\ell_3\leq \ell_1$ and, if $\ell_1=\ell_3$, $\ell_2\leq \ell_4$. The diagonal, $L$, satisfies the triangle conditions $|\ell_1-\ell_2|\leq L\leq \ell_1+\ell_2$ and $|\ell_3-\ell_4|\leq L\leq \ell_3+\ell_4$, due to the $3j$ symbols. Finally, under conjugation and parity-inversion, the trispectrum satisfies
\beq\label{eq: trispectrum-charge-parity}
    \left[t^{\ell_1\ell_2}_{\ell_3\ell_4}(L)\right]^* &=& (-1)^{\ell_1+\ell_2+\ell_3+\ell_4}t^{\ell_1\ell_2}_{\ell_3\ell_4}(L), \qquad \mathbb{P}\left[t^{\ell_1\ell_2}_{\ell_3\ell_4}(L)\right] = (-1)^{\ell_1+\ell_2+\ell_3+\ell_4}t^{\ell_1\ell_2}_{\ell_3\ell_4}(L),
\eeq
respectively; as such, trispectra with even (odd) $\ell_1+\ell_2+\ell_3+\ell_4$ are parity-even (parity-odd) and purely real (imaginary).

\subsection{Binning}\label{subsec: binning}
In this work, we will focus on computing polyspectra in some set of $\ell$-bins, whose formulation we now turn to. An alternative approach would be to project the spectra onto some type of modal decomposition \citep[e.g.,][]{Shiraishi:2014roa,Regan:2010cn,Fergusson:2010gn,2012JCAP...12..032F,2011arXiv1105.2791F}. Our approach has the benefit that the output spectra can be directly compared to theory, in the limit of narrow bins (see \citep{Hivon:2001jp} for techniques going beyond this limit). To include binning, we introduce the (arbitrary) binning function $\Theta_\ell(b)$, which is usually defined to be unity if $\ell$ is in bin $b$ and zero else. For the (isotropic and homogeneous) power spectrum, this leads to the definition
\beq\label{eq: binned-Cl}
    \boxed{\av{a_{\ell_1m_1}a_{\ell_2m_2}}\approx (-1)^{m_1}\delta^{\rm K}_{\ell_1\ell_2}\delta^{\rm K}_{m_1(-m_2)}\sum_b \Theta_{\ell_1}(b)C(b),}
\eeq
where $C(b)$ are the binned quantities we shall construct estimators for.\footnote{Strictly $C(b)$ is related to a sum over the $\ell$-bins rather than being the value at the bin-center. The approximation of \eqref{eq: binned-Cl} is valid for slowly varying $C_\ell$ and suitably narrow bins. Since our focus here is principally on higher-point functions, this is appropriate here.} 

The binned bispectrum, $b(\vec b)$, is similarly defined
\beq\label{eq: binned-bl}
    \boxed{\av{a_{\ell_1m_1}a_{\ell_2m_2}a_{\ell_3m_3}} \approx \mathcal{G}^{\ell_1\ell_2\ell_3}_{m_1m_2m_3}\sum_{b_1b_2b_3}\frac{b(\vec b)}{\Delta_3(\vec b)}\left[\Theta_{\ell_1}(b_1)\Theta_{\ell_2}(b_2)\Theta_{\ell_3}(b_3)+\text{5 perms.}\right],}
\eeq
where $\vec b\equiv \{b_1,b_2,b_3\}$ and the permutations are in $\{\ell_1,\ell_2,\ell_3\}$. To avoid double counting, we restrict the summation to $b_1\leq b_2\leq b_3$ (giving rise to the permutations), and introduce a permutation factor
\beq\label{eq: Delta-3}
    \Delta_3(\vec b) \equiv \begin{cases} 6 & b_1=b_2=b_3 \\ 2 & b_1=b_2\neq b_3 \text{ or } b_1\neq b_2=b_3\\ 1 & \text{else}. \end{cases}
\eeq
This ensures that, in the limit of thin bins, $b(\vec b)$ is equal to the bispectrum evaluated at the bin centers. Note that we can incorporate squeezed triangles into this formalism by allowing a larger $\ell_{\rm max}$ for $\ell_2$ and $\ell_3$ (equivalently $b_2$ and $b_3$) than $\ell_1$.

Finally, we can define the binned trispectrum, $t(\vec b,B)$, by analogy with \eqref{eq: t-l-def}:
\begin{empheq}[box=\widefbox]{align}\label{eq: binned-tl}
  \av{a_{\ell_1m_1}a_{\ell_2m_2}a_{\ell_3m_3}a_{\ell_4m_4}}_c &\approx \sum_{LM}(-1)^Mw^{L(-M)}_{\ell_1\ell_2m_1m_2}w^{LM}_{\ell_3\ell_4m_3m_4}\sum_{\vec b,B}\frac{t(\vec b,B)}{\Delta_4(\vec b)}\Theta_L(B)\\\nonumber
    &\qquad\,\times\,\left[\Theta_{\ell_1}(b_1)\Theta_{\ell_2}(b_2)\Theta_{\ell_3}(b_3)\Theta_{\ell_4}(b_4)+\text{7 perms.}\right]+(2\leftrightarrow3)+(2\leftrightarrow4),
\end{empheq}
summing over the following permutations of $\{1,2,3,4\}\equiv\{b_1,b_2,b_3,b_4\}$: 
\beq
    &\{1,2,3,4\},\{1,2,4,3\},\{2,1,3,4\},\{2,1,4,3\},
    \\\nonumber
    &\{3,4,1,2\},\{3,4,2,1\},\{4,3,1,2\},\{4,3,2,1\},
\eeq
due to the symmetry properties given in \eqref{eq: t-syms}. Here, the trispectrum is defined for all $\vec b\equiv \{b_1,b_2,b_3,b_4\}$ encompassing the external $\{\ell_1,\ell_2,\ell_3,\ell_4\}$ legs and $B$ describing the diagonal $L$. We sum over all $b_1\leq b_2$, $b_3\leq b_4$, $b_1\leq b_3$ and, if $b_3=b_1$, $b_2\geq b_4$, \textit{i.e.} over all independent bins, noting additionally that $\{b_1,b_2,B\}$ and $\{b_3,b_4,B\}$ must satisfy triangle conditions.\footnote{For the parity-odd trispectrum, we can additionally drop bins satisfying $b_1=b_3$ and $b_2=b_4$, which evaluate to zero.} If we wish to include squeezed and doubly-squeezed tetrahedra, we simply extend the $\ell$ ranges to use a larger $\ell_{\rm max}$ for $\ell_2,\ell_4$ and $L$ (due to the triangle conditions), or, if our focus is collapsed tetrahedra, we can use a smaller $\ell_{\rm max}$ for $L$. To avoid double-counting, we introduce the degeneracy factor
\beq
    \Delta_4(\vec b) = \begin{cases} 8 & b_1=b_2=b_3=b_4 \\ 4 & b_1=b_2 \text{ and } b_3=b_4 \\
    2 & b_1 = b_2 \text{ or } b_3=b_4\\
    2 & b_1 = b_3 \text{ and } b_2=b_4\\
    1 & \text{else,}
    \end{cases}
\eeq
which is simply the number of distinct appearances of each term in the above sum over permutations. Finally, we note that we can impose that the trispectrum is parity-even (parity-odd) by adding a factor $[1\pm(-1)^{\ell_{1234}}]/2$ to \eqref{eq: binned-tl}.

\section{Masked Data \& Minimum-Variance Estimators}\label{sec: lik}

We now turn to the problem of estimating the binned polyspectrum coefficients discussed in \S\ref{sec: ideal-binned} from the observed masked data. For this purpose, we will first consider the non-ideal correlators, for which it is useful to work in map-space, rather than harmonic-space. These results may then be used to derive optimal estimators, by maximizing an weakly non-Gaussian likelihood.

\subsection{Non-Ideal Correlators}\label{subsec: nonideal-correlators}
When working with observational data, we may no longer assume rotational symmetry, since the weights, $W$ (encoding the response of the map to the underlying field $a$ ) and the noise, $n$, can be inhomogeneous. In this case, the results of \S\ref{sec: ideal-binned} do not apply. Working in map-space, the two-point correlator of the observed field can be instead written:
\beq\label{eq: tilde-Cij-def}
    \tilde{\mathsf{C}}^{ij} \equiv \av{\tilde a(\hn^i)\tilde a(\hn^j)},
\eeq
where $i,j$ index points on the sky (such as \textsc{healpix} pixels).\footnote{Throughout this work, we will used serif fonts to denote map-space correlators, e.g., $\mathsf{B}$, sans-serif to denote harmonic correlators, e.g., $B$, and lower case for reduced spectra, e.g., $b$. Latin indices $i,j,k,\ldots$ will always denote summation over spatial points. Tildes are added to indicate masked fields.} From the definition of the masked field \eqref{eq: windowed-field-def}, this can be written in terms of the unmasked two-point function, $\mathsf{C}^{ij}\equiv\av{a(\hn^i)a(\hn^j)}$ and the noise $\mathsf{N}^{ij}\equiv \av{n(\hn^i)n(\hn^j)}$:
\beq\label{eq: tilde-Cij-as-Cl}
    \tilde{\mathsf{C}}^{ij} = W(\hn^i)\mathsf{C}^{ij}W(\hn^j)+\mathsf{N}^{ij} = \sum_{\ell m}B_\ell^2C_{\ell}\left[W(\hn^i)Y_{\ell m}(\hn^i)\right]\left[W(\hn^j)Y_{\ell m}^*(\hn^j)\right]+\mathsf{N}^{ij},
\eeq
where we have expanded the true correlator, which \textit{is} rotationally invariant, in terms of \eqref{eq: C-l-def} and additionally included an (isotropic) beam $B_\ell$. When working with discrete data (such as maps in \textsc{HealPix} format), we can additionally include a pixel window function in $B_\ell$ (via $B_\ell\to w_\ell B_\ell$ for window $w_\ell$), to remove the leading dependence on $N_{\rm side}$.

The utility of \eqref{eq: tilde-Cij-as-Cl} is that the windowed correlator is expressed in terms of the quantity we wish to estimate, $C_\ell$ (neglecting binning for now). This is further simplified if one assumes the same window for data and noise: in this case, the masks simply pre- and post-multiply the correlator. In general, the map-level correlator is an $N_{\rm pix}\times N_{\rm pix}$ matrix which is difficult if not impossible to compute explicitly: however, its action on a map can be straightforwardly defined. We will discuss this in \S\ref{subsec: SU-choice}.

The non-ideal three- and four-point correlators take similar forms:
\beq\label{eq: B,T-ijk-def}
    \tilde{\mathsf{B}}^{ijk} \equiv \av{\tilde a(\hn^i)\tilde a(\hn^j)\tilde a(\hn^k)}, \qquad \tilde{\mathsf{T}}^{ijkl} \equiv \av{\tilde a(\hn^i)\tilde a(\hn^j)\tilde a(\hn^k)\tilde a(\hn^l)}
\eeq
As before, these can be straightforwardly written in terms of the map-space ideal correlators ($\mathsf{B}^{ijk}$ and $\mathsf{T}^{ijkl}$), and, via the results of \S\ref{subsec: ideal-correlators}, the binned quantities we wish to measure. Furthermore, whilst they are high-dimensional objects (rank $3$- and $4$-tensors), their action on maps will prove straightforward, due to the internal symmetries in the unwindowed correlators.

\subsection{Optimal Estimators}

Optimal estimators for the binned polyspectra can be derived by maximizing the likelihood of the observed map $\tilde a(\hn)$. Assuming weak non-Gaussianity, this is given by an Edgeworth expansion in terms of the map-space correlators \citep[e.g.,][]{2017arXiv170903452S}:
\beq\label{eq: edgeworth}
    L[\tilde a] \propto \exp\left[-\frac{1}{2}h_i\tilde{\mathsf{C}}^{ij}h_j\right]\left\{1+\frac{1}{3!}\tilde{\mathsf{B}}^{ijk}\mathcal{H}_{ijk}+\frac{1}{4!}\tilde{\mathsf{T}}^{ijkl}\mathcal{H}_{ijkl}+\cdots\right\},
\eeq
where we sum over all repeated indices, such that $\alpha_i\beta^i\equiv \int d\hn\,\alpha(\hn)\beta(\hn)$. Here, we have defined the Wiener-filtered map, $h(\hn)\equiv \left[\tilde{\mathsf{C}}^{-1}\tilde a\right](\hn)$ (recalling that $\tilde{\mathsf{C}}$ contains both signal and noise, and (for now) assuming it to be invertible), as well as the (map-space) Hermite polynomials
\beq\label{eq: hermite}
    \mathcal{H}_{ijk} &\equiv& h_ih_jh_k - (h_i\tilde{\mathsf{C}}^{-1}_{jk}+\text{2 perms.})\\\nonumber
    \mathcal{H}_{ijkl} &\equiv& h_ih_jh_kh_l - (h_ih_j\tilde{\mathsf{C}}^{-1}_{kl}+\text{5 perms.}) + (\tilde{\mathsf{C}}_{ij}^{-1}\tilde{\mathsf{C}}_{kl}^{-1}+\text{2 perms.}).
\eeq

In \eqref{eq: edgeworth}, all cosmology dependence appears through the map-space correlators, $\tilde{\C}$, $\tilde{\B}$, $\tilde{\T}$, which can be related to their binned coefficients, $C(b)$, $b(\vec b)$, $t(\vec b,L)$ using the results of \S\ref{sec: ideal-binned}. To see this, let us consider some binned quantity $x(\vec b)$ arising only in the $N>2$-point correlator, $\tilde{\mathsf{X}}^{i_1\cdots i_N}$. An estimator for $x(\vec b)$ can be obtained by maximizing $\log L[\tilde a](x)$, yielding
\beq\label{eq: general-opt-est}
    \boxed{\widehat{x}(\vec b) \propto \frac{1}{N!}\frac{\partial\tilde{\mathsf{X}}^{i_1\cdots i_N}}{\partial x(\vec b)}\mathcal{H}_{i_1\cdots i_N},}
\eeq
working in the limit of small $x$. The normalization factor (which is, in general, a matrix), can be derived by requiring that the estimator is unbiased, \textit{i.e.} $\mathbb{E}[\widehat{x}(\vec b)]=x(\vec b)$ for expectation operator $\mathbb{E}$. We will refer to its inverse as the \textit{Fisher matrix}, defined as
\beq\label{eq: general-opt-fisher}
    \boxed{\mathcal{F}_N(\vec b, \vec b') = \frac{1}{N!}\frac{\partial\tilde{\mathsf{X}}^{i_1\cdots i_N}}{\partial x(\vec b)}\tilde{\C}^{-1}_{i_1j_1}\cdots \tilde{\C}^{-1}_{i_Nj_N}\frac{\partial\tilde{\mathsf{X}}^{j_1\cdots j_N}}{\partial x(\vec b')},}
\eeq
where we note that all disconnected terms cancel in the expectation of the Hermite tensor $\mathcal{H}_{i_1\cdots i_N}$.

The estimator has the following properties:
\begin{itemize}
    \item \textbf{Unbiased}: This follows from the definition of the Fisher matrix, which ensures $\mathbb{E}[\widehat{x}(\vec b)]=x(\vec b)$. One caveat should be noted: this assumes that the polyspectrum is completely characterized by the set of measured bins $\{x(\vec b)\}$ (which generically include noise contributions), \textit{i.e.} that $\tilde{\mathsf{X}}^{i_1\cdots i_N} = \sum_{\vec b}x(\vec b)\,\left(\partial\tilde{\mathsf{X}}^{i_1\cdots i_N}/\partial x(\vec b)\right)$. Violations of this can occur if there is contribution from modes outside the binning range or unaccounted-for leakage between polyspectra of different parity. For ideal polyspectrum estimators with $N\leq 3$, the Fisher matrix is diagonal, so such effects may be neglected; in the general case, one can ameliorate this \resub{by constructing} the estimator using a slightly larger range of bins than desired in the output data-product, and, if necessary, accounting for leakage between states of different parity (cf.\,\S\ref{subsec: Tl-general}).
    \item \textbf{Window-Free}: Since we consistently include window functions in the $\tilde{\mathsf{X}}$ definitions and take derivatives with respect to the \textit{true} binned correlators, $x(\vec b)$, the measurements are not biased by the window function, \textit{i.e.} the estimators take into account the response of the map to the underlying signal. This lies in contrast to the \textit{pseudo}-spectra measured by direct polyspectrum estimators, and occurs due to the appearance of the mask in the estimator, and the form of the normalization matrix.
    \item \textbf{Optimal}: Since \eqref{eq: general-opt-est} has been derived using maximum-likelihood techniques, it is optimal in the small-correlator limit. As such, the covariance is given by $\mathcal{F}_{N}^{-1}$.\footnote{This is strictly true only for real-valued $x(\vec b)$. Later, we will find that the trispectrum contains an imaginary piece, for which the covariance is $-\mathcal{F}^{-1}_4$. As long as we always take the imaginary part of any such quantities, the above logic holds.}
    \item \textbf{Efficiency} As shown below, the estimator numerators scale at most linearly with the total number of bins in the statistic, $N_{\rm bin}$. When computed using Monte Carlo methods, the rate-determining step in computation of the Fisher matrix is also linear in the number of bins.
\end{itemize}

Whilst the above estimators have significant formal utility, in practice, it will prove useful to consider estimators with a more general choice of weighting, due to the difficulties obtaining accurate noise covariances, $\mathsf{N}^{ij}$, and inverting the covariance $\tilde{\mathsf{C}}^{ij}$. In the below, we will consider a more general choice of weighting to enable efficient computation, defining $h(\hn) = \mathsf{S}^{-1}\tilde a$ instead of $\tilde{\C}^{-1}\tilde a$, where $\mathsf{S}^{-1}$, is some weighting matrix that is not required to be symmetric or invertible. For example, one may wish to project out contaminated areas of a map, which, \textit{a priori}, has a uniform weight; in this case, $\mathsf{S}^{-1}$ would excise regions of the map, and thus be non-invertible, whilst $W$ could be set to the identity operator. Arbitrary linear operations (such as linear inpainting, cf.\,\citep{Gruetjen:2015sta}) can also be included in $\mathsf{S}^{-1}$, as appropriate to the task in question.

Replacing $\mathsf{\tilde C}^{-1}$ with $\mathsf{S}^{-1}$ in \eqref{eq: general-opt-est}\,\&\,\eqref{eq: general-opt-fisher} we obtain an estimator which is always unbiased, and minimum variance in the limit of $\mathsf{S}^{-1}\to{\tilde{\C}}^{-1}$. In the Gaussian regime, the covariance for general (not necessarily invertible) $\mathsf{S}^{-1}$ is given by
\beq
    \mathrm{cov}_N(\vec b,\vec b') = \mathcal{F}_N^{-1}[\mathsf{S}^{-1}]\mathcal{F}_N[\mathsf{S}^{-1}\tilde{\mathsf{C}}\,\mathsf{S}^{-\rm T}]\mathcal{F}_N^{-\rm T}[\mathsf{S}^{-1}],
\eeq
where $\mathcal{F}_N[\mathsf{M}]$ as the Fisher matrix with weighting $\mathsf{M}$ (which may be asymmetric); if $\mathsf{S}^{-1}$ is invertible, the degree of suboptimality is second order in $(\mathsf{S}-\tilde{\C})$ \citep{2011MNRAS.417....2S}, and, if not, an optimal estimator is obtained if $\mathsf{S}^{-1}=\mathsf{S}^{-1}\tilde{\C}\,\mathsf{S}^{-\rm T}$. In the remainder of this work, we consider how such estimators can be efficiently computed.

%\pagebreak
\section{Optimal Power Spectrum Estimation}\label{sec: Cl}
With the above formalism in place, we may now proceed to derive the optimal estimator for the binned full-sky power spectra, $C(b)$, analogous to \citep{1997PhRvD..55.5895T,1998PhRvD..59b7302B,1997ApJ...480...22T}. This is derived in a slightly different manner to the bispectrum and trispectrum estimators discussed below (though ends up taking the same functional form), since the two-point function appears both in the Gaussian likelihood of \eqref{eq: edgeworth}, and in the Wiener filtering. Our estimators can be shown to be equivalent to those of the \textsc{master} formalism in certain limits \citep{Hivon:2001jp} (in particular a uniform mask $W$, and a weighting scheme of the form $[\mathsf{S}^{-1}v](\hn)=s(\hn)v(\hn)$, for some filter $s$, possibly with incomplete support).

Taking derivatives with respect to $C(b)$ (dropping all non-Gaussian correlators) gives
\beq
    \frac{\partial \log L[\tilde a]}{\partial C(b)} &=& \frac{1}{2}\frac{\partial \tilde{\C}^{ij}}{\partial C(b)}h_ih_j-\frac{1}{2}\Tr{\tilde{\C}^{-1}\frac{\partial \tilde{\C}}{\partial C(b)}}\\\nonumber
    \frac{\partial^2 \log L[\tilde a]}{\partial C(b)\partial C(b')} &=& -\left[\frac{\partial \tilde{\C}}{\partial C(b)}\tilde{\C}^{-1}\frac{\partial \tilde{\C}}{\partial C(b)}\right]^{ij}h_ih_j+\frac{1}{2}\Tr{\tilde{\C}^{-1}\frac{\partial \tilde{\C}}{\partial C(b)}\tilde{\C}^{-1}\frac{\partial \tilde{\C}}{\partial C(b')}},
\eeq
noting that $\tilde{\C}^{-1}_{,\alpha} = -\tilde{\C}^{-1}\tilde{\C}_{,\alpha}\tilde{\C}^{-1}$. To derive the optimal estimator, we expand the likelihood to second order around some fiducial spectrum $\overline{C}(b)$, and maximize with respect to the true spectrum $C(b)$, yielding
\beq\label{eq: opt-Cl-1}
    \widehat{C}(b) = \overline{C}(b)+\frac{1}{2}\sum_{b'}\mathcal{F}^{-1}_{2,\rm opt}(b,b')\frac{\partial \tilde{\C}^{ij}}{\partial C(b')}\left[h_ih_j-\tilde{\C}_{ij}^{-1}\right], 
\eeq
defining
\beq
    \mathcal{F}_{2, \rm opt}(b,b')=\frac{1}{2}\Tr{\tilde{\C}^{-1}\frac{\partial \tilde{\C}}{\partial C(b)}\tilde{\C}^{-1}\frac{\partial \tilde{\C}}{\partial C(b')}},
\eeq
and evaluating all quantities at the fiducial spectrum $\overline C$. Our final estimator is formed by replacing $\tilde{\C}^{-1}$ by some generic weighting $\mathsf{S}^{-1}$ (which, as above, need not be symmetric or invertible), and noting that the first and last terms in \eqref{eq: opt-Cl-1} differ only by the noise correlation function, $\mathsf{N}^{ij}$. This gives
\beq\label{eq: opt-Cl-formal}
    \boxed{\widehat{C}(b) = \frac{1}{2}\sum_{b'}\mathcal{F}^{-1}_2(b,b')\left[\frac{\partial \tilde{\C}^{ij}}{\partial C(b')}h_ih_j-\mathrm{Tr}\left(\frac{\partial \tilde{\C}}{\partial C(b')}\mathsf{S}^{-1}\mathsf{N}\mathsf{S}^{-\rm T}\right)\right]\,, \quad \mathcal{F}_2(b,b') =  \frac{1}{2}\Tr{\mathsf{S}^{-\rm T}\frac{\partial \tilde{\C}}{\partial C(b)}\mathsf{S}^{-1}\frac{\partial \tilde{\C}}{\partial C(b')}},}
\eeq
where $h\equiv \mathsf{S}^{-1}\tilde a$ and $\mathsf{S}^{-\rm T}$ is the transpose of $\mathsf{S}^{-1}$. The second term in $\widehat{C}(b)$ subtracts off the estimator bias induced by noise in the data; for the remainder of this work, we will absorb this into $\widehat{C}(b)$, and drop this term. Notably, $\mathcal{F}_2(b,b')$ is only symmetric if $\mathsf{S}^{-1}=\mathsf{S}^{-\rm T}$ (which does not hold if the filtering projects out modes). This is the minimum variance estimator in the limit of a Gaussian likelihood, $\overline C(b) \to C^{\rm true}(b)$, and $\mathsf{S}^{-1}\to\tilde{\C}^{-1}$ (or $\mathsf{S}^{-1}\to\mathsf{S}^{-1}\tilde{\C}\,\mathsf{S}^{-1}$ in general).

\subsection{Idealized Form}\label{subsec: Cl-ideal}
Let us consider \eqref{eq: opt-Cl-formal} in the idealized limit, \textit{i.e.} without a mask or beam and assuming isotropic noise (absorbed into the binned spectrum). Here, the relevant correlator derivative becomes
\beq
    \frac{\partial\C^{ij}}{\partial C(b)} = \sum_{\ell m}\Theta_\ell(b)Y_{\ell m}(\hn^i)Y^*_{\ell m}(\hn^j),
\eeq
moving to harmonic-space and using the binned power spectrum definition \eqref{eq: binned-Cl}. The Wiener-filtered field  can be simply written as $h_{\ell m}=S^{-1}_\ell a_{\ell m}$ (assuming $S$ to be diagonal in harmonic space), thus, following a little algebra, we obtain the estimator
\beq\label{eq: opt-Cl-ideal}
    \boxed{\widehat{C}_{\rm ideal}(b) = \frac{1}{2}\mathcal{F}_{2, \rm ideal}^{-1}(b)\sum_{\ell m}\Theta_\ell(b)\frac{|a_{\ell m}|^2}{S_\ell^2}\,, \qquad \mathcal{F}_{2, \rm ideal}(b) =  \frac{1}{2}\sum_{\ell}\Theta_\ell(b)\frac{2\ell+1}{S_\ell^2}.}
\eeq
This is just the conventional binned power spectrum estimator (summing over all $\ell,m$ allowed by the binning function), albeit including a Wiener-filter weighting. In this case, the normalization is diagonal in the bins, \textit{i.e.} each measurement is independent. Computation of \eqref{eq: opt-Cl-ideal} requires one harmonic transform (to define $a_{\ell m}$), then a simple sum over $\{\ell,m\}$: the latter process scales as $\mathcal{O}(N_{\rm bin})$ for $N_{\rm bin}$ bins in $\{b\}$.

\subsection{General Form}\label{subsec: Cl-general}
In the presence of a mask, we instead simplify \eqref{eq: opt-Cl-formal} by inserting the following two-point function derivative:
\beq\label{eq: c-b-deriv}
    \frac{\partial\tilde{\C}^{ij}}{\partial C(b)} = \sum_{\ell m}B_\ell^2\Theta_\ell(b)\left[W(\hn^i)Y_{\ell m}(\hn^i)\right]\left[W(\hn^j)Y^*_{\ell m}(\hn^j)\right],
\eeq
cf.\,\eqref{eq: tilde-Cij-as-Cl}. The estimator numerator can be written
\beq
    \boxed{\widehat{C}(b)\propto\frac{1}{2}\sum_{\ell m}B_\ell^2\Theta_\ell(b)[Wh]_{\ell m}[Wh]_{\ell m}^*,}
\eeq
where $[Wh]_{\ell m}$ is the harmonic-space representation of $W(\hn)h(\hn)\equiv W(\hn)\left[\mathsf{S}^{-1}\tilde{a}\right](\hn)$. This is straightforward to compute by a direct harmonic-space sum, and scales as $\mathcal{O}(N_{\rm bin})$, as before (with only one invocation of $\mathsf{S}^{-1}$ required).

The Fisher matrix of \eqref{eq: opt-Cl-formal} is more difficult to compute due to the trace, and formally requires $\mathcal{O}(N_{\rm pix}^2)$ operation. One option to compute it is to note that, if the estimator is optimal, it is equal to the covariance of the $\tilde{C}(b)$ numerator. As such, one could compute this quantity for a number of Monte Carlo simulations and form the covariance empirically. However, this is slow to converge (since we require the inverse Fisher matrix), and only exact if $\mathsf{S}^{-1}$ is the true inverse covariance.

To form a practically implementable estimator, we can instead rewrite the Fisher matrix as an expectation over some Gaussian random field (GRF) $u$, as in \citep{2011MNRAS.417....2S,2021PhRvD.103j3504P,Philcox:2021ukg}. This corresponds to writing
\beq
    \mathcal{F}_2(b,b') &=& \frac{1}{2}\av{\left(\frac{\partial \tilde{\C}}{\partial C(b)}\mathsf{S}^{-1}u\right)^{\rm T}\mathsf{S}^{-1}\left(\frac{\partial \tilde{\C}}{\partial C(b')}\mathsf{U}^{-1}u\right)}_u\\\nonumber
\eeq
or, more simply,
\beq
    \boxed{\mathcal{F}_2(b,b') =\frac{1}{2}\av{Q_2^{\rm T}[\mathsf{S}^{-1}u](b)\,\times\,W\mathsf{S}^{-1}W\,\times\,Q_2[\mathsf{U}^{-1}u](b')}_u,}
\eeq
where $\mathsf{U}$ is the (arbitrary, but invertible) covariance of $u$, and given that $\av{\mathsf{U}^{-1}uu^{\rm T}}_u$ is just the identity matrix, is equivalent to the Fisher matrix given in \eqref{eq: opt-Cl-formal}. In the second line, we have defined a filtered map, $Q_2$:
\beq\label{eq: Q2-def}
    Q_2[x](\hn^i;b) \equiv B\cdot\frac{\partial \C^{ij}}{\partial C(b)}[Wx](\hn^j) \quad \Leftrightarrow\quad Q_{2,\ell m}[x](b) = B_\ell^2\Theta_\ell(b)[Wx]_{\ell m},
\eeq
where the second (harmonic-space) definition holds due to \eqref{eq: c-b-deriv}, and the first derivative includes the (optionally pixel-windowed) beam (unlike in \S\ref{subsec: Cl-ideal}). Similar definitions will be used for the higher-order polyspectra. Thus, to form the Fisher matrix, one must compute the $Q_{2}$ filters on a set of $N_{\rm GRF}$ GRFs, $u$,  transform them to map-space, and combine them with a $W\mathsf{S}^{-1}W$ weighting. Each realization is symmetric only if $\mathsf{S}^{-1}=\mathsf{U}^{-1}$ and $\mathsf{S}^{-1}=\mathsf{S}^{-\rm T}$; the average over realizations requires only the latter condition. This can be straightforwardly achieved using repeated spherical harmonic transforms. Notably, it does not require any $\mathcal{O}(N_{\rm pix}^2)$ operations to implement. In practice, we will find that $\mathcal{O}(100)$ GRFs is sufficient for the above calculation, and adds an extra variance to the power spectrum at the $\mathcal{O}(1+N^{-1}_{\rm GRF})$ level. This is much more efficient than the Monte Carlo covariance estimation method discussed above. Computationally it involves two processes: computation of $N_{\rm bin}$ $WQ_2$ and $\mathsf{S}^{-1}WQ_2$ maps, and a summation over all such pairs. The first operation scales as $\mathcal{O}(N_{\rm bin})$ (involving $N_{\rm bin}$ $\mathsf{S}^{-1}$ operations), whilst the second is quadratic in $N_{\rm bin}$. Usually, the first dominates (since each term requires several harmonic transforms, depending on the form of $\mathsf{S}^{-1}$), resulting in a linear computational scaling (in the limit of large memory).

\subsection{Choice of $\mathsf{S}^{-1}$ and $\mathsf{U}$}\label{subsec: SU-choice}
To implement the general estimator described above, we must specify: (a) the mask, $W$, (b) the weighting matrix $\mathsf{S}^{-1}$, which downweights low signal-to-noise or troublesome regions, and (c) the Fisher GRF covariance $\mathsf{U}$. The first is usually the simplest: this is the linear response of the data $d$ to the underlying field $a$, which, for a full-sky map (or an inpainted version thereof), is often unity. For cut-sky data, this indicates which regions are observed and is often \resub{a binary mask}, albeit with some additional smoothing. A variety of additional phenomena can be included here, such as pixel weights and discreteness effects.

The optimal choice for the weighting operator $\mathsf{S}^{-1}$ is the inverse data covariance ${\tilde{\C}}^{-1}$. In realistic scenarios, this is neither diagonal in map- nor harmonic-space, and is thus difficult to invert (though may be possible via approaches such as \citep{Millea:2020iuw}, which supplement the map with additional uncorrelated noise). A simpler choice may be use a diagonal-in-$\ell$ approximation of the covariance to define $\mathsf{S}^{-1}$ (equal to $1/S_\ell$ in harmonic-space). Often, one may wish to downweight or remove specific regions of the map in the analysis before applying such a weight; this can be achieved by first removing areas of the map with some projection matrix $\Pi$, e.g.,
\beq
    \left[\mathsf{S}^{-1}v\right](\hn) = \sum_{\ell m}Y_{\ell m}(\hn)\frac{[\Pi v]_{\ell m}}{S_\ell},
\eeq
for arbitrary map $v$, where $[x]_{\ell m}$ is the harmonic transform of $x$. If $\Pi$ is not of full-rank, $\mathsf{S}^{-1}$ is not invertible.\footnote{Notably, there is a degeneracy between the mask $W$ and the weighting $\mathsf{S}^{-1}$: multiplying the data by some invertible function $f(\hn)$ sends $W\to fW$, resulting in the same estimator if $\mathsf{S}^{-1}\to f^{-1}\mathsf{S}^{-1}f^{-1}$. Note that this also affects the noise correlator also.} One may additionally wish to `inpaint' the map, by filling in small holes with the mean of the surrounding pixels \citep{Gruetjen:2015sta}. Since this is a linear operation it can be included within $\mathsf{S}^{-1}$, and fully accounted for in the normalization (noting that the below estimators require only the action of $\mathsf{S}^{-1}$ on maps $v$, rather than the explicit form of $\mathsf{S}$).

If we wish to use optimal weights, however, some progress can be made using conjugate gradient descent methods. In this framework, we compute the action of the inverse matrix, $\tilde{\C}^{-1}$, on some map $\alpha(\hn)$, using only applications of the uninverted matrix on maps $\beta(\hn)$. These can be computed thus (for arbitrary $\beta$):
\beq
    [\tilde\C\beta](\hn^i) = W(\hn^i)\C^{ij}[W\beta](\hn^j)+\left[\mathsf{N}\beta\right](\hn^i) = W(\hn^i)\left[B_\ell^2C_\ell[W\beta]_{\ell m}\right](\hn^i)+\left[\mathsf{N}\beta\right](\hn^i),
\eeq
where we have written $\tilde\C$ in harmonic space to obtain the second quantity via \eqref{eq: tilde-Cij-as-Cl}, and (as above) denoted forward- and backward harmonic transforms by $[\cdots]_{\ell m}$ and $[\cdots](\hn)$ respectively. To apply $\tilde{\C}$ to a map, our approach is thus: (1) multiply the map by $W$ in map-space, (2) transform to harmonic-space and multiply by $B_\ell^2C_\ell$, (3) transform to map-space and multiply by $W(\hn)$, (4) add on the noise term (which is straightforward if it is diagonal in map-space). Given the above mapping, and an appropriate pre-conditioner (such as the diagonal harmonic-space covariance), we may form $\tilde\C^{-1}\alpha$ iteratively for any given map $\alpha$. We caution that conjugate-gradient-descent inversion is computationally expensive procedure (involving two harmonic transforms per iteration), and we require at least $(N_{\rm bins}+2)$ applications of the inverse map for the full estimator (one for $h$, one for $\mathsf{S}^{-1}u$, and one for each bin in the Fisher matrix). In practice, therefore, we will primarily use a diagonal approximation in this work. Notably, this does not bias any results, but will lead to a slight loss of optimality.

Finally, some care is warranted regarding our choice of the Monte Carlo maps $u$, and their covariance $\mathsf{U}$. Ideally, we require maps that are easy to simulate, \textit{i.e.}\ GRFs. This also simplifies the interpretation, since no higher-point correlators need to be removed (which will be relevant for the bispectrum and beyond). We further require the maps to have a precisely known and simply invertible covariance. The convergence of the Monte Carlo estimates is fastest if $\mathsf{U}^{-1}$ is close to the weighting matrix $\mathsf{S}^{-1}$; in this work, we will fix $\mathsf{U}^{-1}=\mathsf{S}^{-1}$, assuming a diagonal form for both.

%\pagebreak
\section{Optimal Bispectrum Estimation}\label{sec: bl}
We now turn to the window-free bispectrum. As shown in \S\ref{sec: lik}, the general estimator for the binned bispectrum takes the form:
\beq\label{eq: opt-est-B-th}
    \boxed{\widehat{b}(\vec b) = \frac{1}{3!}\sum_{\vec b'}\mathcal{F}_3^{-1}(\vec b,\vec b')\frac{\partial \tilde\B^{ijk}}{\partial b(\vec b)}\bigg[h_ih_jh_k-\left(h_i\av{h_jh_k}+\text{2 perms.}\right)\bigg], \quad \mathcal{F}_{3}(\vec b,\vec b') = \frac{1}{3!}\frac{\partial \tilde\B^{ijk}}{\partial b(\vec b)}\mathsf{S}^{-1}_{il}\mathsf{S}^{-1}_{jm}\mathsf{S}^{-1}_{kn}\frac{\partial \tilde\B^{lmn}}{\partial b(\vec b')}}
\eeq
for $h\equiv \mathsf{S}^{-1}\tilde a$, which is optimal in the limit of weak non-Gaussianity, and $\mathsf{S}^{-1}\to{\tilde{\C}}^{-1}$. In the above, we sum over all bins with $b_1\leq b_2\leq b_3$, and note that the numerator contains both a three- and one-field term. The latter does not affect the mean, but ensures optimality on large scales, and takes a slightly different form from the Hermite tensor definitions \eqref{eq: hermite} since we have introduced a generic weighting $\mathsf{S}^{-1}$. This bears similarities to the estimators of \citep[e.g.,][]{Bucher:2015ura,Komatsu:2001wu,Yadav:2007ny,Komatsu:2003iq,Bucher:2009nm,Fergusson:2010gn,2011MNRAS.417....2S}, but now includes full treatment of masks and weights, and incorporates arbitrary $\ell$-space binning. We discuss its practical implementation below.

\subsection{Idealized Form}
Assuming rotational invariance, a unit beam, and a trivial mask, \eqref{eq: opt-est-B-th} can be simplified by first rewriting the numerator in harmonic space
\beq\label{eq: b-ideal-num}
    \widehat{b}_{\rm ideal}(\vec b) \propto \frac{1}{3!}\sum_{\ell_im_i}\frac{\partial B^{\ell_1\ell_2\ell_3}_{(-m_1)(-m_2)(-m_3)}}{\partial b(\vec b)}\bigg[h_{\ell_1m_1}h_{\ell_2m_2}h_{\ell_3m_3}-\left(h_{\ell_1m_1}\av{h_{\ell_2m_2}h_{\ell_3m_3}}+\text{2 perms.}\right)\bigg],
\eeq
summing over all $\ell_i$ and $m_i$ with $i\in\{1,2,3\}$. Inserting the relation between the ideal harmonic-space bispectrum and the binned form $b(\vec b)$ given in \eqref{eq: binned-bl} yields
\beq\label{eq: opt-b-num1}
    \widehat{b}_{\rm ideal}(\vec b) &\propto& \frac{1}{6\Delta_3(\vec b)}\sum_{\ell_im_i}\mathcal{G}^{\ell_1\ell_2\ell_3}_{m_1m_2m_3}\left[\Theta_{\ell_1}(b_1)\Theta_{\ell_2}(b_2)\Theta_{\ell_3}(b_3)+\text{5 perms.}\right]\\\nonumber
    &&\,\times\,\bigg[h_{\ell_1m_1}h_{\ell_2m_2}h_{\ell_3m_3}-\left(h_{\ell_1m_1}\av{h_{\ell_2m_2}h_{\ell_3m_3}}+\text{2 perms.}\right)\bigg],
\eeq
where $\mathcal{G}$ is the Gaunt factor. Assuming a diagonal choice for $\mathsf{S}^{-1}$, the one-field terms require $\ell_2=\ell_3$, and thus, by the triangle conditions $\ell_1=0$ (or permutations thereof); in the ideal limit, they may thus be dropped. 

To implement \eqref{eq: opt-b-num1} one could perform the $\ell,m$ summation directly, though this has $\mathcal{O}(\ell_{\rm max}^6)$ complexity. A more efficient scheme (first considered in \citep{Komatsu:2003iq}), is to rewrite the Gaunt factor as a spherical harmonic integral using \eqref{eq: gaunt-def}: this separates the three $(\ell_i,m_i)$ terms, yielding
\beq
    \widehat{b}_{\rm ideal}(\vec b) &\propto& \frac{1}{\Delta_3(\vec b)}\sum_{\ell_im_i}\int d\hn\,Y_{\ell_1m_1}(\hn)Y_{\ell_2m_2}(\hn)Y_{\ell_3m_3}(\hn)\Theta_{\ell_1}(b_1)\Theta_{\ell_2}(b_2)\Theta_{\ell_3}(b_3)h_{\ell_1m_1}h_{\ell_2m_2}h_{\ell_3m_3}
\eeq
additionally absorbing the permutation symmetries and dropping a factor of $(-1)^{\ell_1+\ell_2+\ell_3}$, noting that statistically isotropic temperature correlators must be are parity-even. Defining fields 
\beq\label{eq: H-ideal-def}
    H^{\rm ideal}_{b}[x](\hn) = \sum_{\ell m}\Theta_\ell(b)\frac{x_{\ell m}}{S_\ell}Y_{\ell m}(\hn) \qquad \Leftrightarrow \qquad H^{\rm ideal}_{b,\ell m}[x] = \Theta_\ell(b)\frac{x_{\ell m}}{S_\ell},
\eeq
this can be written
\beq
    \boxed{\widehat{b}_{\rm ideal}(\vec b)\propto\frac{1}{\Delta_3(\vec b)}\int d\hn\,H^{\rm ideal}_{b_1}[a](\hn)H^{\rm ideal}_{b_2}[a](\hn)H^{\rm ideal}_{b_3}[a](\hn)}.
\eeq
This is efficient to compute, requiring just one harmonic transform per choice of $b$, and a summation for each choice of $\vec b$; as such, the leading scaling is $\mathcal{O}(N_\ell)$, recalling that $N_\ell$ is the number of one-dimensional $\ell$-bins.

For the Fisher matrix, working in harmonic-space for diagonal $\mathsf{S}^{-1}$, we can write
\beq
    \mathcal{F}_{3, \rm ideal}(\vec b,\vec b') = \frac{1}{6}\sum_{\ell_im_i}\frac{\partial B^{\ell_1\ell_2\ell_3}_{(-m_1)(-m_2)(-m_3)}}{\partial b(\vec b)}S_{\ell_1}^{-1}S_{\ell_2}^{-1}S_{\ell_3}^{-1}\frac{\partial B^{\ell_1\ell_2\ell_3}_{m_1m_2m_3}}{\partial b(\vec b')}.
\eeq
Inserting the binned definition, we will have a sum over two Gaunt factors, which evaluates to
\beq
    \sum_{m_1m_2m_3}\left[\mathcal{G}^{\ell_1\ell_2\ell_3}_{m_1m_2m_3}\right]^2 = \frac{(2\ell_1+1)(2\ell_2+1)(2\ell_3+1)}{4\pi}\tjo{\ell_1}{\ell_2}{\ell_3}^2,
\eeq
and a sum over permutations of binning functions, which evaluates to
\beq
    \Theta_{\ell_1}(b_1)\Theta_{\ell_2}(b_2)\Theta_{\ell_3}(b_3)\left[\Theta_{\ell_1}(b_1')\Theta_{\ell_2}(b_2')\Theta_{\ell_3}(b_3')+\text{5 perms.}\right] = \Delta_3(\vec b)\delta^{\rm K}_{\vec b\vec b'},
\eeq
recalling that bins are ordered and non-overlapping, such that $\Theta_\ell(b)\Theta_\ell(b')= \delta^{\rm K}_{bb'}$. Just as for the power spectrum, the Fisher matrix is diagonal in $\vec b$, and can be evaluate as a triple sum over $\ell_i$, which has $\mathcal{O}(\ell_{\rm max}^3)$ complexity.

Collecting results the ideal bispectrum estimator becomes
\begin{empheq}[box=\widefbox]{align}
    \widehat{b}_{\rm ideal}(\vec b) &=\frac{1}{\Delta_3(\vec b)}\mathcal{F}^{-1}_{3,\rm ideal}(\vec b)\int d\hn\,H^{\rm ideal}_{b_1}(\hn)H^{\rm ideal}_{b_2}(\hn)H^{\rm ideal}_{b_3}(\hn)\\\nonumber \mathcal{F}_{3, \rm ideal}(\vec b) &= \frac{1}{\Delta_3(\vec b)}\sum_{\ell_1\ell_2\ell_3}\frac{(2\ell_1+1)(2\ell_2+1)(2\ell_3+1)}{4\pi}\frac{\Theta_{\ell_1}(b_1)}{S_{\ell_1}}\frac{\Theta_{\ell_2}(b_2)}{S_{\ell_2}}\frac{\Theta_{\ell_3}(b_3)}{S_{\ell_3}}\tjo{\ell_1}{\ell_2}{\ell_3}^2,
\end{empheq}
where $H^{\rm ideal}$ is defined in \eqref{eq: H-ideal-def} and the Fisher matrix is equal to the estimator variance if $S_\ell=C_\ell$.

\subsection{General Form}\label{subsec: Bl-general}
\subsubsection{Numerator}
The general estimator can be derived in a similar manner to the ideal case. Working in harmonic-space, the numerator is akin to \eqref{eq: b-ideal-num}, but includes window functions (due to the $\tilde B$ correlator):
\beq
    \widehat{b}(\vec b) \propto \frac{1}{3!}\sum_{\ell_im_i}\frac{\partial B^{\ell_1\ell_2\ell_3}_{(-m_1)(-m_2)(-m_3)}}{\partial b(\vec b)}\bigg[[Wh]_{\ell_1m_1}[Wh]_{\ell_2m_2}[Wh]_{\ell_3m_3}-([Wh]_{\ell_1m_1}\av{[Wh]_{\ell_2m_2}[Wh]_{\ell_3m_3}}+\text{2 perms.})\bigg],
\eeq
where the derivative includes the beam, $B_{\ell_1}B_{\ell_2}B_{\ell_3}$. Inserting the bispectrum derivative and rewriting the Gaunt factor as an integral, the three-field term takes a similar form to before:
\beq
    \widehat{b}^{(3)}(\vec b)\propto \frac{1}{\Delta_3(\vec b)}\int d\hn\,H_{b_1}[a](\hn)H_{b_2}[a](\hn)H_{b_3}[a](\hn),
\eeq
where $H$ is now defined as
\beq\label{eq: H-def}
    H_b[x](\hn) = \sum_{\ell m}[W\mathsf{S}^{-1}x]_{\ell m}B_\ell\Theta_\ell(b)Y_{\ell m}(\hn) \quad \Leftrightarrow \quad H_{b,\ell m}[x] = [W\mathsf{S}^{-1}x]_{\ell m}B_\ell\Theta_\ell(b).
\eeq
In the presence of a mask, the one-field term is non-trivial, but can be computed via a Monte Carlo average. Defining a set of fields $\{\alpha\}$ with covariance $\tilde{\C}_\alpha$, we can write
\beq
    \widehat{b}^{(1)}(\vec b) = -\frac{1}{\Delta_3(\vec b)}\int d\hn\,H_{b_1}[a](\hn)\av{H_{b_2}[\alpha](\hn)H_{b_3}[\alpha](\hn)}_\alpha+\text{2 perms.},
\eeq
where the average is taken over the random fields. For the estimator to be optimal, $\tilde{\C}_\alpha$ should be equal to the data covariance $\tilde{\C}$; however, given that $\av{a}=0$, the estimator does not become biased if this condition is not satisfied. This is in contrast with the trispectrum estimators of \S\ref{sec: tl}, which require accurate random simulations to remove the disconnected contributions. This has the same computational scalings as the ideal numerator (linear in $N_\ell$), but with runtime additionally proportional to the number of MC simulations, $N_{\rm MC}$.

\subsubsection{Fisher Matrix}
In the non-ideal case, the Fisher matrix is difficult to compute analytically. As for the power spectrum (\S\ref{subsec: Cl-general}), we can use a Monte Carlo procedure, first writing the covariance in real-space:
\beq\label{eq: fish3-wU}
    \mathcal{F}_3(\vec b,\vec b') &=& \frac{1}{6}\frac{\partial\tilde{\mathsf{B}}^{ijk}}{\partial b(\vec b)}\mathsf{S}^{-1}_{il}\mathsf{S}^{-1}_{jm}\mathsf{S}^{-1}_{kn}\frac{\partial\tilde{\mathsf{B}}^{lmn}}{\partial b(\vec b')}\\\nonumber
    &=&\frac{1}{12}\frac{\partial\tilde{\mathsf{B}}^{ijk}}{\partial b(\vec b)}\mathsf{S}^{-1}_{il}\mathsf{S}^{-1}_{jj'}\mathsf{S}^{-1}_{kk'}\left[\mathsf{U}^{j'm'}\mathsf{U}^{k'n'}+\mathsf{U}^{j'n'}\mathsf{U}^{k'm'}\right]\mathsf{U}^{-1}_{m'm}\mathsf{U}^{-1}_{n'n}\frac{\partial\tilde{\mathsf{B}}^{lmn}}{\partial b(\vec b')},
\eeq
inserting two copies of the identity matrix in the second line, for arbitrary invertible matrix $\mathsf{U}$. As for the power spectrum, this is symmetric only if $\mathsf{S}^{-1}=\mathsf{S}^{-\rm T}$. The Fisher matrix can be evaluated by introducing a set of GRFs $\{u\}$ with covariance $\mathsf{U}$, noting that the quantity inside the square brackets is equal to $\av{u^{j'}u^{k'}u^{m'}u^{n'}}-\av{u^{j'}u^{k'}}\av{u^{m'}u^{n'}}$. In this case, the Fisher matrix becomes
\beq\label{eq: fish-3-v1}
    \mathcal{F}_3(\vec b,\vec b') &=& \frac{1}{12}\int d\hn \,d\hn'\,\av{Q_3[\mathsf{S}^{-1}u,\mathsf{S}^{-1}u](\hn;\vec b)[W\mathsf{S}^{-1}W](\hn,\hn')Q_3[\mathsf{U}^{-1}u,\mathsf{U}^{-1}u](\hn';\vec b')}_{u}\\\nonumber
    &&\,-\,\frac{1}{12}\int d\hn \,d\hn'\,\av{Q_3[\mathsf{S}^{-1}u,\mathsf{S}^{-1}u](\hn;\vec b)}_u[W\mathsf{S}^{-1}W](\hn,\hn')\av{Q_3[\mathsf{U}^{-1}u,\mathsf{U}^{-1}u](\hn';\vec b')}_{u},
\eeq
where we have introduced the map (analogous to \eqref{eq: Q2-def} for the power spectrum):
\beq\label{eq: Q3-def}
    Q_3[x,y](\hn^i;\vec b) \equiv \frac{\partial\mathsf{B}^{ijk}}{\partial b(\vec b)}[Wx]_j[Wx]_k.
\eeq
Inserting the bispectrum derivative and converting to harmonic-space, we find
\beq\label{eq: Q-def}
    Q_{3,\ell m}[x,y](\vec b) &=& \frac{1}{\Delta_3(\vec b)}\sum_{\ell_2\ell_3m_2m_3}\mathcal{G}^{\ell\ell_2\ell_3}_{mm_2m_3}[Wx]^*_{\ell_2m_2}[Wy]^*_{\ell_3m_3}B_{\ell}B_{\ell_2}B_{\ell_3}\left[\Theta_{\ell}(b_1)\Theta_{\ell_2}(b_2)\Theta_{\ell_3}(b_3)+\text{5 perms.}\right]\\\nonumber
    &=& \frac{2}{\Delta_3(\vec b)}B_{\ell}\Theta_\ell(b_1)\int d\hn\,Y_{\ell m}^*(\hn)H[x](\hn;b_2)H[y](\hn;b_3)+\text{2 perms.},
\eeq
where we inserted the integral form of the Gaunt factor \eqref{eq: gaunt-def} in the second line, and used the $H$ maps defined in \eqref{eq: H-def}. This is straightforwardly evaluated as a harmonic transform.

Whilst possible to implement \eqref{eq: fish-3-v1} is somewhat unwieldy, since it requires the average of a map, $Q_3(\hn)$, over a set of random fields, which is expensive to store (though \citep{Philcox:2021ukg} took this approach). Instead, one may proceed by introducing \textit{two} (uncorrelated) sets of random fields $\{u_1\}$ and $\{u_2\}$ with covariance $\mathsf{U}$, as in \citep{2015arXiv150200635S}. These can be combined to form the following symmetric combination: 
\beq
    \alpha\left(\av{u_1^{j'}u_1^{k'}u_1^{m'}u_1^{n'}}+\av{u_2^{j'}u_2^{k'}u_2^{m'}u_2^{n'}}\right)+\beta\left(\av{u_1^{j'}u_1^{k'}u_2^{m'}u_2^{n'}}+\av{u_2^{j'}u_2^{k'}u_1^{m'}u_1^{n'}}\right);
\eeq
this is equal to the combination of $\mathsf{U}$ covariances appearing in \eqref{eq: fish3-wU} if $\alpha=-\beta=1/2$.\footnote{For full generality, we could include a third set of terms of the form $\av{u_1^{j'}u_2^{k'}u_1^{m'}u_2^{n'}}$ and permutations thereof. The inclusion of these may lead to a slight reduction in the number of Monte Carlo simulations required, but we neglect them for simplicity here.} Defining 
\beq
    F^{ab,cd}_3(\vec b,\vec b') = \frac{1}{12}\int d\hn \,d\hn'\,\av{Q_3[\mathsf{S}^{-1}u_a,\mathsf{S}^{-1}u_b](\hn;\vec b)[W\mathsf{S}^{-1}W](\hn,\hn')Q_3[\mathsf{U}^{-1}u_c,\mathsf{U}^{-1}u_d](\hn';\vec b')}_{u_a,u_b,u_c,u_d},
\eeq
we can write
\beq\label{eq: F3-def}
    \boxed{\mathcal{F}_3(\vec b,\vec b') = \frac{1}{2}\left(F^{11,11}_3(\vec b,\vec b')+F^{22,22}_3(\vec b,\vec b')\right)-\frac{1}{2}\left(F^{11,22}_3(\vec b,\vec b')+F^{22,11}_3(\vec b,\vec b')\right),}
\eeq
which makes efficient use of the Monte Carlo simulations. Computation requires $N_\ell$ $H$ maps to be computed, which are combined into $N_{\rm bin}$ $Q_{\ell m}$ maps, involving $\mathcal{O}(N_{\rm MC}N_\ell^2)$ harmonic transforms. These are then combined via map-space summation, yielding an estimator for the Fisher matrix that is again linear in $N_{\rm bin}$ (in the large-memory limit), and proportional to the number of Fisher simulations, $N_{\rm fish}$ (which are analyzed independently). Note that there is no scaling with $\ell_{\rm max}$, except for that incurred by the choice of \textsc{healpix} $N_{\rm side}$. 

In summary, the optimal window-free bispectrum estimator is given by 
\beq\label{eq: b-opt-summary}
    \boxed{\widehat{b}(\vec b)= \sum_{\vec b'}\frac{\mathcal{F}^{-1}_3(\vec b,\vec b')}{\Delta_3(\vec b')}\int d\hn\,\bigg[H_{b_1'}[a](\hn)H_{b_2'}[a](\hn)H_{b_3'}[a](\hn)-(\av{H_{b_1'}[\alpha](\hn)H_{b_2'}[\alpha](\hn)}_\alpha H_{b_3'}[a](\hn)+\text{2 perms.})\bigg],}
\eeq
with the Fisher matrix defined in \eqref{eq: F3-def} subject to the $Q$ definitions of \eqref{eq: Q-def}. This is straightforward to implement and optimal in the limit of $\mathsf{S}^{-1}\to\tilde{\C}^{-1}$ and weak non-Gaussianity.

%\pagebreak
\section{Optimal Trispectrum Estimation}\label{sec: tl}
Finally, let us consider optimal estimation of the full-sky trispectrum. Unlike for lower order statistics, this has rarely been considered previously (though see \citep{2015arXiv150200635S,2011MNRAS.412.1993M,Regan:2010cn} for notable examples) and the impact of masks has not been \resub{carefully} assessed. Furthermore, as noted in \S\ref{sec: ideal-binned}, the trispectrum contains both a parity-even and a parity-odd part: the estimators below are the first to measure the latter part. 

As discussed in \S\ref{sec: lik}, the general trispectrum estimator takes the form
\begin{empheq}[box=\widefbox]{align}\label{eq: general-tl-start}
    \widehat{t}(\vec b,B) &= \frac{1}{4!}\sum_{\vec b'}\mathcal{F}^{-1}_4(\vec b,B;\vec b',B')\frac{\partial\tilde{\mathsf{T}}^{ijkl}}{\partial t(\vec b,B')}\bigg[h_ih_jh_kh_l-\left(h_ih_j\av{h_kh_l}+\text{5 perms.}\right)\\\nonumber
    &\qquad\qquad\qquad\qquad\qquad\qquad\qquad\qquad\,+\,\left(\av{h_ih_j}\av{h_kh_l}+\text{2 perms.}\right)\bigg]\\\nonumber
    \mathcal{F}_4(\vec b,B;\vec b',B') &= \frac{1}{4!}\frac{\partial \tilde{T}^{ijkl}}{\partial t(\vec b,B)}\mathsf{S}^{-1}_{im}\mathsf{S}^{-1}_{jn}\mathsf{S}^{-1}_{ko}\mathsf{S}^{-1}_{lp}\frac{\partial \tilde{T}^{mnop}}{\partial t(\vec b',B')},
\end{empheq}
for $h\equiv \mathsf{S}^{-1}\tilde a$. As noted in \S\ref{subsec: binning}, the bins satisfy $b_1\leq b_2$, $b_3\leq b_4$, $b_1\leq b_3$, and, if $b_1=b_3$, $b_2\leq b_4$, as well as a diagonal $L$, binned in some bin $B$ (satisfying triangle conditions on $\{b_1,b_2,B\}$ and $\{b_3,b_4,B\}$). Similarly to before, this is optimal in the limit of vanishing non-Gaussianity and $\mathsf{S}^{-1}\to\tilde{\C}^{-1}$. 

Estimator \eqref{eq: general-tl-start} contains a four-, two-, and a zero-field term; unlike for the bispectrum, all terms are non-trivial, as they subtract off the mean of the signal. One exception to this is the ideal parity-odd trispectrum: since parity-violation only appears at fourth-order for scalars, the disconnected terms vanish in the ideal limit, making this contribution somewhat easier to estimate. In the below, we will consider estimators for both the parity-even and parity-odd trispectra below, which will be denoted $t_{\pm}(\vec b,L)$. We caution that the parity-odd components are purely imaginary, thus their Fisher matrix is negative definite (and equal to the negative of the covariance, if odd- and even-modes are uncorrelated).

\subsection{Idealized Form}
\subsubsection{Four-Field Term}\label{subsubsec: t4-ideal}

In the ideal limit, the trispectrum numerator can be written in harmonic space as
\beq
    \widehat{t}_{\pm, \rm ideal}(\vec b,B) &\propto& \frac{1}{4!}\sum_{\ell_im_i}\frac{\partial T^{\ell_1\ell_2\ell_3\ell_4}_{(-m_1)(-m_2)(-m_3)(-m_4)}}{\partial t_{\pm}(\vec b,B)}\bigg[h_{\ell_1m_1}h_{\ell_2m_2}h_{\ell_3m_3}h_{\ell_4m_4}-\left(h_{\ell_1m_1}h_{\ell_2m_2}\av{h_{\ell_3m_3}h_{\ell_4m_4}}+\text{5 perms.}\right)\nonumber\\
    &&\qquad\qquad\qquad\qquad\qquad\qquad\qquad\qquad\qquad\,+\,\left(\av{h_{\ell_1m_1}h_{\ell_2m_2}}\av{h_{\ell_3m_3}h_{\ell_4m_4}}+\text{2 perms.}\right)\bigg].
\eeq
Inserting the explicit definition of the binned trispectra \eqref{eq: binned-tl}, the four-field term can be written \beq\label{eq: t4-init}
    \widehat{t}^{(4)}_{\pm, \rm ideal}(\vec b, B) &\propto& \frac{1}{\Delta_4(\vec b)}\sum_{\ell_im_i}(-1)^{\ell_{1234}}\sum_{LM}(-1)^Mw_{\ell_1\ell_2m_1m_2}^{L(-M)}w_{\ell_3\ell_4m_3m_4}^{LM}\Theta_L(B)\\\nonumber
    &&\,\qquad\times\,\left[\frac{1\pm(-1)^{\ell_{1234}}}{2}\right]\Theta_{\ell_1}(b_1)\cdots\Theta_{\ell_4}(b_4)h_{\ell_1m_1}\cdots h_{\ell_4m_4},
\eeq
\resub{where we have noted} that all 24 permutations are equivalent (due to the symmetry of the four $h$ fields), and explicitly restricted to even or odd $\ell_{1234}\equiv \ell_1+\ell_2+\ell_3+\ell_4$. By expanding the square bracket, this can be split into two coupled pieces:
\beq\label{eq: t4-ideal-separable}
    \boxed{\widehat{t}^{(4)}_{\pm,\rm ideal}(\vec b, B) \propto \pm\frac{1}{2\Delta_4(\vec b)}\sum_{LM}(-1)^M\Theta_L(B) \left[A^{\rm ideal}_{b_1b_2}(L,-M)A^{\rm ideal}_{b_3b_4}(L,M)\pm\overline A^{\rm ideal}_{b_1b_2}(L,-M)\overline A^{\rm ideal}_{b_3b_4}(L,M)\right],}
\eeq
subject to the definitions
\beq\label{eq: A-def}
    A_{b_1b_2}^{\rm ideal}(L,M) &=& \sum_{\ell_1\ell_2m_1m_2}w_{\ell_1\ell_2m_1m_2}^{LM}\Theta_{\ell_1}(b_1)\Theta_{\ell_2}(b_2)h_{\ell_1m_1}h_{\ell_2m_2}\\\nonumber
    \overline A^{\rm ideal}_{b_1b_2}(L,M) &=& \sum_{\ell_1\ell_2m_1m_2}(-1)^{\ell_1+\ell_2+L}w_{\ell_1\ell_2m_1m_2}^{LM}\Theta_{\ell_1}(b_1)\Theta_{\ell_2}(b_2)h_{\ell_1m_1}h_{\ell_2m_2},
\eeq
which are symmetric under $b_1\leftrightarrow b_2$. The separable form given in \eqref{eq: t4-ideal-separable} is significantly more efficient than a na\"ive estimation using \eqref{eq: t4-init}, with computation scaling as $\ell_{\rm max}^6$ instead of $\ell_{\rm max}^{10}$ for some global maximum scale $\ell_{\rm max}$ (given that each of $(L,M)$ coefficient involves $\mathcal{O}(\ell_{\rm max}^4)$ terms, and there are $\mathcal{O}(N_\ell^2)=\mathcal{O}(\ell_{\rm max}^2)$ such pieces). 

Rather than performing the sum over $\ell_i,m_i$ explicitly, it is preferred to compute $A$ and $\overline{A}$ by first rewriting the weighting function in terms of spin-weighted spherical harmonics, as in \eqref{eq: gaunt-spin-def}. Inserting this relation, we find
\beq
    A^{\rm ideal}_{b_1b_2}(L,M) &=& \int d\hn\,{}_{-2}Y_{LM}(\hn)\left[\sum_{\ell_1m_1}h_{\ell_1m_1}\Theta_{\ell_1}(b_1){}_{+1}Y_{\ell_1m_1}(\hn)\right]\left[\sum_{\ell_2m_2}h_{\ell_2m_2}\Theta_{\ell_2}(b_2){}_{+1}Y_{\ell_2m_2}(\hn)\right]\\\nonumber
    &\equiv& \int d\hn\,{}_{-2}Y_{LM}(\hn)H^{+}_{b_1, \rm ideal}(\hn)H^{+}_{b_2, \rm ideal}(\hn),
\eeq
and similarly 
\beq
    \overline{A}^{\rm ideal}_{b_1b_2}(L,M) = (-1)^{L}\int d\hn\,{}_{-2}Y_{LM}(\hn)\overline{H}^+_{b_1,\rm ideal}(\hn)\overline{H}^{+}_{b_2,\rm ideal}(\hn),
\eeq
defining the spin-weighted fields
\beq
    H_{b,\rm ideal}^{\pm}(\hn) = \sum_{\ell m} h_{\ell m}\Theta_\ell(b){}_{\pm1}Y_{\ell m}(\hn), \qquad 
    \overline{H}^{\pm}_{b,\rm ideal} = \sum_{\ell m}(-1)^{\ell}h_{\ell m}\Theta_\ell(b){}_{\pm1}Y_{\ell m}(\hn).
\eeq
The $H$ fields satisfy the following identity
\beq
    \left[H_{b,\rm ideal}^+(\hn)\right]^* &=& -\sum_{\ell m}h_{\ell m}\Theta_\ell(b){}_{-1}Y_{\ell m}(\hn) \equiv -H^-_{b,\rm ideal}(\hn),
\eeq
(using properties of the spin-weighted spherical harmonics and assuming $a(\hn)$ to be real), implying that $A^{\rm ideal,*}_{b_1b_2}(L,M) = (-1)^{M}\overline{A}^{\rm ideal}_{b_1b_2}(L,-M)$. This has the useful implication that
\beq\label{eq: t4-ideal}
    \widehat{t}^{(4)}_{\pm,\rm ideal}(\vec b, B) &\propto& \pm\frac{1}{\Delta_4(\vec b)}\sum_{LM}\Theta_L(B)\begin{cases}\,\,\,\mathrm{Re}\left[\overline{A}^{\rm ideal,*}_{b_1b_2}(L,M)A^{\rm ideal}_{b_3b_4}(L,M)\right] \\ i\,\mathrm{Im}\left[\overline{A}^{\rm ideal,*}_{b_1b_2}(L,M)A^{\rm ideal}_{b_3b_4}(L,M)\right]\end{cases},
\eeq
which makes clear that parity-even (odd) trispectra are purely real (imaginary). Additionally, it can be used to write the estimator entirely in terms of $M\geq 0$ modes (noting that codes such as \textsc{healpix} generally store only these, by symmetry):
\beq
    \widehat{t}^{(4)}_{\pm,\rm ideal}(\vec b, B) &\propto& \pm\frac{1}{2\Delta_4(\vec b)}\sum_{L,M\geq 0}(1+\delta^{\rm K}_{M>0})\Theta_L(B)\begin{cases}\,\,\,\mathrm{Re}\left[\overline{A}^{\rm ideal,*}_{b_1b_2}(L,M)A^{\rm ideal}_{b_3b_4}(L,M)+ A^{\rm ideal,*}_{b_1b_2}(L,M)\overline A^{\rm ideal}_{b_3b_4}(L,M)\right]\\ i\,\mathrm{Im}\left[\overline{A}^{\rm ideal,*}_{b_1b_2}(L,M)A^{\rm ideal}_{b_3b_4}(L,M)-A^{\rm ideal,*}_{b_1b_2}(L,M)\overline A^{\rm ideal}_{b_3b_4}(L,M)\right]\end{cases},
\eeq
where the factor involving a Kronecker delta gives $2$ if $M>0$ and $1$ else. 

Utilizing these relations, we can compute the four-point term by first assembling all possible $H^{\pm}_{b,\rm ideal}(\hn)$ fields (a total of $N_\ell$), then combining to form each of the $\mathcal{O}(N_{\ell}^2)$ combinations of $A^{\rm ideal}_{b_1b_2}(L,M)$ and performing a pairwise sum over harmonics, restricting to the relevant bin in $L$. In practice, $H^{\pm}_{\rm ideal}$ can be obtained via spin-weighted harmonic transform, since $\pm H^{\pm}_{\rm ideal}(\hn)$ is the map-space spin-$\pm1$ conjugate to the harmonic-space spin-$\pm1$ fields $\pm h_{\ell m}\Theta_\ell(b)$. Similarly, $\overline{A}^{*}_{LM}$ and $A^*_{LM}$ are the harmonic-space spin-$\pm2$ conjugates of the spin-$\pm2$ maps $H^+(\hn)H^+(\hn)$ and $H^-(\hn)H^-(\hn)$ respectively. Thus, the computational cost to form the $A$ fields is $\mathcal{O}(N_{\ell}^2)$, whilst that of the summation is $\mathcal{O}(N_{\rm bin})=\mathcal{O}(N_\ell^4)$, but does not involve harmonic transforms, thus is not likely to be rate limiting.

\subsubsection{Aside: Spin Definitions}
It is interesting to consider why the above decomposition is possible. In the definition of \citep{Regan:2010cn}, the trispectrum coefficients, $T^{\ell_1\ell_2}_{\ell_3\ell_4}(L)$ are defined via
\beq
    T^{\ell_1\ell_2\ell_3\ell_4}_{m_1m_2m_3m_4} = \sum_{LM}(-1)^M\tj{\ell_1}{\ell_2}{L}{m_1}{m_2}{-M}\tj{\ell_3}{\ell_4}{L}{m_3}{m_4}{M}T^{\ell_1\ell_2}_{\ell_3\ell_4}(L)
\eeq
as in \eqref{eq: t-l-def-regan}. With this definition, the trispectrum estimator will involve terms of the form
\beq\label{eq: t-general-spin}
    \sum_{m_1m_2}\tj{\ell_1}{\ell_2}{L}{m_1}{m_2}{M}h_{\ell_1m_1}h_{\ell_2m_2} &=& \resub{\tj{\ell_1}{\ell_2}{L}{-s_1}{-s_2}{s_{12}}^{-1}\sqrt{\frac{4\pi}{(2\ell_1+1)(2\ell_2+1)(2L+1)}}}\\\nonumber
    &&\,\times\,\int d\hn\,{}_{-s_{12}}Y_{LM}(\hn)\left[\sum_{m_1}h_{\ell_1m_1}\,{}_{s_1}Y_{\ell_1m_1}(\hn)\right]\left[\sum_{m_2}h_{\ell_2m_2}\,{}_{s_2}Y_{\ell_2m_2}(\hn)\right].
\eeq
On the RHS, we have inserted the spin-weighted Gaunt factor definition \eqref{eq: gaunt-spin-def} for a general set of spins \resub{$\{s_1,s_2,s_{12}\}$}. This allows the $m_i$ summations to be rewritten as an integral (or equivalently, a set of spin-weighted harmonic transforms); given an appropriate definition for the \textit{reduced} trispectrum coefficients, it also allows us to separate the $\ell_i$ summations.

To perform the above trick, we must carefully choose the spins. In particular, we require the $3j$ symbol to be non-zero for all $\ell$ of interest. Assuming $\ell_i\geq 2$, $|\ell_1-\ell_2|\leq L\leq \ell_1+\ell_2$, and $|s_i|\leq \ell_i$, one might consider $\{s_1,s_2,s_{12}\} = \{0,0,0\}, \{\pm 1, \mp 1,0\}, \{\pm 2,\mp 2,0\}, \{\pm1,\pm1,\mp2\}$. Whilst the former choice matches that used in the bispectrum, it requires even $\ell_1+\ell_2+L=0$, and thus cannot be used for the parity-odd trispectrum. Similarly, the second and third vanish upon symmetrization, thus we here utilize the third, fixing $s_1=s_2=-1$ and $s_{12}=2$.\footnote{Other choices are possible; these will lead to reduced trispectra differing by powers of $\sqrt{\ell}$.} To this end, we absorb the first line on the RHS of \eqref{eq: t-general-spin} into the trispectrum definition, yielding the reduced trispectrum of \eqref{eq: t-l-def}, and allowing separation of the $\ell_i$ summations.

\subsubsection{Two-Field Term}
The two-field term can be obtained by first noting that $\av{h_{\ell m}h_{\ell'm'}} = (-1)^{m}\delta^{\rm K}_{\ell\ell'}\delta^{\rm K}_{m(-m')}C_\ell/S_\ell^2$, assuming uniform weights $S_\ell$. As such, the estimator takes the form
\beq\label{eq: t2-init}
    \widehat{t}^{(2)}_{\pm,\rm ideal}(\vec b, B) &\propto& -\frac{1}{\Delta_4(\vec b)}\sum_{\ell_im_i}(-1)^{\ell_{1234}}\sum_{LM}(-1)^Mw_{\ell_1\ell_2m_1m_2}^{L(-M)}w_{\ell_3\ell_4m_3m_4}^{LM}\Theta_L(B)\\\nonumber
    &&\,\qquad\times\,\left[\frac{1\pm(-1)^{\ell_{1234}}}{2}\right]\Theta_{\ell_1}(b_1)\cdots\Theta_{\ell_4}(b_4)\left[h_{\ell_1m_1}h_{\ell_2m_2}(-1)^{m_3}\delta^{\rm K}_{\ell_3\ell_4}\delta^{\rm K}_{m_3(-m_4)}\frac{C_{\ell_3}}{S_{\ell_3}^2}+\text{5 perms.}\right].
\eeq
Due to the Kronecker deltas, the first two permutations contain the term
\beq
    \sum_{m_3M}(-1)^{m_3+M}\tj{\ell_1}{\ell_2}{L}{m_1}{m_2}{-M}\tj{\ell_3}{\ell_3}{L}{m_3}{-m_3}{M} &\propto& \tj{\ell_1}{\ell_2}{0}{m_1}{m_2}{0}\delta^{\rm K}_{L0}\delta^{\rm K}_{M0}.
\eeq
using properties of Wigner $3j$ symbols \citep{nist_dlmf} and separating out part of the $w^{LM}$ weighting matrices. Since we restrict to $L\geq 2$, this term vanishes always. The other four permutations contain contributions of the form
\beq
    \sum_{m_1M}(-1)^{m_2}\tj{\ell_1}{\ell_2}{L}{m_1}{m_2}{-M}\tj{\ell_3}{\ell_1}{L}{m_3}{-m_1}{M} \propto \delta^{\rm K}_{\ell_2\ell_3}\delta^{\rm K}_{m_2(-m_3)}.
\eeq
In both cases, two pairs of momenta are restricted to be equal, thus $\ell_{1234}$ is even, and any parity-odd contribution to the trispectrum must vanish. For the parity-even part, we find
\begin{empheq}[box=\widefbox]{align}
\label{eq: t2-ideal}
    \widehat{t}^{(2)}_{+,\rm ideal}(\vec b, B) &\propto& -\frac{1}{\Delta_4(\vec b)}(\delta^{\rm K}_{b_1b_4}\delta^{\rm K}_{b_2b_3}+\delta^{\rm K}_{b_1b_3}\delta^{\rm K}_{b_2b_4})\sum_{\ell_1\ell_2L}\frac{(2\ell_1+1)(2L+1)}{4\pi}\tj{\ell_1}{\ell_2}{L}{-1}{-1}{2}^2\\\nonumber
    &&\,\times\,(-1)^{\ell_1+\ell_2+L}\Theta_L(B)\bigg(\Theta_{\ell_1}(b_1)\Theta_{\ell_2}(b_2)+\Theta_{\ell_2}(b_1)\Theta_{\ell_1}(b_2)\bigg)\frac{C_{\ell_1}}{S_{\ell_1}^2}\sum_{m_2}\left|h_{\ell_2m_2}\right|^2,
\end{empheq}
involving the empirical power spectrum estimate $\sum_{m_2}|h_{\ell_2m_2}|^2/(2\ell_2+1)$. This scales as $\mathcal{O}(\ell_{\rm max}^3)$.

\subsubsection{Zero-Field Term}
The zero-field term may be evaluated using a similar prescription. First, we note that this requires two pairs of $\ell_i$ to be equal: due to the $1\pm(-1)^{\ell_{1234}}$ term, the odd-piece must vanish. For the even piece, there are only two non-trivial permutations (due to the above arguments removing the $\ell_1=\ell_2$, $\ell_3=\ell_4$ term):
\beq\label{eq: t0-init}
    \widehat{t}^{(0)}_{+,\rm ideal}(\vec b, B) &\propto& \frac{1}{\Delta_4(\vec b)}\sum_{\ell_im_i}(-1)^{\ell_{1234}}\sum_{LM}(-1)^Mw_{\ell_1\ell_2m_1m_2}^{L(-M)}w_{\ell_3\ell_4m_3m_4}^{LM}\Theta_L(B)\\\nonumber
    &&\,\qquad\times\,\Theta_{\ell_1}(b_1)\cdots\Theta_{\ell_4}(b_4)\left[(-1)^{m_1+m_2}\frac{C_{\ell_1}C_{\ell_2}}{S_{\ell_1}^2S_{\ell_2}^2}\left(\delta^{\rm K}_{\ell_1\ell_3}\delta^{\rm K}_{\ell_2\ell_4}\delta^{\rm K}_{m_1(-m_3)}\delta^{\rm K}_{m_2(-m_4)}+\delta^{\rm K}_{\ell_1\ell_4}\delta^{\rm K}_{\ell_2\ell_3}\delta^{\rm K}_{m_1(-m_4)}\delta^{\rm K}_{m_2(-m_3)}\right)\right].
\eeq
To simplify this, we note that
\beq
    \sum_{m_1m_2M}(-1)^{m_1+m_2+M}\tj{\ell_1}{\ell_2}{L}{m_1}{m_2}{-M}\tj{\ell_1}{\ell_2}{L}{-m_1}{-m_2}{M} = (-1)^{\ell_1+\ell_2+L},
\eeq
thus
\begin{empheq}[box=\widefbox]{align}
\label{eq: t0-ideal}
    \widehat{t}^{(0)}_{+,\rm ideal}(\vec b, L) &\propto \frac{1}{\Delta_4(\vec b)}(\delta^{\rm K}_{b_1b_4}\delta^{\rm K}_{b_2b_3}+\delta^{\rm K}_{b_1b_3}\delta^{\rm K}_{b_2b_4})\sum_{\ell_1\ell_2L}\frac{(2\ell_1+1)(2\ell_2+1)(2L+1)}{4\pi}\tj{\ell_1}{\ell_2}{L}{-1}{-1}{2}^2\Theta_L(B)\\\nonumber
    &\,\qquad\times\,(-1)^{\ell_1+\ell_2+L}\Theta_{\ell_1}(b_1)\Theta_{\ell_2}(b_2)\frac{C_{\ell_1}}{S_{\ell_1}^2}\frac{C_{\ell_2}}{S_{\ell_2}^2},
\end{empheq}
which can be straightforwardly computed in $\mathcal{O}(\ell_{\rm max}^3)$ operations.

\subsubsection{Normalization}
We now turn to the trispectrum Fisher matrix. As we shall find below, this is somewhat more complex than for the power spectrum or bispectrum, since there are off-diagonal correlations even in the ideal case, \textit{i.e.}\ bins with different $\vec b$ can correlate. This arises due to the degeneracy in the quadrilateral definition: there are two choices of diagonal momentum $L$ for any given tetrahedron. As such, the off-diagonal terms will contribute only when $\{b_1',b_2',b_3',b_4'\}$ is some permutation of $\{b_1,b_2,b_3,b_4\}$.

To compute the normalization, we start from \eqref{eq: general-tl-start} and insert the harmonic-space definitions of the binned trispectrum \eqref{eq: binned-tl}, noting that we can absorb a symmetry factor of $24$ since $\mathsf{T}^{ijkl}$ is fully symmetric under index exchange. This gives: 
\beq
    \mathcal{F}^{\rm ideal}_{4\pm}(\vec b,B;\vec b',B')&=& \frac{1}{\Delta_4(\vec b)\Delta_4(\vec b')}\sum_{\ell_im_i}(-1)^{\ell_{1234}}\sum_{LL'MM'}(-1)^{M+M'}\left[\frac{1-(-1)^{\ell_{1234}}}{2}\right]^2S_{\ell_1}^{-1}\cdots S_{\ell_4}^{-1}\\\nonumber
    &&\,\times\,\Theta_L(B)\Theta_{L'}(B')\Theta_{\ell_1}(b_1)\cdots\Theta_{\ell_4}(b_4)w^{L(-M)}_{\ell_1\ell_2m_1m_2}w^{LM}_{\ell_3\ell_4m_3m_4}\\\nonumber
    &&\,\times\,\left\{\left[w^{L'(-M')}_{\ell_1\ell_2m_1m_2}w^{L'M'}_{\ell_3\ell_4m_3m_4}\Theta_{\ell_1}(b_1')\cdots\Theta_{\ell_4}(b_4')+\text{7 perms.}\right]\,+\,(2\leftrightarrow3)+(2\leftrightarrow4)\right\},
\eeq
where the $(-1)^{\ell_{1234}}$ term comes from switching $m_i$ to $(-m_i)$ in one of the trispectrum derivatives. As before, the binning functions satisfy $\Theta_\ell(b)\Theta_{\ell}(b') =\delta^{\rm K}_{bb'}\Theta_\ell(b)$, for contiguous bins; this restricts which bins contribute to the coupling matrix. To proceed it is useful to consider the three permutations in the bottom line separately. The first involves
\beq
    \sum_{m_1m_2}w^{L(-M)}_{\ell_1\ell_2m_1m_2}w^{L'(-M')}_{\ell_1\ell_2m_1m_2} &\propto& \sum_{m_1m_2}\tj{\ell_1}{\ell_2}{L}{m_1}{m_2}{-M}\tj{\ell_1}{\ell_2}{L'}{m_1}{m_2}{-M'}=\frac{1}{2L+1}\delta^{\rm K}_{LL'}\delta^{\rm K}_{(-M)M'},
\eeq
which implies the matrix is diagonal in $L$. Similarly, the binning functions yield a factor $\delta^{\rm K}_{\vec b\vec b'}\Delta_4(\vec b')$ (noting the selection rules on $\vec b$), leading to the final contribution:
\beq\label{eq: fish4-idealA}
    \mathcal{F}^{\rm ideal,(a)}_{4\pm}(\vec b,B;\vec b',B')&=&\pm\frac{\delta^{\rm K}_{\vec b\vec b'}\delta^{\rm K}_{BB'}}{\Delta_4(\vec b)}\sum_{\ell_iL}\left[\frac{1\pm(-1)^{\ell_{1234}}}{2}\right]\Theta_{\ell_1}(b_1)\cdots\Theta_{\ell_4}(b_4)\Theta_L(B)\tj{\ell_1}{\ell_2}{L}{-1}{-1}{2}^2\tj{\ell_3}{\ell_4}{L}{-1}{-1}{2}^2\nonumber\\
    &&\,\times\,\frac{(2\ell_1+1)(2\ell_2+1)(2\ell_3+1)(2\ell_4+1)(2L+1)}{(4\pi)^2}S_{\ell_1}^{-1}\cdots S_{\ell_4}^{-1}.
\eeq
Notably, this factorizes into a piece involving $(\ell_1,\ell_2,L)$ and another involving $(\ell_3,\ell_4,L)$: as such, computation cost is $\mathcal{O}(N_\ell^3)$.

The other permutations do not require $L=L'$, and thus source a (small) mixing between modes. The second involves the following combination of $3j$ symbols (from the modified Wigner symbols):
\beq
    &&\sum_{m_1m_2m_3m_4MM'}(-1)^{M+M'}\tj{\ell_1}{\ell_2}{L}{m_1}{m_2}{-M}\tj{\ell_3}{\ell_4}{L}{m_3}{m_4}{M}\tj{\ell_1}{\ell_3}{L'}{m_1}{m_3}{-M'}\tj{\ell_2}{\ell_4}{L'}{m_2}{m_4}{M'}\nonumber\\
    &=& (-1)^{\ell_2+\ell_3}\begin{Bmatrix}L & \ell_1 & \ell_2\\ L' & \ell_4 & \ell_3\end{Bmatrix}.
\eeq
simplifying in terms of a $6j$ symbol in the second line. Similarly, the third has
\beq
    &&\sum_{m_1m_2m_3m_4MM'}(-1)^{M+M'}\tj{\ell_1}{\ell_2}{L}{m_1}{m_2}{-M}\tj{\ell_3}{\ell_4}{L}{m_3}{m_4}{M}\tj{\ell_1}{\ell_4}{L'}{m_1}{m_4}{-M'}\tj{\ell_3}{\ell_2}{L'}{m_3}{m_2}{M'}\nonumber\\
    &=& (-1)^{L+L'}\begin{Bmatrix}L & \ell_1 & \ell_2\\ L' & \ell_3 & \ell_4\end{Bmatrix}.
\eeq
This leads to the following matrix contributions:
\beq\label{eq: fish4-idealB}
    \mathcal{F}^{\rm ideal,(b)}_{4\pm}(\vec b,B;\vec b',B')&=&\pm\frac{1}{\Delta_4(\vec b)\Delta_4(\vec b')}\left[\delta^{\rm K}_{b_1b_1'}\delta^{\rm K}_{b_2b_3'}\delta^{\rm K}_{b_3b_2'}\delta^{\rm K}_{b_4b_4'}+\text{7 perms.}\right]\sum_{\ell_iLL'}\left[\frac{1\pm(-1)^{\ell_{1234}}}{2}\right]\Theta_{\ell_1}(b_1)\cdots\Theta_{\ell_4}(b_4)\\\nonumber
    &&\,\times\,\Theta_{L}(B)\Theta_{L'}(B')\frac{(2\ell_1+1)(2\ell_2+1)(2\ell_3+1)(2\ell_4+1)(2L+1)(2L'+1)}{(4\pi)^2}(-1)^{\ell_2+\ell_3}\begin{Bmatrix}L & \ell_1 & \ell_2\\ L' & \ell_4 & \ell_3\end{Bmatrix}\\\nonumber
    &&\,\times\,S_{\ell_1}^{-1}\cdots S_{\ell_4}^{-1}\tj{\ell_1}{\ell_2}{L}{-1}{-1}{2}\tj{\ell_3}{\ell_4}{L}{-1}{-1}{2}\tj{\ell_1}{\ell_3}{L'}{-1}{-1}{2}\tj{\ell_2}{\ell_4}{L'}{-1}{-1}{2},
\eeq
and
\beq\label{eq: fish4-idealC}
    \mathcal{F}^{\rm ideal,(c)}_{4\pm}(\vec b,B;\vec b',B')&=&\pm\frac{1}{\Delta_4(\vec b)\Delta_4(\vec b')}\left[\delta^{\rm K}_{b_1b_1'}\delta^{\rm K}_{b_2b_4'}\delta^{\rm K}_{b_3b_3'}\delta^{\rm K}_{b_4b_2'}+\text{7 perms.}\right]\sum_{\ell_iLL'}\left[\frac{1\pm(-1)^{\ell_{1234}}}{2}\right]\Theta_{\ell_1}(b_1)\cdots\Theta_{\ell_4}(b_4)\\\nonumber
    &&\,\times\,\Theta_L(B)\Theta_{L'}(B')\frac{(2\ell_1+1)(2\ell_2+1)(2\ell_3+1)(2\ell_4+1)(2L+1)(2L'+1)}{(4\pi)^2}(-1)^{L+L'}\begin{Bmatrix}L & \ell_1 & \ell_2\\ L' & \ell_3 & \ell_4\end{Bmatrix}\\\nonumber
    &&\,\times\,S_{\ell_1}^{-1}\cdots S_{\ell_4}^{-1}\tj{\ell_1}{\ell_2}{L}{-1}{-1}{2}\tj{\ell_3}{\ell_4}{L}{-1}{-1}{2}\tj{\ell_1}{\ell_4}{L'}{-1}{-1}{2}\tj{\ell_3}{\ell_2}{L'}{-1}{-1}{2}.
\eeq
Computation of this scales as $\mathcal{O}(\ell_{\rm max}^6)$, due to the presence of the Wigner $6j$ symbol.

Combining results, our ideal trispectrum estimators are given by
\begin{empheq}[box=\widefbox]{align}
    \widehat{t}_{+,\rm ideal}(\vec b,B) &= \sum_{\vec b'B'}\mathcal{F}^{{\rm ideal},-1}_{4+}(\vec b,B;\vec b',B')\left[\widehat{t}^{(4)}_{+,\rm ideal}(\vec b',B')+\widehat{t}^{(2)}_{+,\rm ideal}(\vec b',B')+\widehat{t}^{(0)}_{+,\rm ideal}(\vec b',B')\right]\\\nonumber
    \widehat{t}_{-,\rm ideal}(\vec b,L) &= \sum_{\vec b'B'}\mathcal{F}^{{\rm ideal},-1}_{4-}(\vec b,B;\vec b',B')\widehat{t}^{(4)}_{-,\rm ideal}(\vec b',B'),
\end{empheq}
for the parity-even and parity-odd contributions respectively, where the numerators are given in \eqref{eq: t4-ideal},\,\eqref{eq: t2-ideal}\,\&\,\eqref{eq: t0-ideal} and the Fisher matrix is a sum of \eqref{eq: fish4-idealA},\,\eqref{eq: fish4-idealB}\,\&\,\eqref{eq: fish4-idealC}. Note also that there is no correlation between even- and odd-trispectra, since they require even $\ell_{1234}$ and odd $\ell_{1234}$ respectively. As before, the Fisher matrix is equal to the estimator variance if $S_\ell=C_\ell$ in the Gaussian limit (or its negative, for the imaginary parity-odd trispectrum).

\subsection{General Form}\label{subsec: Tl-general}
At the final level of complexity we have the binned trispectrum of a masked field. The numerator of this takes a similar form to the ideal case discussed above, and the Fisher matrix can be computed similarly to that of the bispectrum \S\ref{subsec: Bl-general}. However, we note that, in the general case, the two- and zero-field terms in the parity-odd estimator do not vanish, and, at least in principle, there can be non-trivial mixing between odd- and even-parity trispectra induced by the window function. We show how to account for such effects below, considering each piece of the estimator in turn.

\subsubsection{Four-Field Term}\label{subsubsec: t4-general}
Analogously to \S\ref{subsubsec: t4-ideal}, the four-field component of the full trispectrum numerator is given by
\beq
    \widehat{t}_{\pm}^{(4)}(\vec b,B) &\propto& \frac{1}{24}\sum_{\ell_im_i}\frac{\partial T^{\ell_1\cdots\ell_4}_{(-m_1)\cdots(-m_4)}}{\partial t(\vec b,B)}[Wh]_{\ell_1m_1}\cdots [Wh]_{\ell_4m_4}\\
    \nonumber
    &=&\pm\frac{1}{\Delta_4(\vec b)}\sum_{\ell_im_i}\left[\frac{1\pm(-1)^{\ell_{1234}}}{2}\right]\sum_{LM}(-1)^Mw_{\ell_1\ell_2m_1m_2}^{L(-M)}w_{\ell_3\ell_4m_3m_4}^{LM}B_{\ell_1}B_{\ell_2}B_{\ell_3}B_{\ell_4}\Theta_L(B)\Theta_{\ell_1}(b_1)\cdots\Theta_{\ell_4}(b_4)\\\nonumber
    &&\qquad\qquad\,\times\,[Wh]_{\ell_1m_1}\cdots [Wh]_{\ell_4m_4},
\eeq
inserting the binned trispectrum definition in the second line. This differs only from the ideal case by the replacement $h\to Wh\equiv W\mathsf{S}^{-1}\tilde a$, and is similar to the even-parity estimator of \citep{2015arXiv150200635S}. Following similar logic to before, the estimator separates into a more straightforwardly computable form:
\beq
    \boxed{\widehat{t}^{(4)}_{\pm}(\vec b, B) \propto \pm\frac{1}{2\Delta_4(\vec b)}\sum_{LM}(-1)^M\Theta_L(B) \left[A_{b_1b_2}(L,-M)A_{b_3b_4}(L,M)\pm\overline A_{b_1b_2}(L,-M)\overline A_{b_3b_4}(L,M)\right],}
\eeq
which could be expressed as a real or imaginary part as in \eqref{eq: t4-ideal}. This uses the (mask-dependent) definitions
\beq\label{eq: A-def-nonideal}
    A_{b_1b_2}[x,y](L,M) &=& \int d\hn\,{}_{-2}Y_{LM}(\hn)H^+_{b_1}[x](\hn)H^+_{b_2}[y](\hn)\\\nonumber
    \overline{A}_{b_1b_2}[x,y](L,M) &=& (-1)^L\int d\hn\,{}_{-2}Y_{LM}(\hn)\overline{H}^{+}_{b_1}[x](\hn)\overline{H}^{+}_{b_2}[y](\hn)\\\nonumber
    H^{\pm}_{b}[x](\hn) &=& \sum_{\ell m}[Wx]_{\ell m}B_\ell\Theta_\ell(b){}_{\pm1}Y_{\ell m}(\hn)\\\nonumber
     \overline{H}^\pm_{b}[x](\hn) &=& \sum_{\ell m}(-1)^\ell [Wx]_{\ell m}B_\ell\Theta_\ell(b){}_{\pm1}Y_{\ell m}(\hn).
\eeq
These may be computed via weighted spherical harmonic transforms, as discussed in \S\ref{subsubsec: t4-general}, and differ only by the mask $W$ and the beam $B_\ell$. As such, the four-field term can be computed as a set of forward and reverse harmonic transforms, and finally a harmonic space sum in some bin $B$. As for the ideal case, the computational scaling is $\mathcal{O}(N_\ell^2)$ for the $A$ fields, and $\mathcal{O}(N_{\rm bin})$ for the overall summation.

\subsubsection{Two-Field Term}
As mentioned above, the two-field term is not guaranteed to vanish in the general parity-odd estimator, nor does it take a simple form in the general parity-even estimator. This is due to multipole mixing induced by the mask: even $\ell_{1234}$ in the true map does not necessarily correspond to even $\ell_{1234}$ in the windowed map.\footnote{See \citep{Coulton:2019bnz} for further discussion of this in the context of the parity-odd bispectrum.} For the parity-odd case, however, the two-field term is likely to be small, assuming a relatively well-behaved window function. %This is useful since it implies that biases obtained from an incorrectly assumed fiducial spectrum are much smaller in the parity-odd spectrum than one is used to in the parity-even case.

In general, the two-field term is equal to the four-field term but with two of the Monte Carlo fields contracted, \textit{i.e.} with the replacement $h_ih_j\to\av{h_ih_j}$. As for the bispectrum (\S\ref{subsec: Bl-general}), we will compute this by averaging over a set of simulations, $\{\alpha\}$, with covariance $\tilde{\C}_\alpha$.\footnote{Note that there is no requirement for the simulations to have accurate statistics beyond $\tilde{\C}$: this is discussed in \S\ref{subsubsec: t4-general}.} In this case, however, the estimator will be biased if $\tilde\C_\alpha$ is not equal to the data covariance $\tilde\C$, though, the bias is expected to be small in the parity-odd case, given that the term vanishes in the ideal limit. Furthermore, in the weakly non-Gaussian regime, the disconnected terms are large compared to the connected ones, thus we may require a substantial number of simulations to compute this contribution, to avoid additional sources of variance. We can write the two-field term in the following manner:
\begin{empheq}[box=\widefbox]{align}
\label{eq: t2-nonideal}
    \widehat{t}^{(2)}_{\pm}(\vec b, B) &\propto \mp\frac{1}{2\Delta_4(\vec b)}\sum_{LM}(-1)^M\Theta_L(B) \left\{A_{b_1b_2}[h,h](L,-M)\av{A_{b_3b_4}[\mathsf{S}^{-1}\alpha,\mathsf{S}^{-1}\alpha](L,M)}_\alpha\right.\\\nonumber
    &\qquad\qquad\qquad\qquad\qquad\left.\pm\,\overline A_{b_1b_2}[h,h](L,-M)\av{\overline A_{b_3b_4}[\mathsf{S}^{-1}\alpha,\mathsf{S}^{-1}\alpha](L,M)}_\alpha\right\}+\text{5 perms.},
\end{empheq}
where the permutations are over positions of the $\alpha$ mocks, arising due to the permutations contained within the binned trispectrum definition. To implement \eqref{eq: t2-nonideal}, we must compute both $\av{A_{bb'}[\mathsf{S}^{-1}\alpha,\mathsf{S}^{-1}\alpha](L,M)}$ and $\av{A_{bb'}[h,\mathsf{S}^{-1}\alpha](L,-M)A_{b''b'''}[h,\mathsf{S}^{-1}\alpha](L,M)}$; in practice, computation is dominated by the latter, since we must combine the data with each of $N_{\rm MC}$ simulations, with each requiring a harmonic transform per bin pair.

\subsubsection{Zero-Field Term}\label{subsubsec: t0-general}
The general zero-field term may be computed analogously, and takes the form
\begin{empheq}[box=\widefbox]{align}
    \widehat{t}^{(0)}_{\pm}(\vec b, B) &\propto \pm\frac{1}{4\Delta_4(\vec b)}\sum_{LM}(-1)^M\Theta_L(B) \left\{\av{A_{b_1b_2}[\mathsf{S}^{-1}\alpha_1,\mathsf{S}^{-1}\alpha_2](L,-M)A_{b_3b_4}[\mathsf{S}^{-1}\alpha_1,\mathsf{S}^{-1}\alpha_2](L,M)}_{\alpha_1,\alpha_2}\right.\\\nonumber
    &\qquad\qquad-\left.\av{\overline A_{b_1b_2}[\mathsf{S}^{-1}\alpha_1,\mathsf{S}^{-1}\alpha_2](L,-M)\overline A_{b_3b_4}[\mathsf{S}^{-1}\alpha_1,\mathsf{S}^{-1}\alpha_2](L,M)}_{\alpha_1,\alpha_2}\,+\,\text{5 perms.}\right\},
\end{empheq}
where $\{\alpha_1\}$ and $\{\alpha_2\}$ are two independent sets of simulations with the same covariance, and we sum over their possible locations. If the simulations were Gaussian, one could use only a single set and compute the four-point average via $\av{\alpha^4}\sim \tilde\C_\alpha\tilde\C_\alpha$; here, we allow for non-Gaussianities (for example from lensing), thus use only two-point averages.\footnote{One may also utilize non-Gaussian simulations to remove unwanted trispectra (arising from lensing or noise, for example); this is detailed in \citep{2015arXiv150200635S}.}

\subsubsection{Normalization}
Mask-induced multipole mixing can lead to non-trivial leakage between even- and odd-parity trispectra. As such, the general Fisher matrix contains even-even correlations (denoted $\mathcal{F}_{4++}$), odd-odd correlations ($\mathcal{F}_{4--}$) and even-odd correlations ($\mathcal{F}_{4+-}$ and $\mathcal{F}_{4-+}$). Thanks to the optimal estimator formalism, the full trispectrum estimates obtained should be free from this mixing, \textit{i.e.}\ the measured parity-odd modes should not contain a parity-even contribution. This is important if one is searching for a signal in the former, and wants to avoid, for example, lensing-based contributions to the latter. 

To compute the Fisher matrix, we start from the general relation given in \eqref{eq: general-tl-start}, and denote the two parity states by $\lambda,\lambda'\in\{\pm1\}$:
\beq\label{eq: norm-general-nonideal}
    \mathcal{F}_{4\lambda\lambda'}(\vec b,B;\vec b',B') = \frac{1}{24}\frac{\partial \tilde{\mathsf{T}}^{ijkl}}{\partial t_{\lambda}(\vec b,B)}\mathsf{S}^{-1}_{im}\mathsf{S}^{-1}_{jn}\mathsf{S}^{-1}_{ko}\mathsf{S}^{-1}_{lp}\frac{\partial \tilde{\mathsf{T}}^{mnop}}{\partial t_{\lambda'}(\vec b',B')}.
\eeq
As with the bispectrum, this must be significantly simplified to avoid a heinously expensive sum. Whilst one could compute $\mathcal{F}_4$ as the covariance of the unnormalized $\widehat{t}$ estimator applied to a set of GRFs, this requires a large number of Monte Carlo simulations to converge and is accurate only in the limit of $\mathsf{S}^{-1}\to\tilde{\C}^{-1}$. Instead (following the bispectrum logic, and \citep{Philcox:2021ukg,2011MNRAS.417....2S,2015arXiv150200635S}), we can use the following identity:
\beq\label{eq: gaussian-trick}
    \mathsf{S}^{-1}_{im}\left[\mathsf{S}^{-1}_{jn}\mathsf{S}^{-1}_{ko}\mathsf{S}^{-1}_{lp}+\text{5 perms.}\right] &=& \frac{1}{6}\mathsf{S}^{-1}_{im}\mathsf{S}^{-1}_{jj'}\mathsf{S}^{-1}_{kk'}\mathsf{S}_{ll'}^{-1}\left[\mathsf{U}_{j'n'}\mathsf{U}_{k'o'}\mathsf{U}_{l'p'}+\text{5 perms.}\right]\mathsf{U}^{-1}_{n'n}\mathsf{U}^{-1}_{o'o}\mathsf{U}^{-1}_{p'p}\\\nonumber
    &=& \frac{1}{6}\mathsf{S}^{-1}_{im}\mathsf{S}^{-1}_{jj'}\mathsf{S}^{-1}_{kk'}\mathsf{S}_{ll'}^{-1}\mathsf{U}^{-1}_{n'n}\mathsf{U}^{-1}_{o'o}\mathsf{U}^{-1}_{p'p}\av{u_{j'}u_{k'}u_{l'}u_{n'}u_{o'}u_{p'}}_{\rm fc},
\eeq
where we have inserted three copies of the unit matrix, for symmetric invertible matrix $\mathsf{U}$ and GRFs $u$, which satisfy $\av{uu^{\rm T}} = \mathsf{U}$. The correlator has the subscript `fc' corresponding to `fully-connected', \textit{i.e.} we consider only two-point contractions when each of $\{j',k',l'\}$ contracted with one of $\{n',o',p'\}$.

With the decomposition \eqref{eq: gaussian-trick}, the Fisher matrix can be split into two pieces, connected only by a known matrix, $\mathsf{S}^{-1}$. Explicitly, each takes the form
\beq\label{eq: Q4-intro}
    \frac{\partial \tilde{\mathsf{T}}^{ijkl}}{\partial t_{\lambda}(\vec b,B)}x_jy_kz_l &=& W(\hn^i)\sum_{\ell_1\cdots\ell_4m_1\cdots m_4}Y^*_{\ell_1 m_1}(\hn^i)\frac{\partial T^{\ell_1\cdots\ell_4}_{(-m_1)\cdots(-m_4)}}{\partial t(\vec b,B)}[Wx]_{\ell_2m_2}[Wy]_{\ell_3m_3}[Wz]_{\ell_4m_4}\\\nonumber
    &\equiv& W(\hn^i)Q_{4\lambda}[x,y,z](\hn^i;\vec b,B)
\eeq
for some $\{x,y,z\}$, converting the trispectrum to harmonic space, and introducing $Q_{4\pm}$ functions, akin to \eqref{eq: Q3-def}. This function is just a real-space map for each choice of $\vec b$ and $L$. Using the above definition, the coupling matrix can be written
\beq\label{eq: coupling-nonideal}
    \mathcal{F}_{4\lambda\lambda'}(\vec b,B;\vec b',B') &=& \frac{1}{144}\int d\hn\,d\hn'\,\left\langle Q_{4\lambda}[\mathsf{S}^{-1}u,\mathsf{S}^{-1}u,\mathsf{S}^{-1}u](\hn;\vec b,B)\left[W\mathsf{S}^{-1}W\right](\hn,\hn')\right.\nonumber\\
    &&\,\qquad\qquad\qquad\qquad\left.\times\,Q_{4\lambda'}[\mathsf{U}^{-1}u,\mathsf{U}^{-1}u,\mathsf{U}^{-1}u](\hn',\vec b',B')\right\rangle_{\rm fc},
\eeq
which is a Monte Carlo average over realizations of $u$, akin to \eqref{eq: fish-3-v1} for the bispectrum. 

To compute the fully-connected correlator, we must subtract off the unwanted correlations. This can be done by introducing multiple sets of GRFs, denoted, $\{u_n\}$, and computing expressions of the form
\beq
    F_{4\lambda\lambda'}^{abc,def} &\equiv& \frac{1}{144}\int d\hn\,d\hn'\, \left\langle Q_{4\lambda}[\mathsf{S}^{-1}u_a,\mathsf{S}^{-1}u_b,\mathsf{S}^{-1}u_c](\hn;\vec b,B)\left[W\mathsf{S}^{-1}W\right](\hn,\hn')\right.\nonumber\\
    &&\,\qquad\qquad\qquad\times\,\left.Q_{4\lambda'}[\mathsf{U}^{-1}u_d,\mathsf{U}^{-1}u_e,\mathsf{U}^{-1}u_f](\hn',\vec b',B')\right\rangle_{u_a,u_b,u_c,u_d,u_e,u_f},
\eeq
analogous to those in \S\ref{subsec: Bl-general}.
Most simply, we could use three such sets, giving $\mathcal{F}_{4\lambda\lambda'}(\vec b,B;\vec b',B') = 6F_{4\lambda\lambda'}^{123,123}$, such that only fully-connected terms can contribute. As shown in \citep{2015arXiv150200635S}, a more efficient way is to instead use two sets of GRFs, and compute the Fisher matrix as
\beq
    \boxed{\mathcal{F}_{4\lambda\lambda'}(\vec b,B;\vec b',B') =\frac{1}{8}\left[\left(F_{4\lambda\lambda'}^{111,111}+F_{4\lambda\lambda'}^{222,222}\right)+9\left(F_{4\lambda\lambda'}^{112,112}+F_{4\lambda\lambda'}^{122,122}\right)-6\left(F_{4\lambda\lambda'}^{111,122}+F_{4\lambda\lambda'}^{222,112}\right)\right],}
\eeq
with coefficients chosen to minimize the variance of the $\mathcal{F}_4$ estimate, \textit{i.e.} reduce the number of Monte Carlo simulations required.

We now turn to the computation of $Q_{4\pm}$ maps. First, we insert the explicit trispectrum of \eqref{eq: binned-tl} into \eqref{eq: Q4-intro}, finding
\beq
    Q_{4\pm}[x,y,z](\hn^i;\vec b,B) &=& \pm\frac{1}{\Delta_4(\vec b)}\sum_{\ell_im_i}Y^*_{\ell_1m_1}(\hn^i)\sum_{LM}(-1)^Mw^{L(-M)}_{\ell_1\ell_2m_1m_2}w^{LM}_{\ell_3\ell_4m_3m_4}B_{\ell_1}B_{\ell_2}B_{\ell_3}B_{\ell_4}\\\nonumber
    &&\,\times\,[Wx]_{\ell_2m_2}[Wy]_{\ell_3m_3}[Wz]_{\ell_4m_4}\Theta_L(B)\left[\frac{1\pm(-1)^{\ell_{1234}}}{2}\right]\left[\Theta_{\ell_1}(b_1)\cdots\Theta_{\ell_4}(b_4)+\text{7 perms.}\right]\\\nonumber
    &&\,+\,(2\leftrightarrow3)+(2\leftrightarrow4).
\eeq
Na\"ive computation of this expression is highly expensive, due to the large number of coupled $\ell$ summations. To simplify, we insert the definitions of $A$ and $\bar A$ given in \eqref{eq: A-def-nonideal} and expand the first weighting matrix in terms of spin-weighted spherical harmonics. For the first permutation, this gives
\beq
    Q^{(a)}_{4\pm}[x,y,z](\hn^i;\vec b,B) &=& \pm\frac{1}{2\Delta_4(\vec b)}\sum_{\ell_1\ell_2m_1m_2}Y^*_{\ell_1m_1}(\hn^i)\sum_{LM}B_{\ell_1}B_{\ell_2}\Theta_L(B)\Theta_{\ell_1}(b_1)\Theta_{\ell_2}(b_2)[Wx]_{\ell_2m_2}(-1)^M\\\nonumber
    &&\,\times\,\int d\hn\,{}_{-2}Y_{LM}(\hn){}_{+1}Y_{\ell_1m_1}(\hn){}_{+1}Y_{\ell_2m_2}(\hn)\bigg[A_{b_3b_4}[y,z](L,-M)\pm(-1)^{\ell_1+\ell_2+L}\overline{A}_{b_3b_4}[y,z](L,-M)\bigg].
\eeq
Next, the $\ell_1$ and $L$ summations can be written as spin-weighted spherical harmonic transforms. Denoting
\beq
    {}_s[X]^b(\hn) = \sum_{\ell m}{}_sY_{\ell m}(\hn)\Theta_\ell(b)X_{\ell m},
\eeq
and using the relation $(-1)^\ell{}_sY_{\ell m}(\hn) = {}_{-s}Y_{\ell m}(-\hn)$, we can write the harmonic-space $Q_{4\pm}$ as
\beq
    Q^{(a)}_{4\pm,\ell_1 m_1}[x,y,z](\vec b,B) &=& \mp\frac{B_{\ell_1}\Theta_{\ell_1}(b_1)}{2\Delta_4(\vec b)}\left[\int d\hn\,{}_{-1}Y^*_{\ell_1m_1}(\hn)H^{+}_{b_2}[x](\hn){}_{-2}[\overline{A}^*_{b_3b_4}[y,z]]^{B}(\hn)\right.\nonumber\\
    &&\qquad\qquad\qquad\,\pm\,\left.\int d\hn\,{}_{+1}Y^*_{\ell_1m_1}(\hn)H^{-}_{b_2}[x](\hn){}_{+2}[A^*_{b_3b_4}[y,z]]^{B}(\hn)\right],
\eeq
recalling that ${}_sY_{\ell m}^*=(-1)^{s+m}{}_{-s}Y_{\ell(-m)}$. Here, the $\hn$ integral can be evaluated as a further spin-$1$ spherical harmonic transform. Summing over permutations, we find the final form:
\beq
    Q_{4\pm,\ell m}[x,y,z](\vec b,B) &=& \left[Q^{(a)}_{4\pm,\ell m}[x,y,z](\{b_1,b_2,b_3,b_4\},B)+\text{7 perms.}\right]\\\nonumber
    &&\,+\,(x\leftrightarrow y)+(x\leftrightarrow z),
\eeq
where the permutations preserve the $\{b_1,b_2\}$ and $\{b_3,b_4\}$ pairs. This involves $\mathcal{O}(N_\ell^4)$ harmonic transforms (since only $\Theta_\ell(b)$ can be separated from the above expression), giving a slightly more favorable scaling than $\mathcal{O}(N_{\rm bin})=\mathcal{O}(N_\ell^5)$.

Given $Q_{4\pm,\ell m}$ we can compute $Q_{4\pm}(\hn)$, and thus compute the coupling via \eqref{eq: coupling-nonideal}, multiplying by the filter $W\mathsf{S}^{-1}W$, which requires straightforward transformations between real and harmonic space, with a computational cost of $\mathcal{O}(N_{\rm bin})$. Computing the above factors is likely the most labor-intensive section of optimal trispectrum estimator, but, thanks to the above simplifications, still scales favorably with the number of Monte Carlo simulations (linearly, and much faster than simply using them to numerically estimate the covariance of the unnormalized estimator \citep{2011MNRAS.417....2S}), and the number of bins \resub{(technically quadratically, but with the rate-limiting pieces (computation of $W\mathsf{S}^{-1}Q_{4\pm}$) scaling linearly)}.

To summarize, our estimator of the full-sky binned trispectrum is given by
\begin{empheq}[box=\widefbox]{align}
    \widehat{t}_{\lambda}(\vec b,B) &=& \sum_{\vec b'B'\lambda'}\mathcal{F}^{-1}_{4\lambda\lambda'}(\vec b,B;\vec b',B')\left\{\tau_{\lambda'}[h,h,h,h](\vec b',B')-6\av{\tau_{\lambda'}[h,h,\mathsf{S}^{-1}\alpha,\mathsf{S}^{-1}\alpha](\vec b',B')}_\alpha\right.\\\nonumber
    &&\,\qquad\qquad\qquad\qquad\qquad\left.\,+\,3\av{\tau_{\lambda'}[\mathsf{S}^{-1}\alpha_1,\mathsf{S}^{-1}\alpha_1,\mathsf{S}^{-1}\alpha_2,\mathsf{S}^{-1}\alpha_2](\vec b',B')}_{\alpha_1,\alpha_2}\right\},
\end{empheq}
where the unnormalized estimator can be written explicitly as
\beq
    \tau_{\pm}[\alpha,\beta,\gamma,\delta](\vec b,B) &=& \pm\frac{1}{48\Delta_4(\vec b')}\sum_{LM}(-1)^M\Theta_L(B)\left\{A_{b_1b_2}[\alpha,\beta](L,-M)A_{b_3b_4}[\gamma,\delta](L,M)\right.\\\nonumber
    &&\qquad\qquad\qquad\qquad\qquad\qquad\left.\pm\,\overline{A}_{b_1b_2}[\alpha,\beta](L,-M)\overline{A}_{b_3b_4}[\gamma,\delta](L,M)\right\}+\text{23 perms.}
\eeq
where the permutations are over the positions of $\{\alpha,\beta,\gamma,\delta\}$, and $\alpha$ are random fields satisfying $\av{\alpha\alpha^{\rm T}} = \tilde{\mathsf{C}}$. The $A$ and $\bar{A}$ fields are defined in \eqref{eq: A-def-nonideal} and the general Fisher matrix is given in \eqref{eq: coupling-nonideal}. If one wishes to ignore the even-odd coupling in the estimator, one just evaluates the above expression fixing $\lambda'=\lambda$.

\section{Validation}\label{sec: testing}
In the above sections, we have derived optimal and ideal estimators for the full-sky power spectrum, bispectrum, and (parity-even and odd) trispectrum. To demonstrate their efficacy, we will now consider a variety of tests on synthetic data, both for Gaussian and non-Gaussian maps, optionally including a non-trivial mask. This section makes extensive use of the 
public \textsc{PolyBin} code,\footnote{\href{https://github.com/oliverphilcox/PolyBin}{GitHub.com/oliverphilcox/PolyBin}} which implements the above estimators in \textsc{python}, with harmonic manipulations performed using \textsc{healpix} \citep{Gorski:2004by}. Spectra can be computed using arbitrary binning schemes, with the option of different binning for squeezed and collapsed configurations. For this purpose, we will specialize to CMB applications, though we note that the tools developed above apply much more generally.

\subsection{Practicalities}\label{subsec: test-practical}
\begin{figure}[!t]
    \centering
    \includegraphics[width=0.48\textwidth]{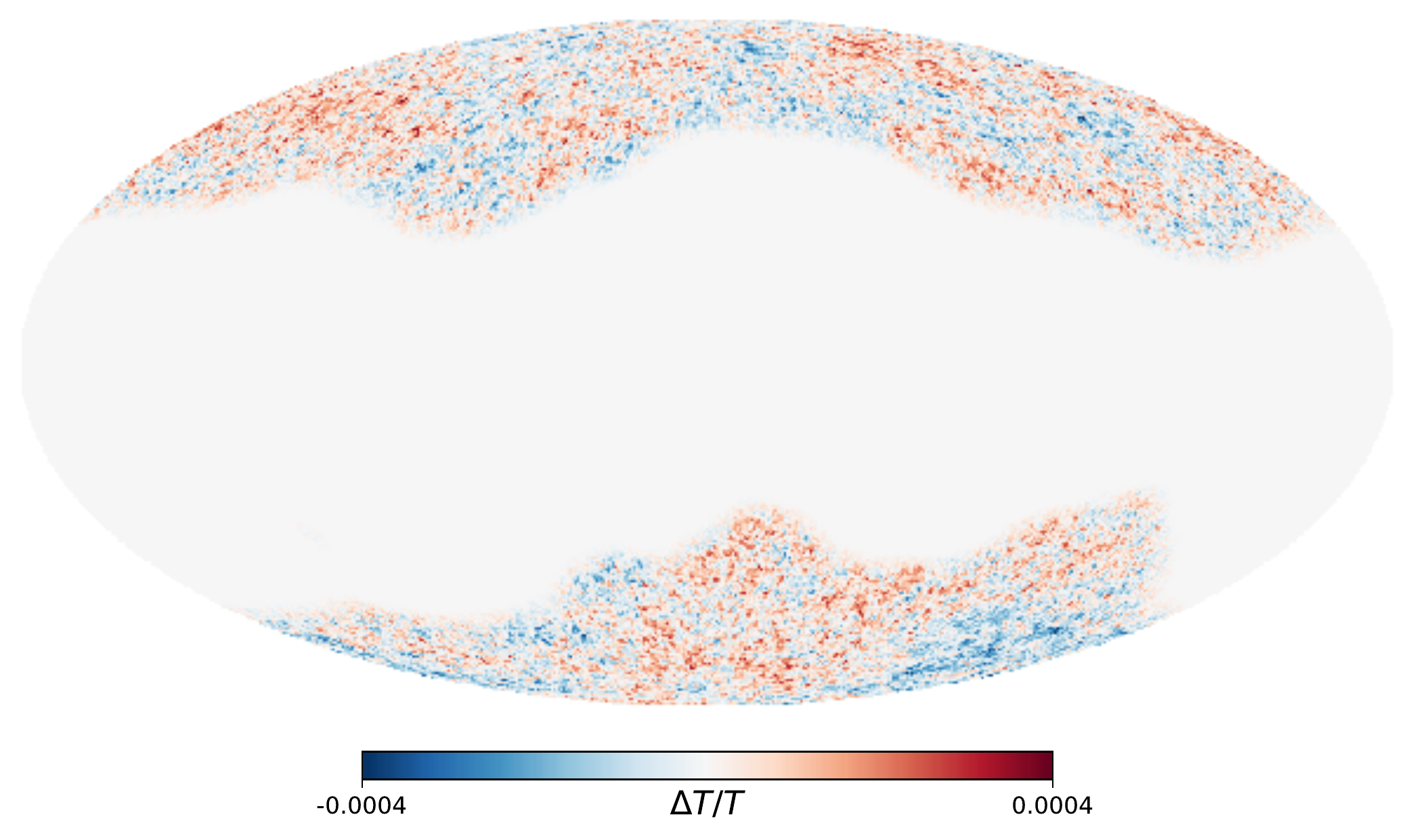}
    \includegraphics[width=0.48\textwidth]{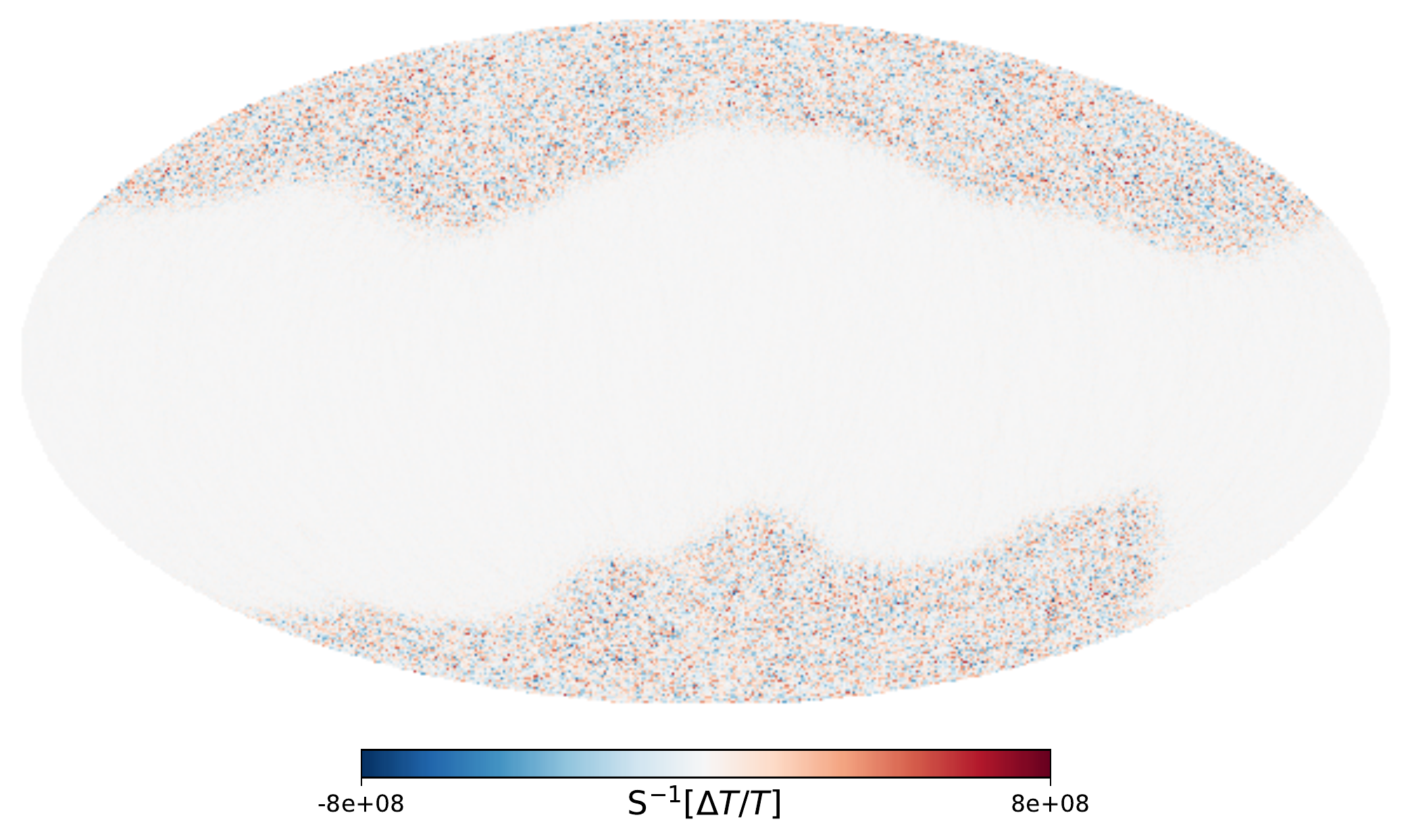}
    \caption{Examples of Gaussian full-sky maps used to test the polyspectrum estimators. The left panel shows a dataset constructed with the \textit{Planck} power spectrum and noise parameters and an apodized sky mask. In the right panel, we show the map after application of the quasi-optimal $\mathsf{S}^{-1}$ weighting.}
    \label{fig: maps}
\end{figure}

To test our estimators, we will primarily use synthetic Gaussian random fields (GRFs) created using \textsc{healpix}. These are constructed using the following (statistically isotropic) correlator:
\beq\label{eq: alm-ideal}
    \av{a_{\ell m}a_{\ell'm'}} = (-1)^{m}\delta^{\rm K}_{\ell\ell'}\delta^{\rm K}_{m(-m')}\left[B_\ell^2C_\ell^{TT} + N_\ell\right],
\eeq
where $C_\ell^{TT}$ is the CMB temperature power spectrum predicted by \textsc{class} with the \textit{Planck} best-fit parameters \citep{2020A&A...641A...6P}, and we set the beam, $B_\ell$, to unity. Since we generate and analyze simulations with the same \textsc{HealPix} $N_{\rm side}$ we do not include a pixel beam. The noise model is given by
\beq
    N_\ell = \Delta_T^2\,\mathrm{exp}\left(\frac{\ell(\ell+1)\theta^2_{\rm FWHM}}{8\log 2}\right),
\eeq
where we fix $\Delta_T = 1\mu\,K$-$\mathrm{arcmin}$ and $\theta_{\rm FWHM}=5\,\mathrm{arcmin}$. Though we will usually work with Gaussian fields, we also consider simulations with a synthetic bispectrum injected. These can be obtained following \citep{2011MNRAS.417....2S}, via the transformation on a GRF $a_{\ell m}$
\beq
    a_{\ell m}\to a_{\ell m} + \frac{1}{6}\mathcal{G}^{\ell\ell_2\ell_3}_{mm_2m_3}b_{\ell\ell_2\ell_3}^{\rm theory}h_{\ell_2m_2}^*h_{\ell_3m_3}^*,
\eeq
where $b_{\ell_1\ell_2\ell_3}^{\rm theory}$ is the desired bispectrum (see also \citep[Eq.\,1.3]{Shiraishi:2014roa}) \resub{and $h_{\ell m} \equiv [\mathsf{C}^{-1}a]_{\ell m}$}. Here, we will use the factorized form $b_{\ell_1\ell_2\ell_3}=\prod_{i=1}^3\beta_{\ell_i}$, where, for definitiveness, we set  $\beta_\ell=2\,\mathrm{exp}\left[(\ell-2)/40\right]$ (in $\mu\rm{K}^3$ units). Due to the factorization, this can be written
\beq
    a_{\ell m} \to a_{\ell m} + \frac{1}{6}\beta_\ell \int d\hn\,Y^*_{\ell m}(\hn)[\beta h]^*(\hn)[\beta h]^*(\hn),
\eeq
using \eqref{eq: gaunt-def} and writing $[\beta h](\hn) = \sum_{\ell m}\beta_\ell h_{\ell m}Y_{\ell m}(\hn)$, which can be evaluated as a harmonic transform.\footnote{We also subtract off the mean of the signal, to ensure that $\av{a_{\ell m}}=0$.} Finally, we will often consider windowed fields: for this, we utilize a \textit{Planck} $40\%$ Galactic sky mask, with $2\degree$ Gaussian apodization, denoted $W$;\footnote{Available at \href{http://pla.esac.esa.int/pla}{pla.esac.esa.int/pla}.} this is akin to a (highly anisotropic) window that would be used in a realistic \textit{Planck} analysis, though we pick a somewhat severe example for the sake of demonstration. The full field is given by $d(\hn)=W(\hn)a(\hn)$, an example of which is shown in Fig.\,\ref{fig: maps}.

To form the window-free estimators we require the random fields $u$ and their covariance, $\mathsf{U}$. As noted in \S\ref{subsec: SU-choice}, the estimator is unbiased for any $\mathsf{U}$, however, the Monte Carlo variance can be reduced if $\mathsf{U}^{-1}$ is close to the weighting matrix $\mathsf{S}^{-1}$. Here, we will assume a diagonal weighting, such that
\beq
    \av{u_{\ell m}u_{\ell'm'}} = (-1)^{m}\delta^{\rm K}_{\ell\ell'}\delta^{\rm K}_{m(-m')}\left[B_\ell^2C_\ell^{TT} + N_\ell\right] \equiv (-1)^{m}\delta^{\rm K}_{\ell\ell'}\delta^{\rm K}_{m(-m')}U_\ell,
\eeq
as in \eqref{eq: alm-ideal}: if the synthetic data is unwindowed, this matches the true covariance $\C$. With the definition, the action of the $\mathsf{U}^{-1}$ weighting on a map $\beta$ is given by
\beq
    \mathsf{U}^{-1}_{ij}\beta^j =\sum_{\ell m}\frac{\beta_{\ell m}}{U_{\ell}}Y_{\ell m}(\hn_i),
\eeq
which is straightforwardly computed as a harmonic transform. In practice, one does not deal with continuous maps on the two-sphere, but discrete \textsc{healpix} pixels: this affects things only by introducing a factor $A_{\rm pix}=4\pi/N_{\rm pix}$ whenever a summation over pixels is involved.\footnote{Here, we neglect discrete pixel weights, which could be included to ensure that $\mathsf{U}$ is the exact covariance of $u$. Assuming that the $\ell$ range in question is sufficiently small compared to $\ell_{\rm max}$, this approximation is justified.} Finally, we must choose a form for the $\mathsf{S}^{-1}$ optimality weighting. Here, we use a diagonal approximation (neglecting the window function, except for some rescaling, which cancels), fixing $\mathsf{S}^{-1}=\mathsf{U}^{-1}$, which we expect to be close to optimal on the scales considered herein. An example of the $\mathsf{S}^{-1}$-filtered data is shown in Fig.\,\ref{fig: maps}. As noted in \S\ref{subsec: SU-choice}, an alternative approach would be to omit the window from $\tilde a$ (and its correlators), and instead include it as a projection in $\mathsf{S}^{-1}$ (\textit{i.e.}\ treating the true map as the input, and zero-weighting bad regions).\footnote{See \citep{Millea:2020iuw} for an alternative approach that allows for invertible covariances.} This may be more appropriate for real analyses with complex window functions and inpainted maps, and will be the approach used in \citep{PhilcoxCMB}.

\subsection{Power Spectrum}

We begin by validating the power spectrum estimators of \S\ref{sec: Cl}. For this, we apply both the ideal (\S\ref{subsec: Cl-ideal}) and optimal (\S\ref{subsec: Cl-general}) estimators to a suite of GRF simulations created as described above, optionally including a \textit{Planck} mask. For this test, we focus on comparatively large scales (where the impact of the mask is largest), considering the binned power spectrum in $n_{\ell}=100$ linearly spaced bins of width $\Delta \ell=4$ from $\ell_{\rm min}=2$ to $\ell_{\rm max}=402$, though we drop the last bin in all cases to mitigate correlations of the extremal bins with their neighbors. Data are constructed using a \textsc{healpix} grid of $N_{\rm side}=256$, giving $\ell_{\rm max}^{\rm HEALPIX}=767$, far above the scales of interest here. To construct the Fisher matrices required in the optimal estimator, we use $N_{\rm fish}=100$ simulations. Computation required $\approx 30$ CPU-seconds per Fisher realization, and $\approx 0.4$ CPU-seconds for each estimator numerator (both for the ideal and optimal approaches); the optimal estimator thus required $\approx 1$ CPU-hour of additional time to compute, though we note that this is independent of the number of simulations analyzed.

To compare theory and data, we require some procedure for estimating the binned models from the unbinned spectra $C_\ell^{\rm th}$. An appropriate choice is the following:
\beq
    C^{\rm th}(b) = \left[\sum_{\ell}\Theta_{\ell}(b)(2\ell+1)\frac{C^{\rm th}_{\ell}}{S_{\ell}^2}\right] \slash \left[\sum_{\ell}\Theta_{\ell}(b)(2\ell+1)\frac{1}{S_{\ell}^2}\right],
\eeq
derived from considering the expectation of the ideal estimator. This matches the approach of \citep{Bucher:2015ura} for the bispectrum, but includes our custom weighting $S_\ell$ (or the diagonal part thereof).

\begin{figure}
    \centering
    \includegraphics[width=\textwidth]{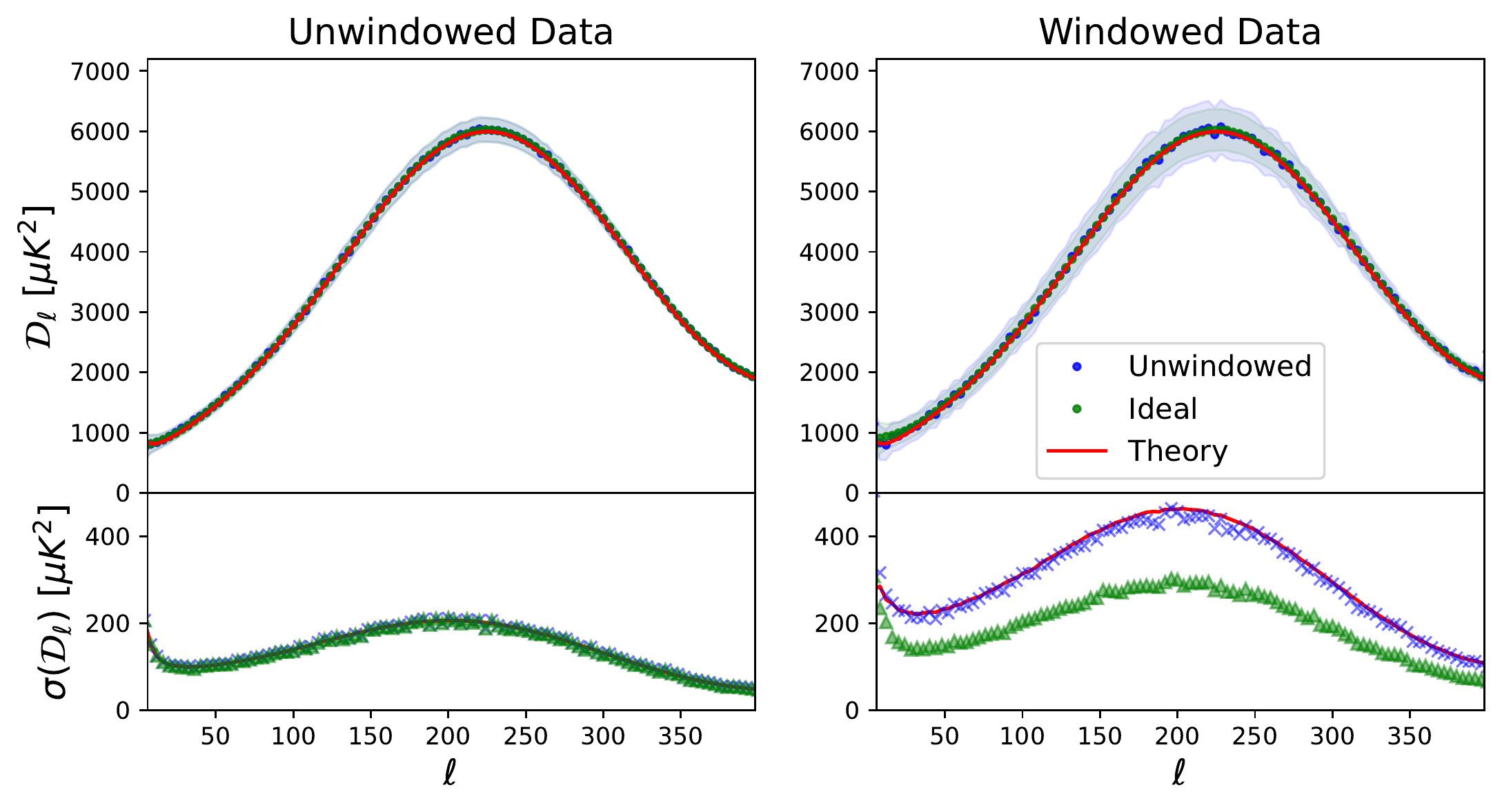}
    \caption{Comparison of binned power spectrum estimators for unwindowed (left) and windowed (right) data. In each case, we plot the binned power spectrum obtained from the ideal (\S\ref{subsec: Cl-ideal}, blue) and maximum-likelihood (\S\ref{subsec: Cl-general}, green) estimators, alongside the true theory model obtained from \textsc{class}. The top panels show the raw measurements, in bins of width $\Delta\ell=4$ (normalized by $\ell(\ell+1)/(2\pi)$), whilst the bottom panels show the errors. Data is obtained from 1000 Gaussian random field simulations, with the Fisher matrix of the optimal estimators constructed using 100 Monte Carlo realizations, using quasi-optimal weights (cf.\,\S\ref{subsec: test-practical}). In all cases the estimators appear unbiased, and the unwindowed variances are almost identical to the inverse Fisher matrix (red lines). For the windowed data, the ideal estimator appears to have lower variance than the optimal estimator: this is due to significant correlations between neighbouring bins, as seen in Fig.\,\ref{fig: Cl-cov}.}
    \label{fig: Cl-plot}
\end{figure}

Fig.\,\ref{fig: Cl-plot} shows the measured binned power spectrum from the unmasked and masked simulations alongside the true injected power spectrum, averaging over $1000$ simulations. In both cases, we find excellent agreement between data and theory, as expected. When the synthetic data does not include a window, the two estimators agree precisely; when a mask is included, the means are consistent, but the variance properties differ. In the latter case, the variance is significantly increased (by a factor of approximately $\av{W^4}/\av{W^2}^2$, due to the reduced area observed), and the ideal estimator seems to considerably outperform the optimal one. This appears paradoxical: however, it occurs since the various bins are correlated in the ideal estimator, but anti-correlated in the optimal approach (with both estimators yielding similar signal-to-noise).

\begin{figure}
    \centering
    \includegraphics[width=0.9\textwidth]{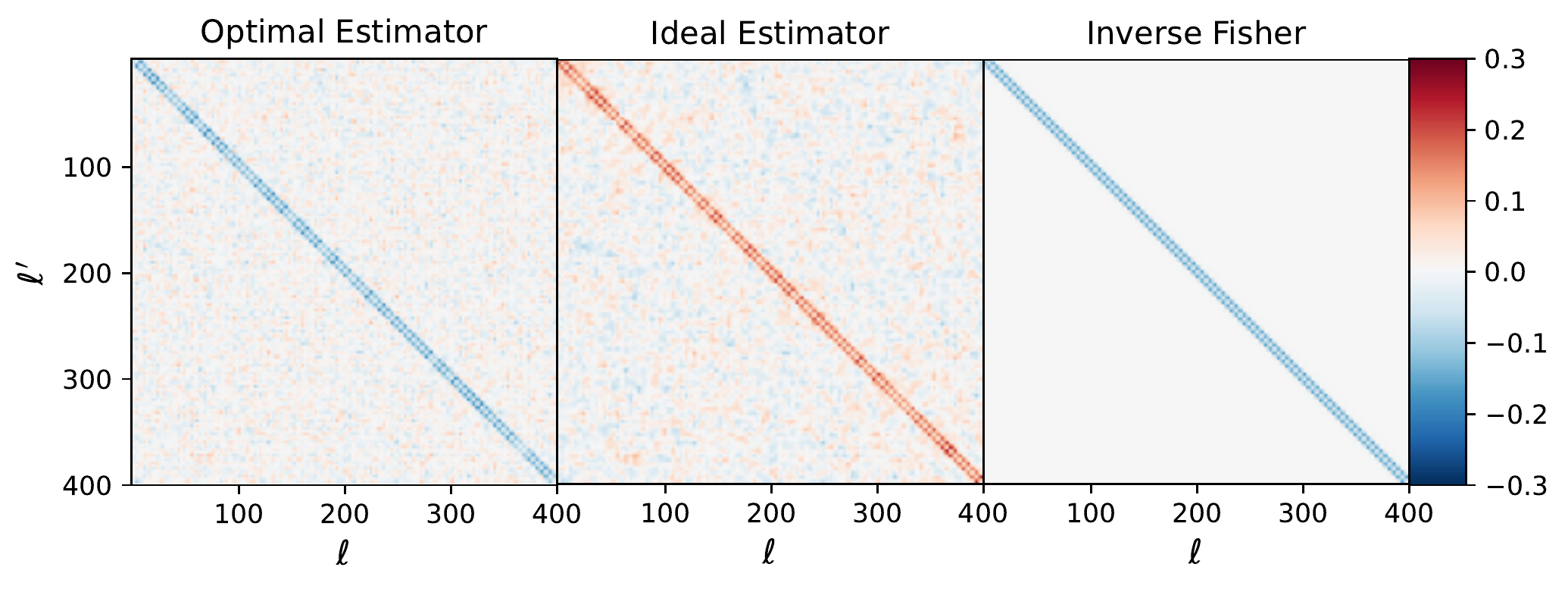}
    \caption{Correlation matrices for the ideal and unwindowed power spectrum measurements plotted in Fig.\,\ref{fig: Cl-plot} (right panel). In all cases, the underlying data contains a \textit{Planck} sky mask, which, for the conventional (ideal) estimator, gives a clear correlation between neighbouring bins. In the optimal estimator, we see an anticorrelation, which is closely matched by the inverse Fisher matrix, as expected. The correlation matrix is defined by $\mathcal{C}_{ij}/\sqrt{\mathcal{C}_{ii}\mathcal{C}_{jj}}$ for covariance $\mathcal{C}_{ij}$, and we subtract off the leading diagonal for clarity.}
    \label{fig: Cl-cov}
\end{figure}

In Fig.\,\ref{fig: Cl-cov}, we plot the correlation matrices for the two estimators applied to the windowed data-set (noting that the unwindowed case is trivially diagonal). If the optimal estimator is, as the name would suggest, optimal, its covariance should be equal to the inverse of the Fisher matrix, $\mathcal{F}_2$. From Fig.\,\ref{fig: Cl-cov} and the lower part of Fig.\,\ref{fig: Cl-plot}, this is exactly what is observed on all scales, implying that our choice of weighting, $\mathsf{S}^{-1}$, is appropriate.\footnote{In practice, we find little dependence of the power spectrum measurements on the choice of weighting scheme, which occurs since the data is mostly uncorrelated and the $\ell$-bins are narrow \citep[cf.][]{2021PhRvD.103j3504P}.} As noted above, we observe different correlation properties for the optimal and ideal estimators, with a positive correlation between neighboring bins seen in the latter case. One feature of the optimal prescription is that we can naturally form a quantity of unit variance, with no cross-correlations: $\sum_{b'}\mathcal{F}_2^{1/2}(b,b')\widehat{C}(b')$ \citep[e.g.,][]{Hamilton:1999uw}; we have verified that the optimal estimator correlation matrix of this object shows no obvious departures from the identity matrix beyond that expected from noise fluctuations. Finally, we consider the dependence on the number of Monte Carlo simulations used to define the Fisher matrix (the limiting step in the estimator). Reducing to just ten realizations ($N_{\rm fish}=10$) changes the power spectrum predictions by at most $0.25\sigma$, thus we conclude that the above choice of $N_{\rm fish}=100$ is both sufficient and conservative.

\subsection{Bispectrum}

Next, we turn to the three-point function. Here, we will consider two scenarios: (1) pure GRFs without a window function (to test optimality), and (2) simulations with an injected bispectrum and a mask (to test bias). Due to the higher dimensionality of the three point function, we consider broader (linear) bins, using $\ell_{\rm min}=2$, $\Delta\ell=10$, and $n_{\ell}=15$.\footnote{In practice, it may be preferable to use non-linearly spaced bins, such that the signal-to-noise is more evenly distributed across bins.} To avoid edge effects, we will drop any bin containing the largest $\ell$ values: this reduces the total number of elements in the data-vector from $372$ to $308$. Given the lower $\ell_{\rm max}$ used in this test, we fix $N_{\rm side}=128$, which significantly reduces run-time. Here, we require $240$ CPU-seconds to compute the Fisher matrix using a single pair of GRF realizations, and $50$ CPU-seconds to compute the estimator numerators. The runtime is dominated by the $100$ Monte Carlo simulations (used to compute the one-field term, cf.\,\S\ref{subsec: Bl-general}), but greatly reduced if one analyzes multiple datasets in series (since maps relating to the Monte Carlo simulations do not need to be recomputed). The ideal estimator (which does not include a one-field term) requires only $0.2$ CPU-seconds, though with another $35$ CPU-seconds to compute the (diagonal) normalization. In this case, one should bin the theory model in the following manner \citep[cf.,][]{Bucher:2015ura}:
\beq
    b^{\rm th}(\vec b) \propto \sum_{\ell_{123}}\Theta_{\ell_1}(b_1)\Theta_{\ell_2}(b_2)\Theta_{\ell_3}(b_3)\frac{(2\ell_1+1)(2\ell_2+1)(2\ell_3+1)}{4\pi}\tjo{\ell_1}{\ell_2}{\ell_3}^2\frac{b^{\rm th}_{\ell_1\ell_2\ell_3}}{S_{\ell_1}S_{\ell_2}S_{\ell_3}},
\eeq
where the normalization factor takes the same form but without $b^{\rm th}_{\ell_1\ell_2\ell_3}$.

\begin{figure}
    \centering
    \includegraphics[width=0.9\textwidth]{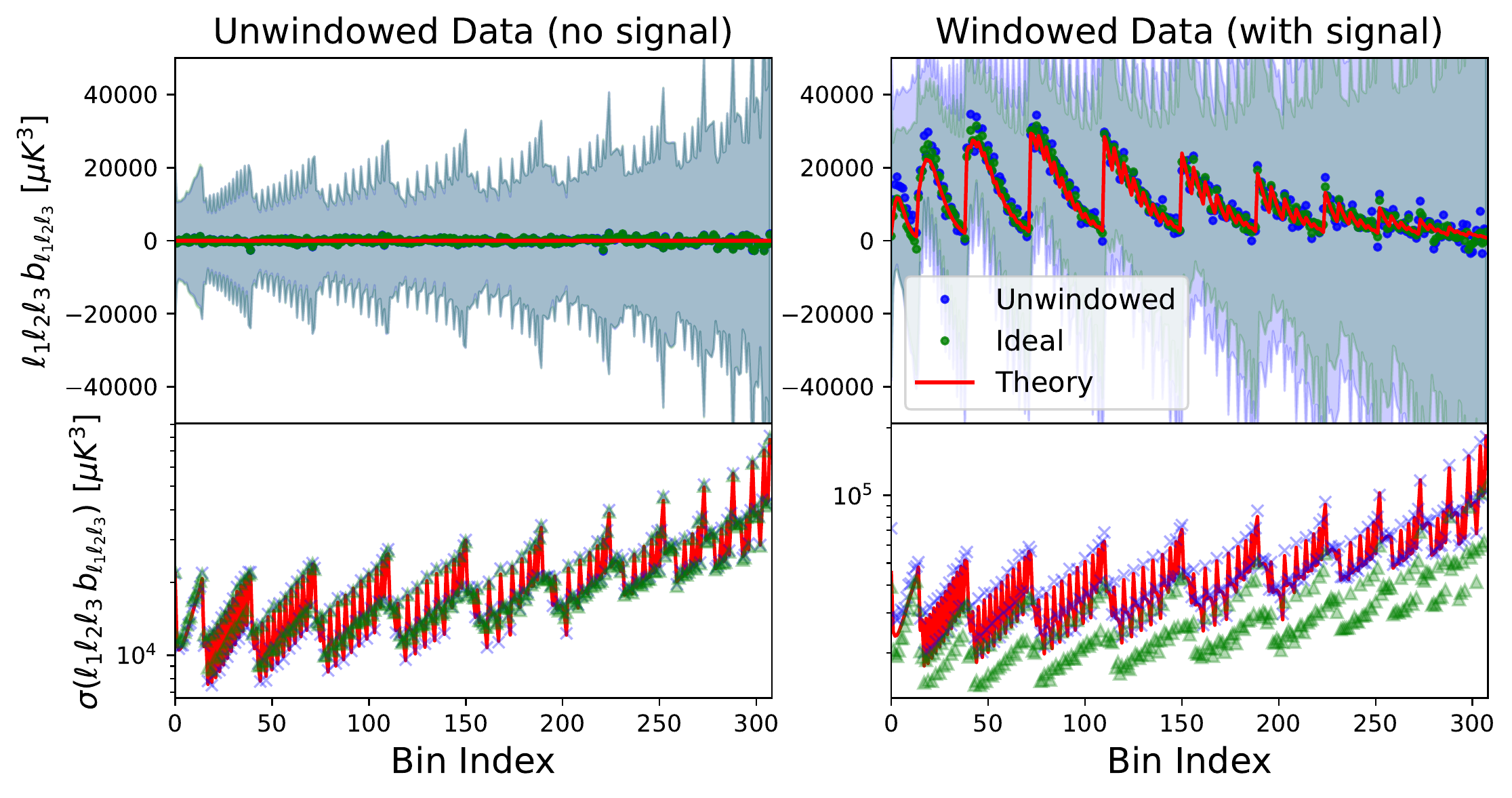}
    \caption{Comparison of binned bispectrum estimators for unwindowed zero-signal (left) and windowed non-zero-signal (right) data. The measurements are akin to those in Fig.\,\ref{fig: Cl-plot}, but use broader bins with $\Delta \ell=10$ and $\ell\in[2,142]$, and we weight the data by $\ell_1\ell_2\ell_3$, averaging over $1000$ simulations. Here, we plot all bispectrum bins $\{b_1,b_2,b_3\}$, which satisfying the triangle conditions (at the bin centers) and $b_1\leq b_2\leq b_3$. These are collapsed into one dimension for visualization, starting from the lowest $\ell$ bins on the LHS, and sequentially updating $b_3$, $b_2$, then $b_1$. We see that the estimator is unbiased in both cases, and that the variance of the optimal estimator matches its theoretical prediction (red lines), implying that it is close to minimum variance. The corresponding correlation matrix is shown in Fig.\,\ref{fig: Bl-cov}.}
    \label{fig: Bl-plot}
\end{figure}

Fig.\,\ref{fig: Bl-plot} shows our measurements of the reduced bispectrum. In the absence of a signal, we recover null detections (as expected), and find a similar (though not identical) variance between the optimal and ideal estimators, with the optimal estimator performing somewhat better on large scales. When a signal is included, we find unbiased results from both estimators, and, as before, note that the variance of the optimal estimator lies very close to the inverse Fisher matrix (and somewhat higher than the ideal estimator variance, due to bin anticorrelations). This again indicates that the optimal estimator is close to minimum variance.\footnote{Note that this is not guaranteed in this case even if $\mathsf{S}^{-1}={\tilde{\C}}^{-1}$, since the field is non-Gaussian, thus the covariance strictly contains a piece proportional to $\mathsf{B}^2$.}

\begin{figure}
    \centering
    \includegraphics[width=0.9\textwidth]{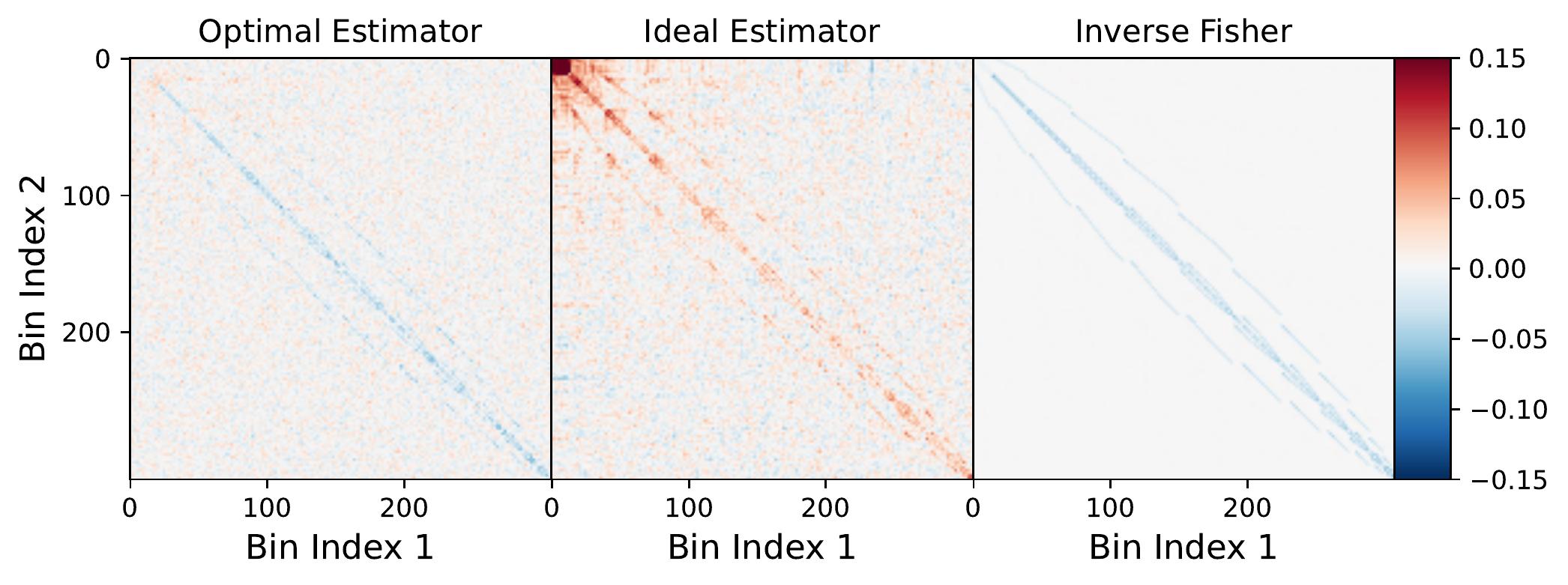}
    \caption{As Fig.\,\ref{fig: Cl-cov}, but showing the correlation of the reduced bispectrum measurements displayed in Fig.\,\ref{fig: Bl-plot}. As before, the covariance of the optimal estimator matches the inverse Fisher matrix to high accuracy; in this case, the correlation structure is more complex due to window-function induced mode coupling. The ideal estimator shows strong correlations between bins, particularly at the top left, corresponding to the lowest bins. These would need to be modeled in any analysis.}
    \label{fig: Bl-cov}
\end{figure}

The correlation matrices shown in Fig.\,\ref{fig: Bl-cov} confirm the above results. Here, the mask induces non-trivial correlations between the various bins (the size of which depend on the ratio of the bin width and the characteristic scale of the mask), particularly those with $\vec b'=\vec b\pm\{1,0,0\}$, or some permutation thereof, extending up to high $\ell$ (large scales). This structure is well captured by the Fisher matrix, and differs significantly from the (generally positive) correlations of the ideal estimator. In particular, the low-$\ell$ region shows strong correlations between a variety of bins, up to $\sim 20\%$. These may be difficult to model, and are not found in the optimal estimator, due to its particular choice of weighting scheme. As before, we find that the Fisher matrix is well-converged: reducing to $N_{\rm fish}=10$ biases the bispectrum measurements by at most $0.2\sigma$.

\begin{figure}
    \centering
    \includegraphics[width=0.5\textwidth]{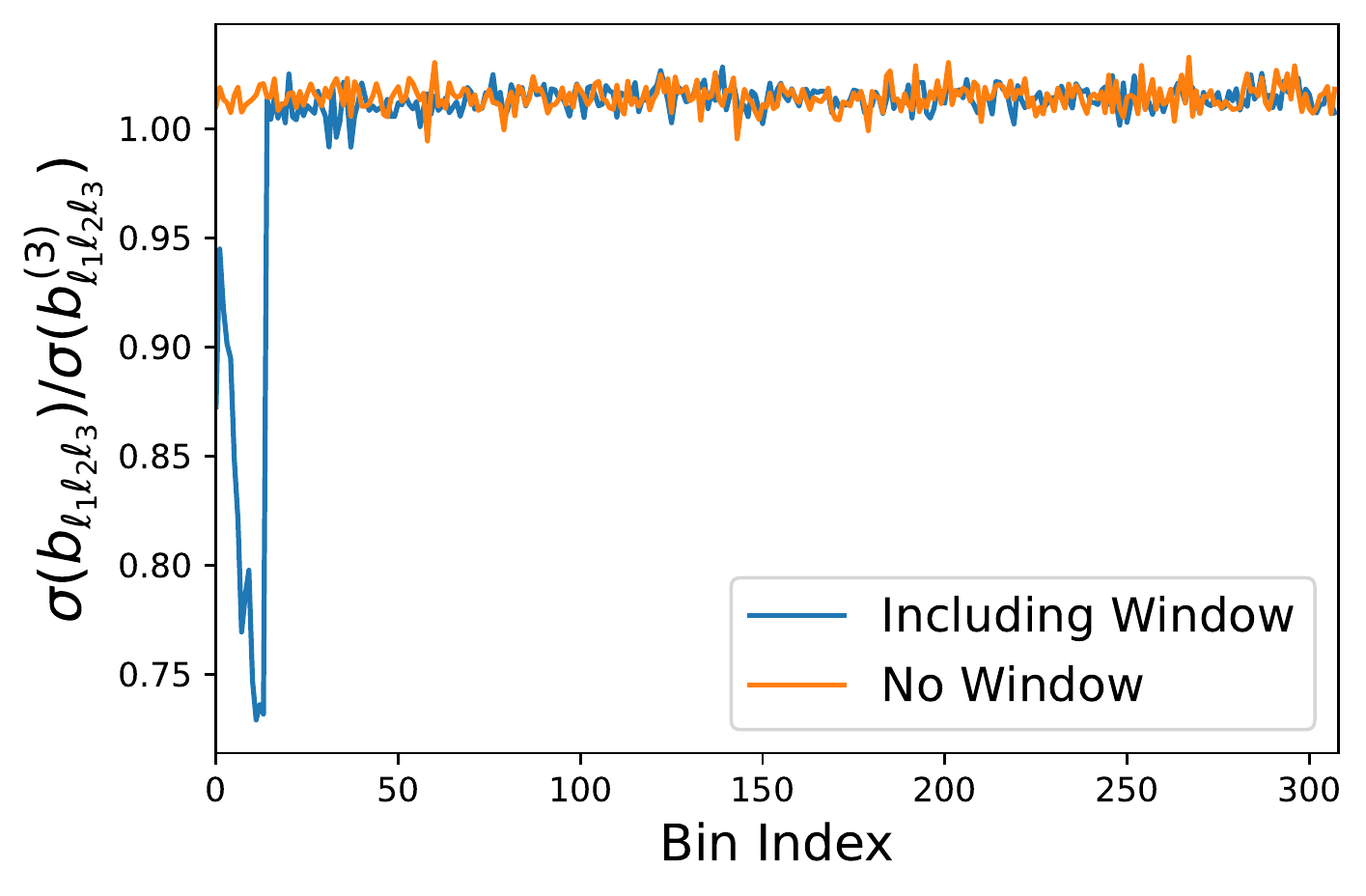}
    \caption{Impact of the linear term in the bispectrum estimators. We plot the ratio of the errorbars between bispectra estimated including and excluding the one-field term (cf.\,\ref{eq: b-opt-summary}), for realizations with (blue) and without (orange) a mask. Significant differences are observed for the first few bins of the windowed data; these correspond to configurations including modes in the lowest $\ell$ bin.}
    \label{fig: B-onefield}
\end{figure}

Finally, it is interesting to consider the impact of the linear term in the bispectrum estimator, \textit{i.e.}\ that proportional to $\av{\alpha \alpha}a$. As noted above, this term does not contribute to the mean of the signal, but can have non-trivial impacts on its covariance. In Fig.\,\ref{fig: B-onefield} we compare the estimator variance both with and without the one-field term. For the first fifteen or so bins, we find a notable reduction in the errorbar from including the linear term for windowed data, up to $\approx 25\%$, but essentially no change for the other bispectrum components, nor for unwindowed data. Noting that the affected bins are the only ones to contain the lowest $\ell$ modes (here with $\ell_1\in[2,12]$ and free $\ell_2,\ell_3$), we conclude that the linear term significantly reduces the measurement uncertainty on large scales if a window is present. This matches previous results \citep[e.g.,][]{2011MNRAS.417....2S,2012JCAP...12..032F} (noting that many non-ideal estimators include such a term) and is important if one wishes to constrain large-scale signals, such as those of primordial non-Gaussianity.

\subsection{Trispectrum}

Finally, we validate the trispectrum estimator. As discussed in \S\ref{sec: tl} the trispectrum contains two contributions (of even- and odd-parity); here we will measure both simultaneously, and, for the sake of plotting, work with the imaginary part of $t_{-}$. Generating realizations with injected non-Gaussianity is non-trivial, especially for the parity-odd terms (though see \citep{2015arXiv150200635S}), though for the parity-even terms, one may consider using lensed simulations, which include a known four-point function. However, to verify the estimators it is sufficient to check that (a) before subtraction of the disconnected terms, the parity-even estimator recovers the Gaussian expectation (\textit{i.e.}\ that of the form $C_\ell^2$), (b) after subtraction, the estimator is consistent with zero when applied to Gaussian realizations, (c) the estimator variance matches the Fisher prediction. For the parity-odd case, it is usually sufficient to restrict to comparatively large scales, since (if the underlying theory is statistically isotropic), any parity-violating trispectrum must vanish in the small-scale regime.\footnote{This occurs since, at high-$\ell$, the trispectrum is approximately plane-parallel. On $\mathbb{R}^2$, a parity flip is equivalent to a rotation in $\mathbb{R}^3$, and thus trivial if the theory is invariant under rotations. This strictly requires \textit{all} the $\ell$ modes to be small: in practice, one may wish to include parity-odd modes in the squeezed configuration (depending on the physical models of interest). These can be included by allowing for larger $\ell_2$, $\ell_4$ and (by the triangle conditions) $L$.}

\begin{figure}
    \centering
    \includegraphics[width=0.9\textwidth]{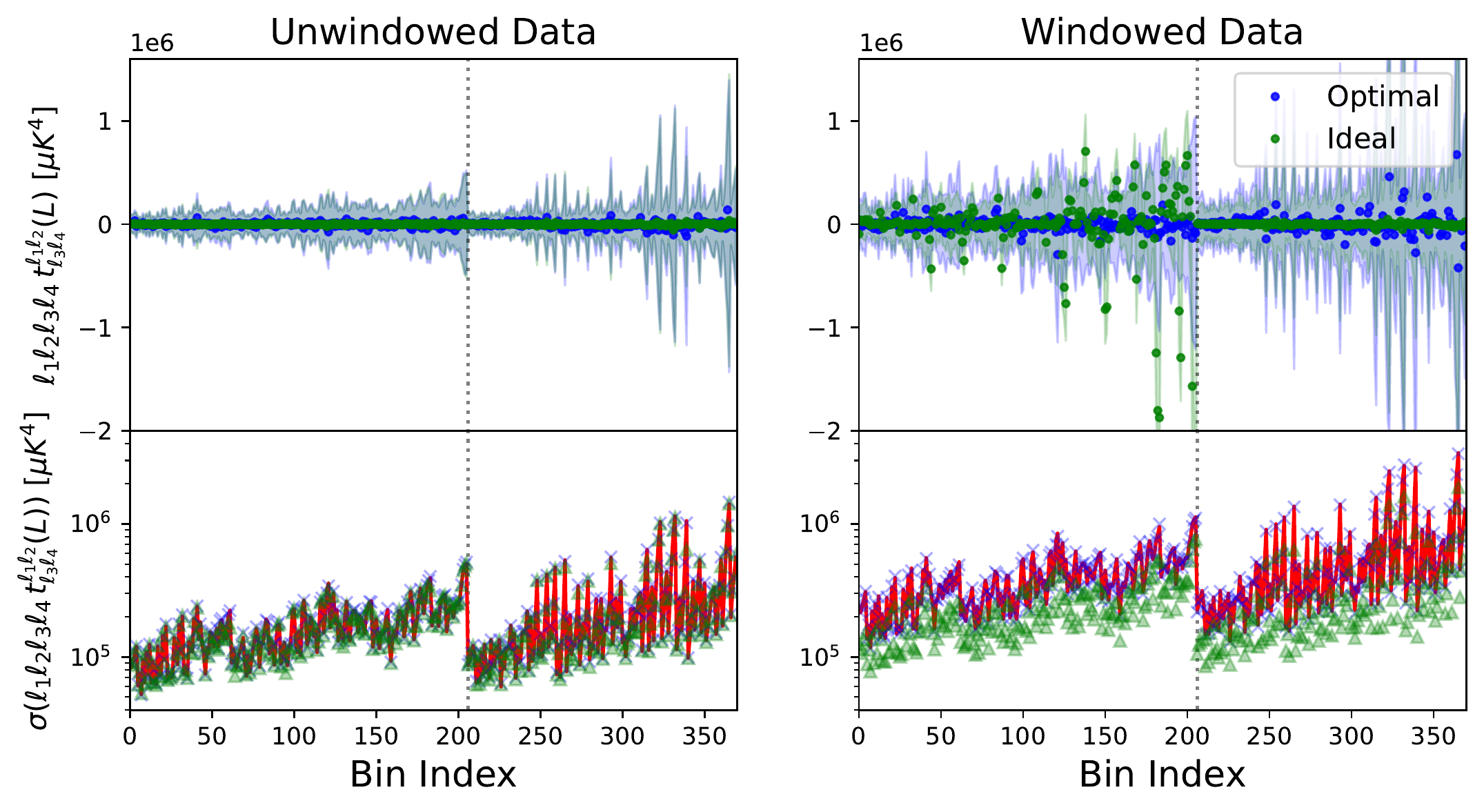}
    \caption{Comparison of binned trispectrum estimators applied to 1000 unwindowed (left) and windowed (right) Gaussian simulations. The measurements are similar to those of Fig.\,\ref{fig: Bl-plot}, except with the binning $\Delta\ell=20$, $\ell\in[2,102]$ and weighting the data by $\ell_1\ell_2\ell_3\ell_4$. Both parity-even and parity-odd measurements are plotted, demarcated by the vertical dotted lines, condensing all allowed trispectra in bins $\{b_1,b_2,b_3,b_4,B\}$ into a single dimension for visualization. As before, the characteristic $\ell$ values in the bin gradually increase in size from the left to the right. The mean of both the ideal and optimal estimators appears consistent with zero, and, for the optimal case, the variance matches the inverse Fisher matrix (red lines), as expected. Correlation matrices for the windowed field are shown in Fig.\,\ref{fig: Tl-cov}.}
    \label{fig: Tl-plot}
\end{figure}

As for the bispectrum, the trispectrum is a high-dimensional object, containing $\mathcal{O}(n_\ell^{5})$ elements (for $n_\ell$ $\ell$-bins). To keep the computation tractable, we will consider the following binning parameters (using linear bins for simplicity, noting that other choices may be more efficient): $\ell_{\rm min}=2$, $\Delta\ell=20$, $n_\ell=6$, and drop the largest $\ell$-bin to avoid edge effects. We again work at $N_{\rm side}=128$, which is appropriate for these large-scale modes. In total, we estimate 455 even-parity and 386 odd-parity configurations, which reduces to 249 and 222 when removing the final $\ell$ bin.\footnote{We recall that the parity-odd estimator vanishes if $b_1=b_3$ and $b_2=b_4$, unlike the parity-even case.}  Here, we apply our estimator to the $1000$ GRF simulations described above, with the Fisher matrix computed using $N_{\rm fish}=100$ realizations. We also utilize $100$ GRFs to compute the disconnected two- and zero-field terms in the optimal estimator. Each Fisher realization requires $40$ CPU-minutes to analyze, with the data piece taking $\approx 10$ CPU-minutes per simulation, again dominated by the Monte Carlo computations. The ideal numerator is significantly faster, since it does not involve Monte Carlo simulations, and requires only $\approx 1$ CPU-second per iteration. We caution however, that the ideal normalization is non-trivial for trispectra, due to its off-diagonal correlators and $6j$ symbols. For the binning parameters discussed above, \resub{the ideal Fisher matrix} required $24$ CPU-hours to compute (after removing the largest $\ell$ bin); this scales as $\mathcal{O}(\ell_{\rm max}^6)$, which is prohibitive for large $\ell_{\rm max}$ (unlike the optimal schemes).

\begin{figure}
    \centering
    \includegraphics[width=0.9\textwidth]{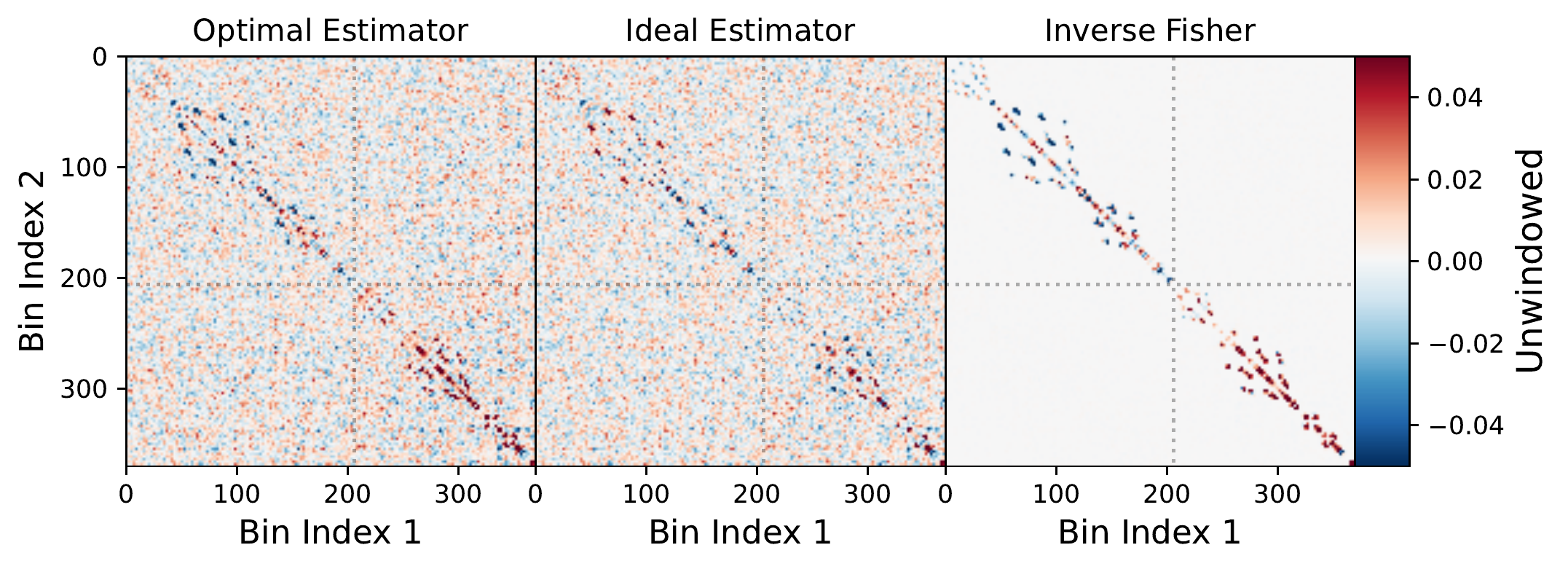}
    \includegraphics[width=0.9\textwidth]{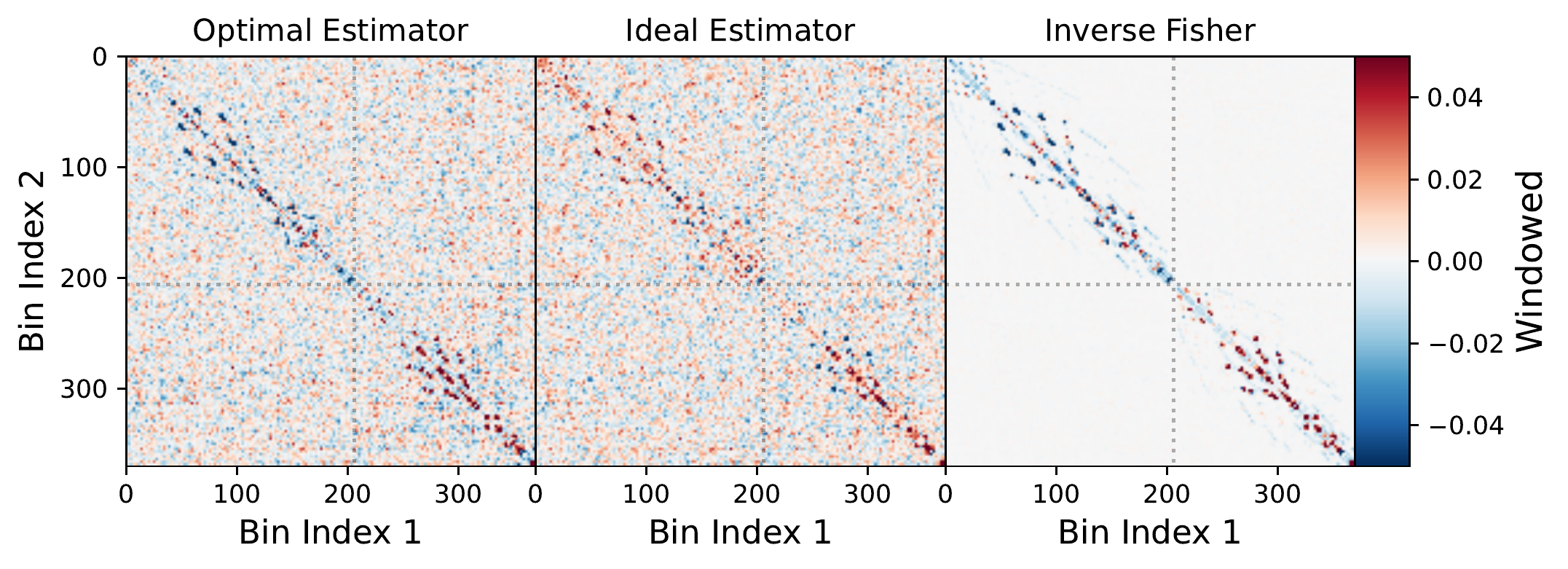}
    \caption{As Fig.\,\ref{fig: Bl-cov}, but showing the correlation of the reduced trispectrum measurements displayed in Fig.\,\ref{fig: Tl-plot}. We utilize the same binning strategy as before, with the parity-even modes shown in the top left and the parity-odd in the bottom right of each matrix. We show results both for unwindowed (top) and windowed (bottom) data, noting that the correlation structure is non-trivial in both cases,  due to degeneracies within the trispectrum definition. Regardless of the mask, the inverse Fisher matrix closely matches the optimal estimator covariance, indicating that the estimator is approaching the maximum likelihood solution. The mask is seen to induce non-negligible correlations on small scales, though we do not find significant mixing between the parity-even and parity-odd trispectra.}
    \label{fig: Tl-cov}
\end{figure}

To compare theory and observations for the trispectrum, we should bin the underlying reduced trispectrum $t^{\ell_1\ell_2,\rm th}_{\ell_3\ell_4}(L)$ in the following manner:
\beq
    t^{\rm th}(\vec b,B) &\propto& \sum_{\ell_{1234}L}\Theta_{\ell_1}(b_1)\Theta_{\ell_2}(b_2)\Theta_{\ell_3}(b_3)\Theta_{\ell_4}(b_4)\frac{(2\ell_1+1)(2\ell_2+1)(2\ell_3+1)(2\ell_4+1)(2L+1)}{(4\pi)^2}\\\nonumber
    &&\,\times\,\tj{\ell_1}{\ell_2}{L}{-1}{-1}{2}^2\tj{\ell_3}{\ell_4}{L}{-1}{-1}{2}^2\frac{t^{\ell_1\ell_2, \rm th}_{\ell_3\ell_4}(L)}{S_{\ell_1}S_{\ell_2}S_{\ell_3}S_{\ell_4}},
\eeq
with an appropriate normalization factor. This is again derived from the expectation of the idealized estimator, but we drop a $6j$ term (which mixes different $L$ and $L'$ modes), which is subdominant, and prevents efficient factorization in $\{\ell_1,\ell_2,L\}$ and $\{\ell_3,\ell_4,L\}$.

In Fig.\,\ref{fig: Tl-plot} we show the trispectrum measurements extracted from the GRF realizations. Though detailed interpretation of this plot is hampered by the statistic's high dimensionality, it is clear that both estimators return amplitudes consistent with zero (though there may be some outliers in the ideal windowed scenario, due mask-induced effects). This indicates that the subtraction of the disconnected terms is working as expected. Furthermore, the variances of the optimal estimator are consistent with those predicted by the inverse Fisher matrix, for both the parity-even and parity-odd components (with a ratio of $1.021\pm0.003$). This is shown further in Fig.\,\ref{fig: Tl-cov}, where we observe that the complex correlation structure of the Fisher matrix matches the covariance of the simulated realizations, implying that the estimator is close to optimal, and that we have used sufficient number of simulations to compute the disconnected terms. When the mask is included, the variance of the estimator increases significantly (roughly by a factor of $\av{W^8}/\av{W^2}^4$), and the correlation structure changes, seen particularly in the low-$\ell$ modes. Unlike the power spectrum and bispectrum estimators, the covariance of the unmasked fields is non-diagonal: this is in accordance with the discussion of \S\ref{sec: tl}, and is due to the labelling degeneracy, where the diagonal of the quadrilateral $ABCD$ can be placed between sides $A$ and $C$ or $B$ and $D$. Finally, we note that $N_{\rm fish}=100$ is sufficient for Fisher matrix convergence, as before; reducing to $N_{\rm fish}$ gives a (stochastic) bias of at most $0.04\sigma$. 

\begin{figure}
    \centering
    \includegraphics[width=0.48\textwidth]{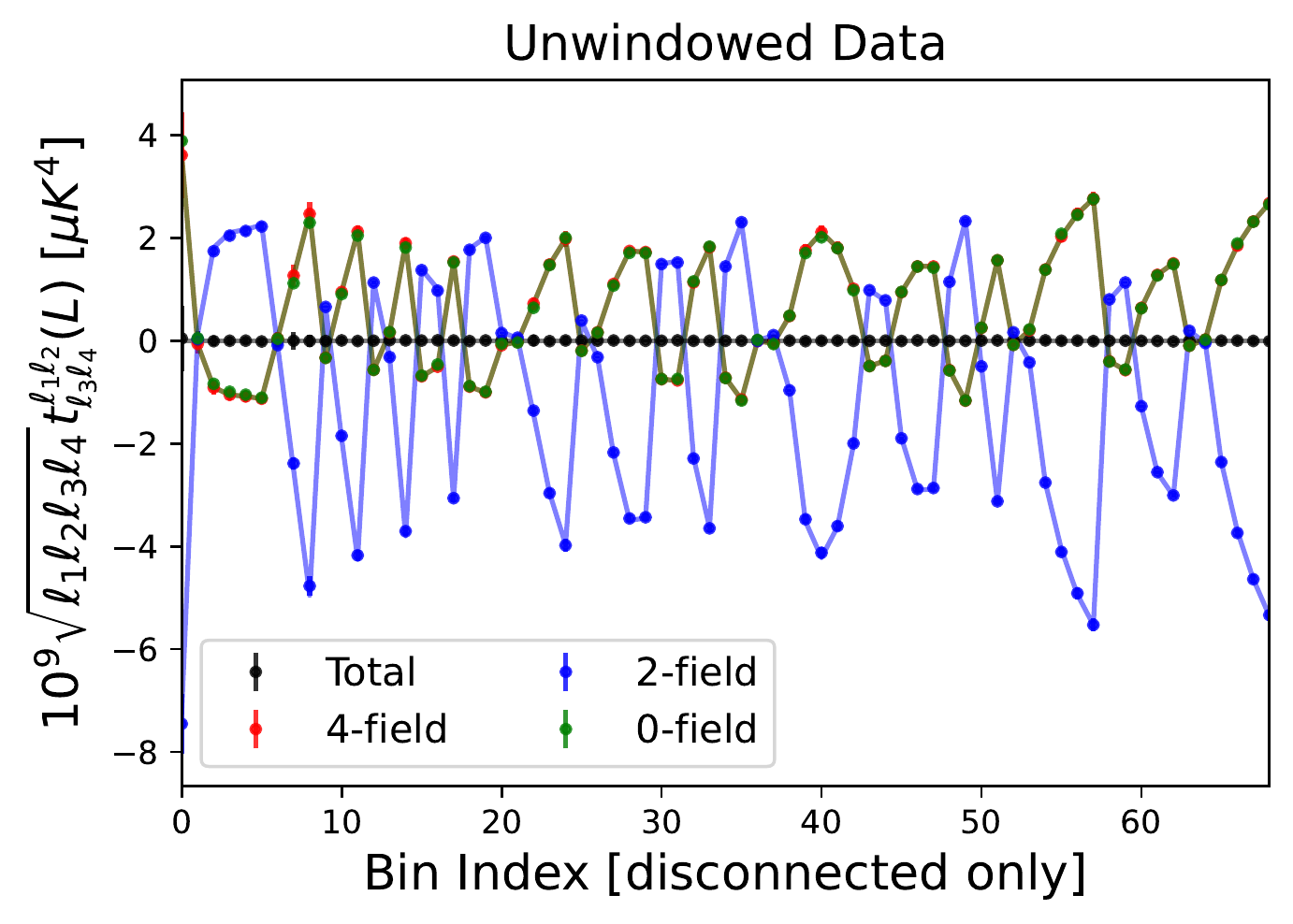}
    \includegraphics[width=0.48\textwidth]{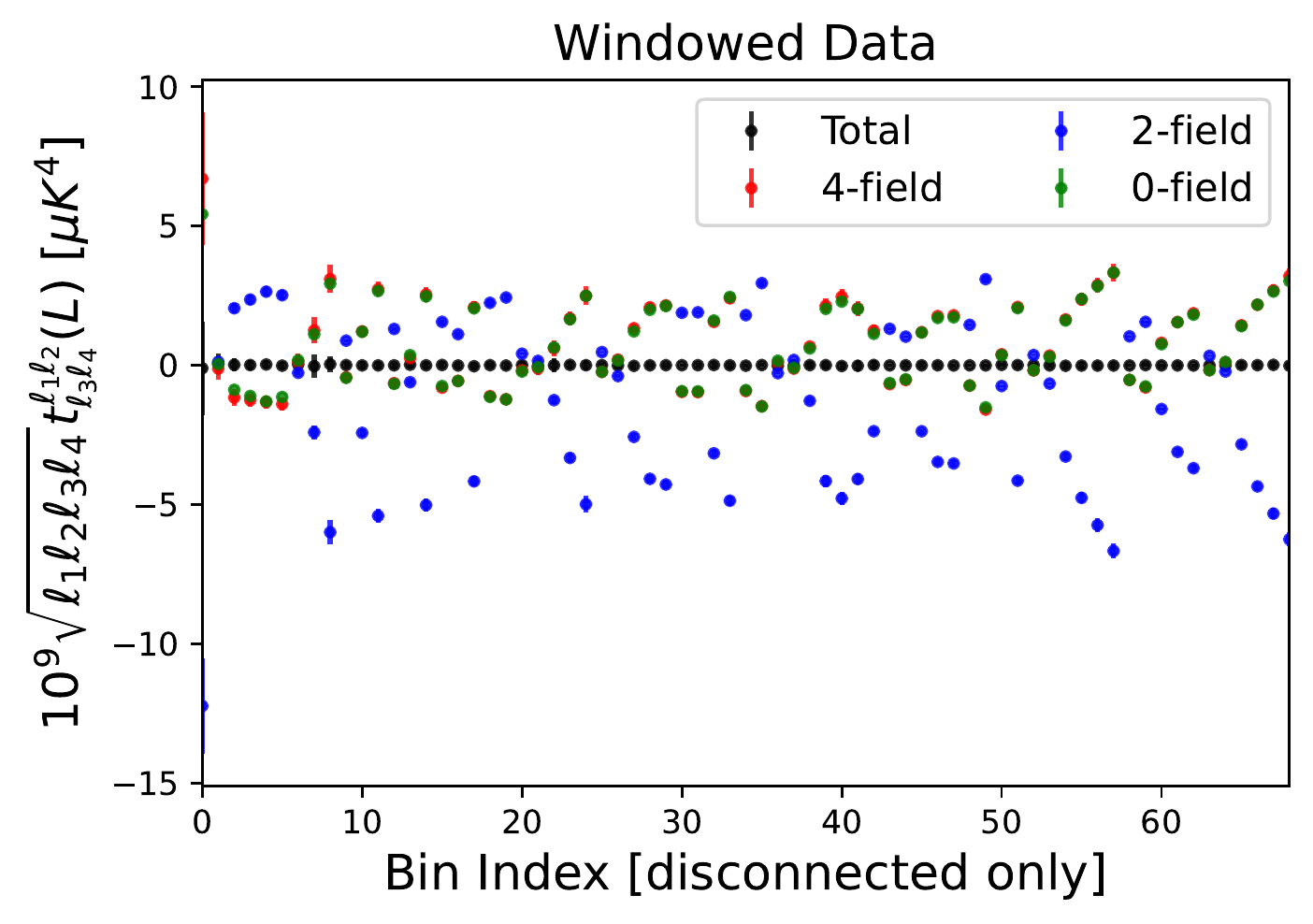}
    \caption{Comparison of the optimal trispectrum estimators, specializing to the bins with a non-trivial contribution to the disconnected (two- and zero-field) terms. The left and right panels show results for unwindowed and windowed data respectively, with colors discriminating the various terms. In each case, we plot the relevant term in the trispectrum numerator, normalized by the unwindowed Fisher matrix for visualization. For the left panel, we plot also the theoretical prediction, as given in \eqref{eq: t-disc-exp}. The disconnected terms closely match their expectations and, as expected, cancel when combined to yield a zero detection of the connected trispectrum. \resub{In most cases, the errorbars are too small to discern.}}
    \label{fig: Tl-gaussian}
\end{figure}

To check the parity-even estimator in more detail it is useful to examine the disconnected terms. In the ideal Gaussian limit, $\mathbb{E}[t_4]=-(1/2)\mathbb{E}[t_2]=\mathbb{E}[t_0]$, and the unnormalized estimators satisfy
\beq\label{eq: t-disc-exp}
	\mathbb{E}[t_0(\vec b,L)] \propto \frac{1}{\Delta_4(\vec b)}\sum_{\ell_1\ell_2}\frac{(2\ell_1+1)(2\ell_2+1)(2L+1)}{4\pi}\begin{pmatrix}\ell_1 & \ell_2 & L\\ -1 & -1 & 2\end{pmatrix}^2\frac{(-1)^{\ell_1+\ell_2+L}}{C_{\ell_1}C_{\ell_2}}\left(\delta_{b_1b_3}^{\rm K}\delta_{b_2b_4}^{\rm K}+\delta_{b_1b_4}^{\rm K}\delta_{b_2b_3}^{\rm K}\right)
\eeq
In Fig.\,\ref{fig: Tl-gaussian}, we plot the various terms entering the trispectrum numerators, for both the unwindowed and windowed estimators. In the former case, we observe excellent agreement between the disconnected pieces and \eqref{eq: t-disc-exp}, whilst for the latter (for which theoretical predictions are non-trivial), we see similar behavior as a function of scale. Crucially, whilst the disconnected terms themselves are large, their sum is negligible; this indicates that the estimators are performing as expected, and do not yield a false detection.

\begin{figure}
    \centering
    \includegraphics[width=0.95\textwidth]{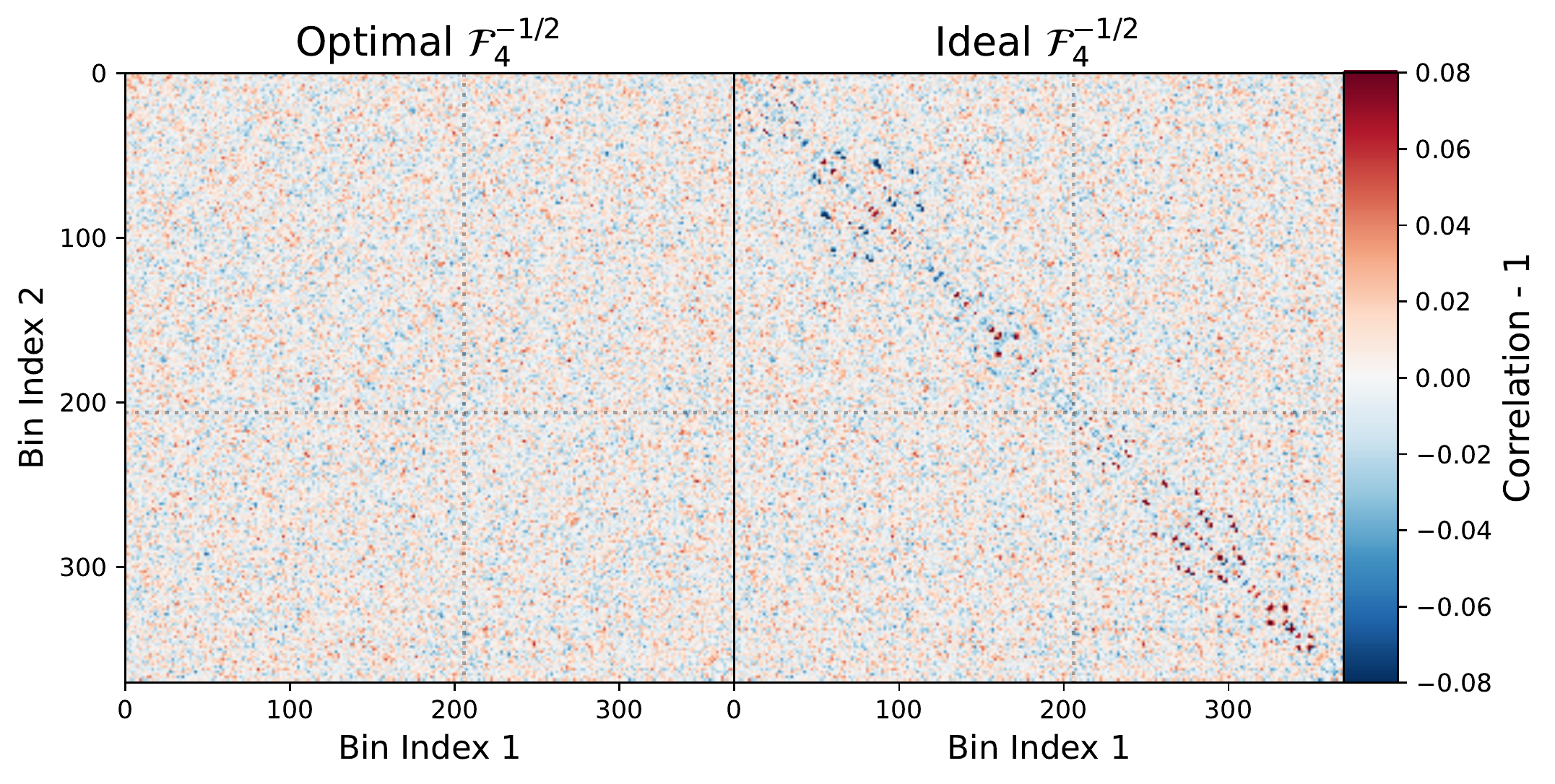}
    \caption{Correlation matrix of the windowed trispectrum dataset, weighted by the Cholesky factorization of $\mathcal{F}_4$. The left and right panels show the results using $\mathcal{F}_4$ matrices obtained from optimal and ideal estimators, subtracting the leading diagonal in each case. If the estimator is ideal, this matrix should be consistent with unity. Here, we find good results for the optimal estimators (with a variance of $0.98\pm0.04$) but clear structure in the ideal case (with a variance of $5.8\pm1.4$). This again indicates that the optimal estimators are close to minimum variance, and sources a useful projection of the data.}
    \label{fig: Tl-F-half}
\end{figure}

Our final consistency check is shown in Fig.\,\ref{fig: Tl-F-half}. Here, we consider the quantity $\widehat{t}^{\rm uncorr}\equiv \mathcal{F}_4^{-{\rm T}/2}\widehat{t}$, which, as noted in \citep{Hamilton:1999uw}, follows a unit Gaussian distribution if the estimator is optimal. In this limit, its covariance would be simply be equal to the identity matrix. From the figure, we find that, using the Fisher matrix obtained from the optimal estimator, the covariance of $\widehat{t}^{\rm uncorr}$ (on the masked dataset) is indistinguishable from a unit normal, and thus the estimator is close to optimal. If one instead uses the `ideal' Fisher matrix $\mathcal{F}_4$ (which can be obtained without Monte Carlo methods, albeit with large computational costs), we find clear structure to the correlation matrix, due to the impact of the window functions on the Fisher matrix. This decomposition also provides a useful projection scheme; for realistic scenarios (including non-Gaussian effects such as CMB lensing), the various bins of $\widehat{t}^{\rm uncorr}$ are expected to remain almost uncorrelated. 

\section{Conclusions} \label{sec: conclusion}

Through the measurement and interpretation of random processes, we can understand the physics of a wide variety of phenomena. Correlation functions, or polyspectra, are a key tool with which to do this, allowing for the rich statistics of a stochastic field to be expressed in terms of low-dimensional functions. In this work, we consider the measurement of such quantities for fields on the two-sphere, relevant to a range of disciplines including cosmology and geophysics. In particular, we derive estimators for the two-, three- and four-point correlators (power spectra, bispectra, and trispectra, respectively), and discuss how they may be efficiently applied to isotropic data. We consider two classes of estimators: `ideal' and `optimal'. The first match standard definitions in the literature, and are derived under ideal assumptions, \textit{i.e.} assuming isotropic noise without masks. In contrast, our optimal estimators defined by maximizing the theoretical likelihood for the masked data (including beams), which yields a number of useful properties. These include:
\begin{itemize}
	\item \textbf{Optimality}: Assuming that field is close to Gaussian, the variance of the optimal estimators takes its minimum value. Strictly, this is true only if the data is optimally weighted: we have additionally considered close-to-optimal weighting schemes that come close to saturating this bound in realistic scenarios.  
	\item \textbf{Bias}: The optimal estimators are unbiased, such that their expectation is equal to the true underlying statistic, regardless of the survey mask and (isotropic) beam. This allows the measurements to be directly compared to data, unlike for the ideal estimators, for which the window should be included in the theory model, requiring a complex convolution.
	\item \textbf{Separability}: Since we specialize to binned polyspectra, the estimators can be efficiently computed through a set of spherical harmonic transforms. The accompanying Fisher matrices may be estimated via Monte Carlo methods, which are shown to converge quickly. 
    \item \textbf{Computational Efficiency}: Computation of the estimator numerators involves sets of harmonic transforms scaling as $N_\ell$ (for the power spectrum and bispectrum) and $N_\ell^2$ (for the trispectrum), as well as a summation scaling as $N_{\rm bin}$, for $N_\ell$ $\ell$-bins and $N_{\rm bin}$ total bins. Similarly, the rate limiting step of the optimal Fisher matrix estimator has the scaling $\mathcal{O}(N_{\rm bin}N_{\rm fish})$, unlike na\"ive $\mathcal{O}(N_{\rm bin}^2)$ expectations, utilizing $N_{\rm fish}\sim 10-100$ simulations.
\end{itemize}

To facilitate general use, we have implemented the above estimators in a publicly available \textsc{Python} package, which has been extensively tested in \S\ref{sec: testing}. These could be used for a number of applications, including general (model-independent) non-Gaussian analyses of the cosmic microwave background (CMB) or cosmic shear. A particularly exciting prospect concerns the parity-odd trispectrum. Utilizing these estimators, we robustly measure the statistic, taking into account subtleties such as the leakage of disconnected terms and parity-even modes, and thus place the first CMB-derived constraints on scalar parity-violation in the Universe. This will be discussed in \citep{PhilcoxCMB}. Naturally, many other applications are possible.

\acknowledgments
{\small
\noindent We thank Will Coulton, Cyril Creque-Sarbinowski, Colin Hill, \resub{Masahiro Takada}, and, in particular, Adri Duivenvoorden for insightful discussions. \resub{We are additionally grateful to the anonymous referee for their careful reading of the manuscript.} OHEP is a Junior Fellow of the Simons Society of Fellows and thanks the inhabitants of Bukit Lawang for inspiration. The author is pleased to acknowledge that the work reported in this paper was substantially performed using the Princeton Research Computing resources at Princeton University, which is a consortium of groups led by the Princeton Institute for Computational Science and Engineering (PICSciE) and the Office of Information Technology's Research Computing Division.
}
%\clearpage

\appendix

\bibliographystyle{apsrev4-1}
\bibliography{refs}% Produces the bibliography via BibTeX.

\end{document}